\documentclass{article}               %     _     _     
\usepackage[utf8]{inputenc}           %  2 | |   | |   0 
\usepackage{amsmath}                  %    | |   | |  __  
\usepackage{graphicx}                 %    | |   | | / / 
\usepackage{tikz}                     %   _| |___| |/ /  
\usepackage{array}                    %  | | | | |   <     
\usepackage{longtable}                %  |___| __|_/\_\  
\usepackage{natbib}                   %      | |          
\usepackage{pgfplots}                 %      | |          
\usepackage{units}                    %      | |        
\setcitestyle{aysep={}}               %  2   |_|       3 
\usepackage{authblk}

\graphicspath{{./kuvat}} % Kuvahakemisto

\begin{document}

\title{Ensembles in Urban Large Eddy Simulations with Changing Wind Direction
}

\author{Jukka-Pekka Keskinen}
\author{Antti Hellsten}

\affil{Finnish Meteorological Institute, Helsinki, Finland}

\date{April 28, 2025}

\maketitle

\begin{abstract}

  Differences between time-averaged and ensemble-averaged wind are
  studied for the case of changing wind direction. We consider a flow
  driven by a temporally turning pressure gradient in both an
  idealized case of a staggered cube array and a realistic urban
  environment. The repeating structure of the idealized case allows us
  to construct a large ensemble of 3\,240 members with a reasonable
  compute time. The results indicate that the use of plain
  time averaging instead of an ensemble average can severely reduce
  the accuracy of both the mean and variance. These errors are the
  largest when the averaging time is of the same order as the time
  scale associated with the turning. Utilizing Taylor diagrams, we
  show that a reasonable compromise between ensemble size and accuracy
  can be achieved by calculating the ensemble statistics from
  temporally averaged results with an averaging time that is clearly
  smaller than the characteristic time scale. This allows the use of
  reasonably-sized ensembles with 10--50 members. By applying this
  approach to the realistic urban geometry, we identify building wakes
  as the regions most severely affected by the incorrectly use of
  time averaging.

\end{abstract}

\section{Introduction}

Cities and other urban areas are the primary environment for the majority of the world's population.
Wind flow through the urban landscape significantly affects the health and comfort of the residents of cities.
Wind influences urban air quality by transporting and mixing pollutants, whether emitted within the city or arriving from upstream locations.
Moreover, airflow mitigates the urban heat island effect by removing excess heat.
On the other hand, strong winds at the street level can be uncomfortable for inhabitants.
It is therefore important to understand the details of the airflow in and around urban areas.

Large eddy simulation (LES) is a numerical approach for the solution of fluid flows (see, e.g., \citet{pope_turbulent_2000}).
It is based on the idea of explicitly resolving the larger turbulent motions of the flow while applying parametrizations for the smaller scales.
These small scales, commonly referred as subgrid scales (SGS), are assumed to possess a certain level of universality and to contain only a small part of the total kinetic energy of the flow.
Usually the scale separation is based on the utilized computational grid.
LES offers a compromise between accuracy and computational costs.
It has been applied in a wide variety of different topics, ranging from clouds \citep{ahola_technical_2022} to engine parts \citep{keskinen_large_2011}.
With the continual increase in available computational power, LES is expected to become an attractive option for an even wider variety of flows.

LES has been used successfully to simulate the flow and dispersion in cities and city-like geometries for more than 15 years \citep{hanna_detailed_2006, tseng_modeling_2006, letzel_high_2008, karttunen_large-eddy_2020, giometto_effects_2017, park_impacts_2015, letzel_les_2012}.
LES is well-suited for the task, as it resolves a large part of the turbulence and can therefore be used to study the intermittent interaction between wind and buildings.
Unlike experimental approaches, LES also provides access to the three-dimensional, time-dependant flow field thereby offering a more comprehensive view of the wind field.

Due to its time-resolved nature, LES is well-suited for the simulation of statistically nonstationary flows.
In the context of urban flows, nonstationarity can enter the simulation, for example, through dynamic boundary conditions \citep{kurppa_sensitivity_2020,lin_wrf4palm_2021,park_impacts_2015}, often due to coupling with a mesoscale numerical weather prediction (NWP) model.
Such simulations are expected to become more common in the future and have been designated as a response to a Grand Challenge in environmental fluid mechanics \citep{dauxois_confronting_2021}.

In statistically nonstationary flows, time averages are not invariant under shifts in time (see, e.g., \citet{pope_turbulent_2000}) and simple time averaging is not a valid approach anymore.
This introduces additional difficulties in interpreting output data.
In earlier studies on statistically nonstationary urban flows, statistical values have been calculated primarily using short averages over periods from one minute to one hour \citep{khan_development_2021,park_impacts_2015,lin_wrf4palm_2021}, spatial averages in the horizontal direction \citep{kurppa_sensitivity_2020, lin_wrf4palm_2021}, or combinations of both \citep{park_impacts_2015,kurppa_sensitivity_2020}.
Using short time averages implies an assumption of approximate statistical stationarity during the averaging period.
Horizontal averaging, on the other hand, is strictly appropriate only if the flow is homogeneous in the horizontal direction.

To obtain reliable statistical quantities from a statistically nonstationary flow, ensemble averaging is required.
An ensemble can be created by conducting a set of statistically nonstationary simulations, each perturbed in some way, for example by starting them with different initial conditions.
However, it is important that the perturbations do not modify the underlying statistics of the simulations.
Ensemble averaging is widely used in NWP \citep{leutbecher_ensemble_2008} and climate modelling \citep{eyring_overview_2016}.
Although clearly less common in atmospheric LES, ensemble approaches have been employed previously for topics such as forest edge flow \citep{kanani_what_2014}, flow past an isolated hill \citep{chow_evaluation_2009}, and the convective boundary layer \citep{maronga_large-eddy_2013}.

The authors of this article are aware of only three other publications reporting LES ensembles in urban or urban-like environments.
In the earliest one, \citet{patnaik_large_2007} studied the dispersion of a contaminant in Downtown Los Angeles, USA, using an eight-member LES ensemble and compared the results against field measurements.
Comparison with the measurements was carried out using statistical approaches, with each experimental release considered as a single realisation of the same dispersion case.
The variation between the ensemble members appeared significant, which the authors attributed to vortex shedding by the buildings.

A larger LES ensemble of 60 members, together with corresponding wind tunnel measurements, was created by \citet{harms_validating_2011} to study the dispersion of puffs in the Central Business District of Oklahoma City, USA.
Based on the experiments, an ensemble of at least 200 members is required for statistically reliable results, depending on the distance to the release point.
\citet{harms_validating_2011} performed statistical comparisons between the experiments and the simulations based on the peak time of the puffs for two measurement locations.
The comparison yielded conflicting results, which were attributed to both strong spatial variability of the flow and to the choice of comparison metric.

More recently, \citet{li_mean_2023} considered an idealized urban flow consisting of cube array driven by a pulsatile forcing.
The periodic nature of the forcing allowed the accumulating sufficient statistics within a single simulation.
Depending on the frequency of the forcing, averages were calculated over 80--1000 forcing cycles.
Together with spatial averaging over 192 repeating units of the cube array, \citet{li_mean_2023} conducted their analysis using ensembles with 15\,360--192\,000 members.
They observed that flow unsteadiness had a significant impact on time-averaged statistics, although certain quantities, such as displacement height and shear stress, were unaffected.

Several factors may account for the scarcity of further urban LES ensemble studies, particularly those involving real urban areas.
First of all, LES in not employed operationally for forecasting, which is a clear motivation for ensembles in NWP.
Models used in NWP often also contain a large number of parametrizations and submodels, potentially amplifying simulation variability.
However, the most probable reason is the computational cost.
Resolving flow-building interactions requires a grid resolution down to the order of metres.
On the other hand, cities are large entities so that the computations have to be made with grids containing tens or hundreds of millions of grid points.
Performing tens of such simulations is a formidable task and may still not be enough as, in the case of Oklahoma City \citep{harms_validating_2011}, a sufficient ensemble size was estimated to be 200 members.
Properly converged higher-order statistics might require even more members, perhaps something similar to the 15\,360--192\,000 member ensembles used by \citet{li_mean_2023}.

It is obvious that statistically nonstationary urban flows will be studied using LES in the future, and it is expected that many of these studies will be done without a sufficiently large ensemble.
Time-averaged results can provide useful information, but it is unclear how reliable this kind of averaging is in the case of statistically nonstationary flows.
This approach may be especially problematic especially when one is considering the flow in the roughness sublayer, close to buildings and other obstacles.
It is therefore important to assess how accurately one can study urban flows, particularly within the roughness sublayer.
Furthermore, as computational resources available to LES continue to increase and ensembles become more accessible, it is important to obtain an estimate for a sufficient ensemble size for urban LES.

In this article, we study the difference between time averaging and ensemble averaging in the case of urban flows with nonstationary forcing, as well as the requirements for ensemble size.
We consider two cases: an idealized case of a staggered cube array and a real city.
In both cases, we limit the study to neutral stratification and use a rotating pressure gradient is to drive the flow.
By comparing time-averaged and ensemble-averaged quantities, we observe clear differences, especially in fluctuating velocities.
The numerical approaches, including the ensemble averaging approach, are described next in
Sect. \ref{menetel}.
The results are given then in Sect. \ref{tulokset}, followed by the conclusions in Sect. \ref{johtop}.

\section{Numerical Methods} \label{menetel}

We used the PALM model to simulate the flow past a staggered cube array and a real urban environment.
Details of the applied LES approach are given in Sect. \ref{lesl}, and a description of the model used are given in Sect. \ref{palml}.
The set-up of the flow, including the details of the cube array and the considered urban environment, is given in Sect. \ref{laskul}, while the details of the applied ensemble approach are described in Sect. \ref{parvil}.

\subsection{Large Eddy Simulation} \label{lesl}

In LES, only the large scale motions of the flow are solved.
The smaller scales, the SGS, are unsolved and their effects on the resolved are parametrized.
Formally, this scale separation is achieved by a filtering operation.
In most cases, including the present work, the filtering is performed implicitly by the computational grid.

In the case of a neutrally stratified, incompressible flow without molecular viscosity, the filtered continuity and  Navier-Stokes equations read:
\begin{align}
  \frac{\partial \overline{u}_j}{\partial x_j} & = 0, \quad \textrm{and} \\
  \frac{\partial \overline{u}_i}{\partial t} & = -
  \overline{u}_j \frac{\partial \overline{u}_i}{\partial x_j} -
  \frac{\partial \overline{p}}{\partial x_i}
  - \frac{\partial \tau_{ij}^r}{\partial x_j} - \overline{f}_i,
\end{align}
where $x_i$ is the $i$th Cartesian coordinate, $t$ is time, $u_i$ is the fluid velocity, $p$ is the modified perturbation pressure divided by density, and $f_i$ contains all body forces acting on the fluid.
In the present case, the body force $f_i$ is the pressure gradient that is used to drive the flow.
The filtered quantities are indicated using the overbar.
The lower index denotes the coordinate direction in a three-dimensional Cartesian coordinate system, and a repeated index within a term implies summation over the coordinate directions according the Einstein summation convention.
The effects of SGS on the resolved scales are represented through the anisotropic residual stress tensor $\tau_{ij}^r$.
The isotropic part of the residual stress tensor is included in the modified perturbation pressure.

The SGS can be parametrized using a variety of approaches.
In atmospheric LES, Deardorff's 1.5-order parametrisation \citep{deardorff_stratocumulus-capped_1980,sagaut_large_2006} is a common choice and we apply it according to the implementation of \citet{maronga_parallelized_2015}.
The parametrisation is of eddy-viscosity type where
\begin{equation}
  \tau_{ij}^r = -K_m\left( \frac{\partial \overline{u}_i}{\partial x_j}
  + \frac{\partial \overline{u}_j}{\partial{ x_i}} \right)
\end{equation}
and the SGS eddy diffusivity is given as
\begin{equation}
  K_m = c_m l \sqrt{e}
\end{equation}
where $c_m$ is a model constant, here set as $0.1$, the SGS mixing length $l$, and the SGS turbulent kinetic energy (TKE) $e$.
In the case of neutral stratification $l=\min(1.8z,\Delta)$, where $z$ is the distance from the bottom wall.
Here, $\Delta = \sqrt[3]{\Delta x_1 \Delta x_2 \Delta x_3}$ characterizes the local grid spacing using the grid spacings $\Delta x_i$ in all coordinate directions.
The SGS TKE is obtained by solving an additional transport equation:
\begin{equation}
  \frac{\partial e}{\partial t} = - \overline{u}_j \frac{\partial e}{\partial x_j}
  - \tau^r_{ij} \frac{\partial \overline{u}_i}{\partial x_j}
  - \frac{\partial \gamma_j}{\partial x_j}  - \epsilon,
\end{equation}
where the turbulent diffusion term is parametrized using the gradient hypothesis
\begin{equation}
  \gamma_j = -2 K_m \frac{\partial e}{\partial x_j}
\end{equation}
and the turbulent dissipation term is given by
\begin{equation}
  \epsilon = \left( 0.19 + 0.74 \frac{l}{\Delta} \right)
  \frac{e^{3/2}}{l}.
\end{equation}

In the current simulations, the effects of the Coriolis force and the buoyancy have been neglected.
These simplifications were introduced to focus the study solely to the effects of changing wind direction.
The reduction in realism is hence compensated by better control of the studied phenomena.
Similar simplifications have been made earlier in various studies concerning both simplified \citep{li_mean_2023, tseng_modeling_2006, letzel_high_2008} and realistic urban configurations \citep{hanna_detailed_2006,  tseng_modeling_2006, karttunen_large-eddy_2020,  giometto_effects_2017, letzel_les_2012}.
The effects of changing forcing under non-neutral stratification, together with the Coriolis force, are beyond the scope of the present paper, and should be addressed in future studies.

\subsection{Computational Set-Up} \label{palml}

In this study, we use the PALM model system, version 6.0 \citep{maronga_overview_2020,maronga_parallelized_2015}.
It is a highly parallelisable code that has been utilized for over 20 years, mainly for the simulation of various atmospheric boundary layer flows.
The model can be used to perform both LES and Reynolds-Averaged Navier-Stokes simulations.
In order to consider the effects of buoyancy, either the Boussinesq or the anelastic approximations can be applied.
PALM contains a variety of model components allowing consideration of effects due to e.g. radiation and microphysics.

We use PALM to carry out LES and to solve the filtered, incompressible Navier-Stokes equations at an infinitely high Reynolds number.
The equations are solved using finite differences with a fifth-order upwind-biased numerical scheme by \citet{wicker_time-splitting_2002} for the
advective terms.
A third-order Runge-Kutta method by \citet{williamson_low-storage_1980} is used for time advancement.
To ensure the incompressibility of the flow field, a predictor-corrector method is used in PALM.
Divergences in the initial flow field are attributed solely to the pressure term.
The Poisson equation for the modified perturbation pressure is solved after each Runge-Kutta step, using an iterative multi-grid scheme (see, e.g., \citet{patrinos_numerical_1977, hackbusch_MG_1985}).
PALM is parallelized using MPI \citep{gropp_MPI_1999} and utilizes a two-dimensional domain decomposition in the horizontal directions.

The computations are carried out on the Puhti supercomputer provided by CSC – IT Center for Science Ltd., Finland.
Puhti is an Atos Bullsequana X400 platform where each node contains two Intel Xeon processor (Cascade Lake), each with 20 cores at \unit[2.1]{GHz}. 

\subsection{Case Set-Up} \label{laskul}

We have simulated two different flows.
The first is a flow past a staggered cube array and the second is a flow past a realistic urban environment.
The cube array is relatively inexpensive to simulate, which allows us to obtain a large number of ensemble members.
Although the cube array can be viewed as an idealized urban environment, cities rarely exhibit such a high level of regularity.
For this reason we also study the flow past the city of Turku, Finland.
In both cases, the viscous stresses are neglected, and hence the corresponding Reynolds numbers are infinite.
We have chosen to limit the study to the specific case of pure neutral stratification.
As the nonstationary forcing, we consider a turning pressure gradient.

\subsubsection{Staggered Cube Array}

\begin{figure}[tp]
  \includegraphics[width=\textwidth]{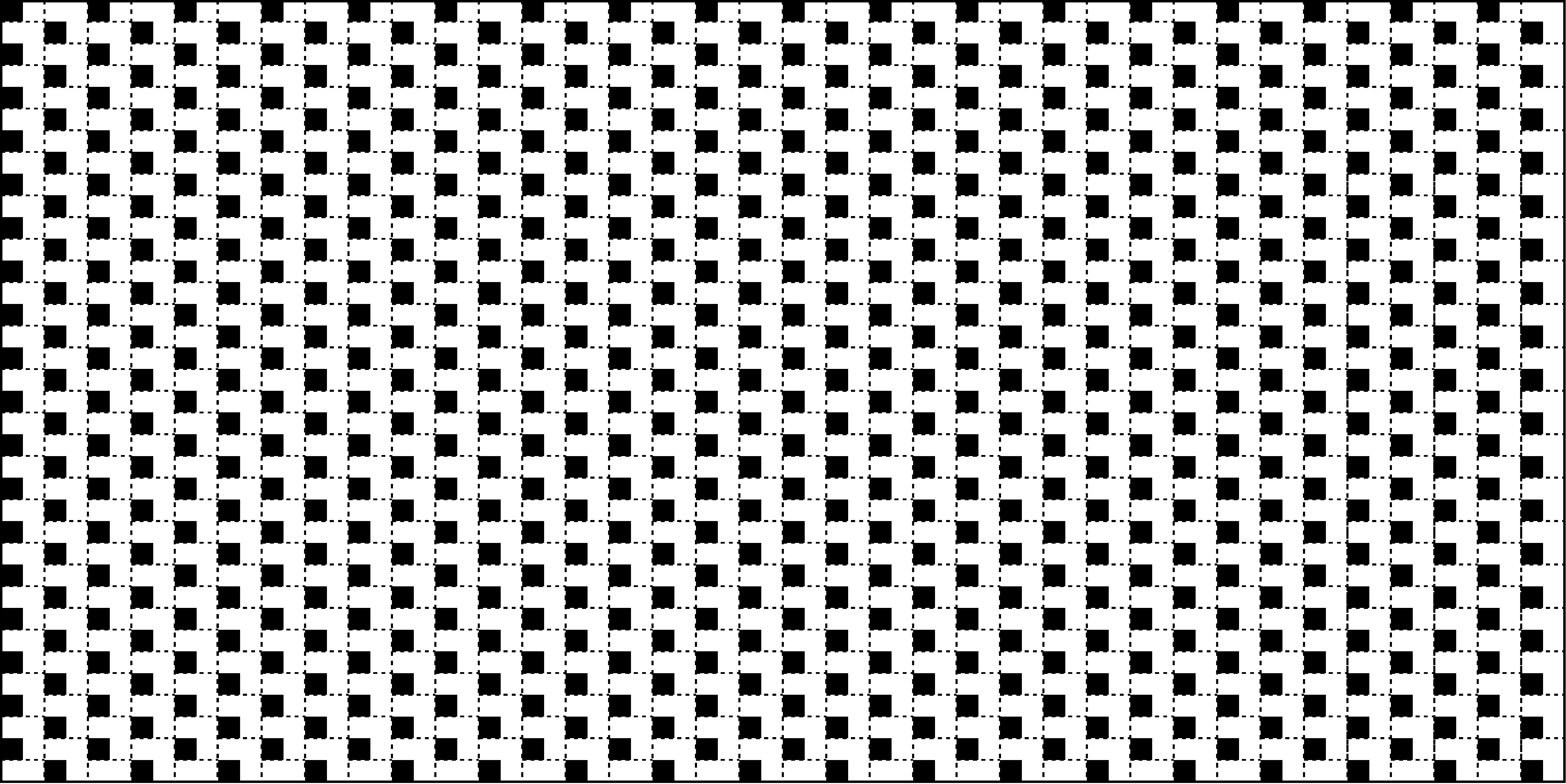} 
  \caption{\small Staggered cube array as viewed from top.
    All cubes are of the same size, and with equal separation.
    The repeating pattern, consisting of a cube and its adjoining empty space, is used for calculating averages and is indicated for each cube using a dashed line.
    In every second column, the repeating pattern is wrapped from top to bottom due to cyclic boundary conditions.
    The full extent of the computational domain is shown with a solid line}
  \label{kuutiokuva}
\end{figure}

We investigate the flow past an array of 648 cubes of edge length $h$, arranged in a staggered formation along the $x$ direction and aligned in the $y$ direction.
The distance between the cubes in both the $x$ and $y$ directions is $h$.
The arrangement in the horizontal ($xy$) plane is shown in Fig.\,\ref{kuutiokuva}.
It contains a repeating pattern with one cube and adjoining empty space, denoted in the figure
with a dashed line.
The size of the domain is $72h$ in the $x$ direction, which is also the main flow direction.
In the $y$ direction the domain spans $36h$ and in the vertical ($z$) direction the domain is $6h$.
The size of the computational domain was chosen according to the recommendations by \citet{munters_shifted_2016}.
The array can be characterized using the plan area index $\lambda_p$, defined as the horizontal area of the cubes divided by the horizontal area of the domain, and the frontal area index $\lambda_f$, defined as the windward area of the cubes divided by the horizontal area of the domain.
In the current case of a regular array of cubes, we have $\lambda_p = \lambda_f = \tfrac{1}{4}$ with no variation with the wind direction.
Flows with similar staggered cube arrays have been studied extensively in the past, both numerically (e.g. \citet{leonardi_channel_2010, cheng_adjustment_2015,ahn_statistics_2013}) and experimentally (e.g. \citet{blackman_effect_2015,hagishima_aerodynamic_2009}).

The simulations on the staggered cube array were carried out using an equidistant mesh with 1\,152 grid points in the $x$ direction, 578 grid points in the $y$ direction, and 96 grid points in the $z$ direction.
This results in a total of approximately 64 million grid points.
The mesh resolves each obstacle with 16 grid points per direction, conforming to the recommendations of \citet{xie_les_2006}.
We have set $h=\unit[32]{m}$ and hence the grid spacing is 2 metres.

All lateral boundaries are treated as cyclic.
However, during initial tests with a smaller cube array, we observed very large streamwise structures in the flow when the pressure gradient was directed along the $x$ axis.
For this reason, we have applied a shift of one repeating unit in the $y$ direction between the $yz$ domain boundaries in order to weaken these structures, following  \citet{munters_shifted_2016}.
The lower boundary is applied using PALM's rough-wall boundary condition with the roughness length set to zero.
Since the near-wall resolution is the same as elsewhere in the domain, and because we simulate a flow without viscosity, typical smooth-wall phenomena will not be resolved correctly.
However, as the focus of this paper is in the effects of the roughness elements, we do not consider the under-resolved smooth-wall features to be a concern.
At the upper boundary, the zero-gradient boundary condition was applied.
The flow is driven by a pressure gradient of constant magnitude, $\unit[0.0008]{m^2s^{-2}}$.

First, a spinup simulation was carried out, during which the direction of the pressure gradient was kept constant at \unit[240]{$^\circ$}.
Turbulence was allowed to develop for 18 hours, after which the flow was considered fully developed.
Simulations with a turning pressure gradient were branched off from the spinup.
The pressure gradient was rotated at a constant rate, $\Omega = \unit[15]{^\circ h^{-1}} \approx \unit[7.3 \times 10^{-5}]{s^{-1}}$.
Thus, at the end of the four hour simulation, the pressure gradient was directed at \unit[300]{$^\circ$}.

The flow with a turning pressure gradient is characterized by the (modified) Rossby number $Ro = U_b/(H\Omega) \approx 210$, where $H=\unit[194]{m}$ is the domain height, and $U_b = \unit[2.9]{ms^{-1}}$ is the bulk velocity of the flow.
The Rossby number can be interpreted as the ratio of the time scales of the turning speed of the pressure gradient $T_\Omega = \Omega^{-1} \approx \unit[230]{min}$ and that of the bulk flow $T_b = H/U_b \approx \unit[1.1]{min}$.
To generalize the results, nondimensionalized quantities will be used in the remainder of the paper when discussing the staggered cube array.

\subsubsection{Realistic Urban Environment}

In addition to the cube array, we study also a flow past a real urban environment.
The considered urban area is the city of Turku, located in southwestern Finland, on the coast of the Baltic Sea.
The terrain in Turku is relatively flat: its height ranges from zero at sea level to approximately 55 meters.
We concentrate on a residential area characterized by low-rise buildings, houses, and trees, located just outside central Turku.
The same area has been studied earlier with a similar computational set-up by \citet{ITM39} for the dispersion
of a passive scalar.

\begin{figure}[tp]
\includegraphics[trim=0 0 0 0,width=\textwidth,clip]{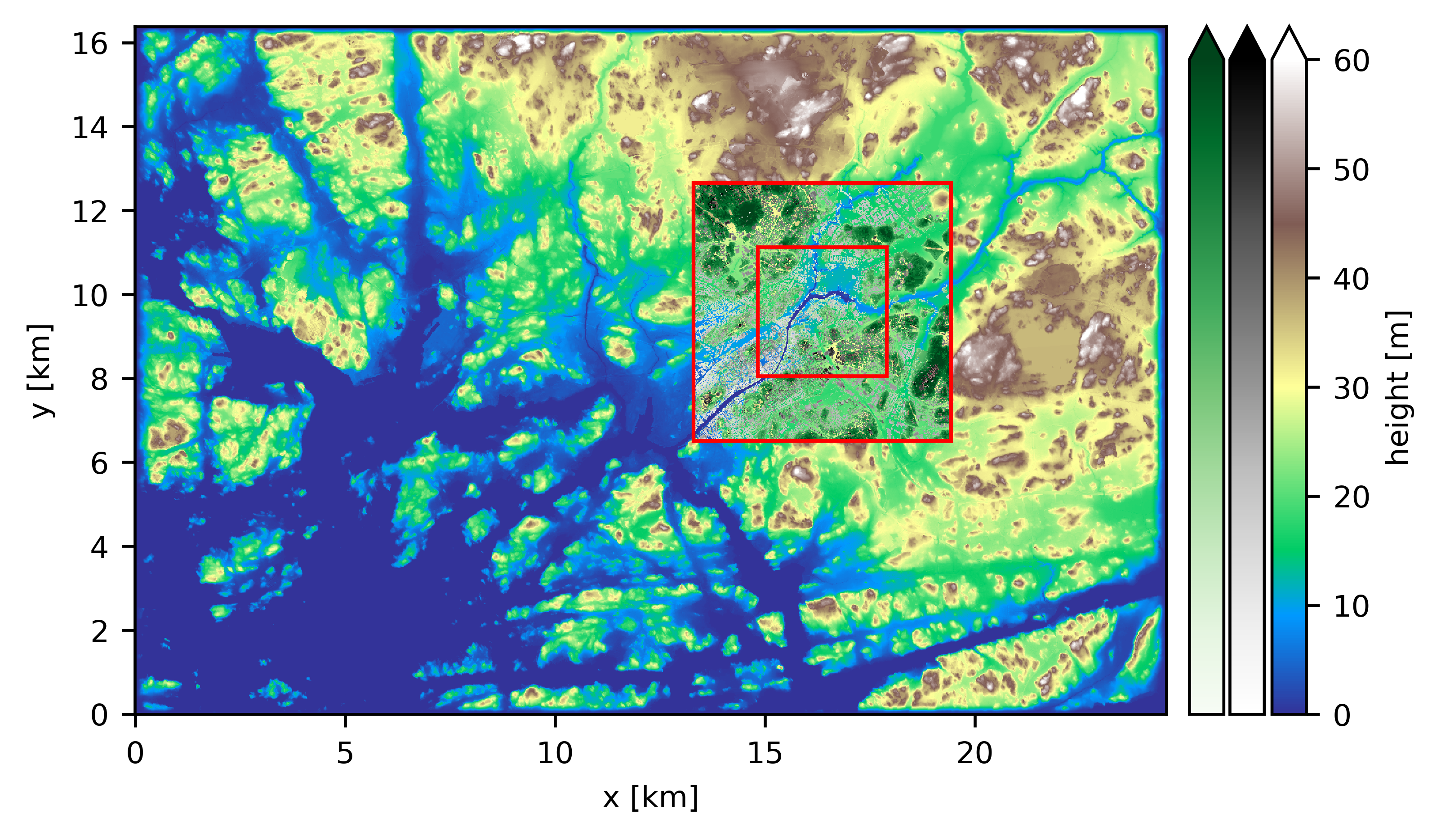}
\caption{\small Topography height in the simulated area.
  The green colour bar indicates areas with vegetation, the gray colour bar the areas with buildings, and the multi-coloured bar indicates terrain only.
  The outlines of the intermediate and the innermost domain are indicated with red.
  Buildings and vegetation are not included in the largest domain.
  North is up.
  The coordinates indicate the distance to local origin at 60°22'27"N 21°59'35"E.}
\label{kartta}       % Give a unique label
\end{figure}

For the most part, our simulations are set up using height information from existing LIDAR data, provided by the city of Turku.
These data provide separate height information for terrain, buildings, and vegetation.
For areas extending beyond the borders of the city of Turku, the digital terrain model of the National Land Survey of Finland was used.
We also added two buildings that were constructed after the collection of the LIDAR data using vector files provided by the city of Turku.
An overview of the simulated area and its topographical height is shown in Fig.\,\ref{kartta}. 

Our simulations are carried out using three, two-way coupled computational domains that are nested within each other and with increasing resolution using PALM's nesting feature \citep{hellsten_nested_2021}.
The largest domain spans approximately $\unit[25]{km} \times \unit[16]{km} \times \unit[670]{m}$, the intermediate domain $\unit[6]{km} \times \unit[6]{km} \times \unit[260]{m}$, and the smallest domain $\unit[3]{km} \times \unit[3]{km} \times \unit[190]{m}$.
Figure\,\ref{kartta} shows the full extent of the largest domain with the intermediate and innermost domains being highlighted with red squares.
The largest domain has $1\,536 \times 1\,024$ grid points in the horizontal directions with grid spacing of 16 metres.
The two inner domains have $768 \times 768$ grid points each, with a horizontal spacing of 8 metres in the intermediate domain and 4 metres in the innermost domain.
In the vertical direction, the largest domain has grid spacing of 8 metres up to 266 metres.
Beyond this height, the grid spacing is stretched by a factor of 1.03, up to a maximum spacing of 16 meters.
There are 64 grid points in total in the vertical direction in the largest domain.
The intermediate and the innermost domain have 64 and 96 equally spaced grid points in the vertical direction with grid spacings of 4 metres and 2 metres.

Our main interest is in the innermost domain.
For this reason, simplifying modelling choices have been made in the description of the outermost domain.
Firstly, only the orographical features are accounted for in the outermost domain i.e. buildings and vegetation are not considered.
Secondly, we apply cyclic boundary conditions on the lateral boundaries.
In order to facilitate this, the orography has been forced to zero at the boundaries using an artificial slope.
Thirdly, an irregular array of elongated porous objects have been placed on the uppermost part of the domain in order to break up unphysical persistent large-scale flow structures maintained by the cyclic boundary conditions and thus to reduce the time required to achieve a statistically steady flow.
The outermost domain has been limited in the vertical direction to contain the lower range of boundary-layer heights.

The sizes and relative positions of the domains have been chosen so that they mitigate the effects of the simplifying modelling choices above.
The use of cyclic boundary conditions is a departure from reality and, in the current case, it introduces a sudden transition from inland environments with higher terrain to coastal environments with low terrain.
This change is very clear in the $x$ (east-west) direction.
The placement of the intermediate domain was chosen so that there's a clear separation between the borders of the intermediate domain and the cyclic boundaries of the outermost domain.
Because the outermost domain is elongated in the $x$ direction, southwesterly and westerly winds have ample time to adjust for the terrain shape before encountering the intermediate domain.

In the two inner domains, a higher level of detail is applied.
Solid obstacles, such as orography and buildings, are resolved using the masking method as described by \citet{maronga_parallelized_2015}.
Trees and other vegetation are treated as porous objects (momentum sinks).
The inner domains obtain their lateral and top boundary conditions from the larger domain they are nested in.
The two-way coupling of the nested domains is active at height above 150 meters.
The innermost domain is centred within the intermediate domain, with a distance of approximately \unit[1.5]{km} between their respective borders.
This is more than two times the height of the outermost domain and the implied boundary layer height, which should be a sufficient distance for the flow to adjust to both the presence of the buildings and vegetation, as well as to the higher resolution.
In this way, the incoming flow and boundary conditions for the innermost domain should realistically represent the typical near-neutral conditions for the chosen location.

On the lower boundary, the rough-wall boundary condition based on the Monin–Obukhov similarity theory is used \citep{maronga_overview_2020}, while the slip condition is applied on the upper boundary.
In the same manner as with the cube array, the flow was driven using a pressure gradient with a constant magnitude of $5.0 \times \unit[10^{-3}]{m^2 s^{-2}}$ and directed at \unit[240]{$^\circ$}.
The body force is applied in the upper parts of the outermost domain, above 350 metres.
The resulting flow has a bulk velocity $u_b \approx \unit[10.5]{m s^{-1}}$, calculated using the mean horizontal wind speed in the largest domain.
We performed a spinup simulation with a stationary pressure gradient and an ensemble of turning pressure gradient simulations.
In the latter case, we matched the conditions of the cube array, using the same turning speed: $\Omega = \unit[15]{^\circ h^{-1}} \approx \unit[7.3 \times 10^{-5}]{1 s^{-1}}$.
The temporal extent of the simulation was likewise four hours, with the pressure gradient pointing at \unit[300]{$^\circ$} at the end of the simulations.

The nondimensional number characterising the flow with a turning pressure gradient is the (modified) Rossby number $Ro =U_b/(H\Omega) \approx 210$, calculated using values for the largest domain as this is where the turning pressure gradient is applied.
The time scales of the turning pressure gradient $T_\Omega = 1/\Omega \approx \unit[230]{min}$ and the bulk flow $T_b = H/U_b \approx \unit[1.1]{min}$ were identical to those in the cube array.
Results are presented using nondimensional time and velocity, but lengths are given in dimensional units to preserve the spatial correspondence to the real urban area of Turku, Finland.

The simulations are started by initialising a laminar flow state with an approximately realistic wind profile in the largest domain.
The flow is then allowed to adjust and develop large-scale turbulence for $t/T_\Omega = 2.10$ without the smaller domains.
After this first spinup simulation, we perform a second spinup of the same duration with two-way coupling enabled in order to properly initialize the inner domains with fully developed turbulence.
The final simulations are then carried out for $t/T_\Omega = 1.05$ with the turning pressure gradient, branching off from the spinup.

\subsection{Ensemble Approach} \label{parvil}

We constructed our ensembles of turning pressure gradient simulations by branching from a fully developed, constant pressure gradient simulation.
In other words, we sampled full flow fields from the constant pressure gradient simulation at fixed time intervals.
These fields are then used as initial conditions for the simulations with a turning pressure gradient.
This approach is used in climate modelling and it was indicated as a ``common way to do'' ensembles in the sixth coupled climate model comparison project (CMIP6) \citep{eyring_overview_2016}.

The sampled initial conditions should possess a representative amount of variation.
In the case of turbulent flow, this means that one should not sample the flow at intervals smaller than the integral time scale $T_I$ (see, e.g. \citet{pope_turbulent_2000}).
Using time series of horizontal wind speed from the upper parts of the domain, where the roughness elements do not distort the turbulent flow, we calculated the integral length scale by numerically integrating the correlation function up to the point at which it reaches zero.
For the cube array, this results in $T_I \approx 0.0023 T_\Omega$ and for the realistic urban environment in the largest simulated domain, $T_I \approx 0.0019 T_\Omega$. 

In many cases, the flow possesses structures that are much larger than the integral scales.
Sampling should be carried out such that the different phases of the larger structures are captured.
The scales contained in the simulations are limited by the size of the simulated domain hence the sampling rate should be comparable to a corresponding time scale, the flow-through time.
For the cube-array, the flow-through-time is approximately $0.0580 T_\Omega$.
In the outermost domain of the realistic urban environment, the flow-through time is approximately $0.174 T_\Omega$. 

We have chosen to sample both flows at intervals of $0.0876 T_\Omega$ each.
This interval is approximately 1.5 times the flow-through-time in
the case of the staggered cube-array, and approximately half a flow-through time in the outermost domain of the realistic urban environment.
For both cases, the integral length-scale is two orders of magnitude smaller.
In the case of cube array, we also exploit the repeating pattern by averaging over each as if it were a separate ensemble member.
This approach is expected to introduce some spatial correlation among the ensemble members.
However, we expect that the time separation between initial conditions mitigates the effect of using the repeating pattern in the ensemble calculation.

The ensemble mean for the velocity $u_i$ is calculated using the arithmetic mean on the ensemble members with
\begin{equation}
  \langle u_i(t) \rangle_e = \frac{1}{N}\sum_{j=1}^N u_{i,j}(t),
\end{equation}
where $j$ indicates the ensemble member and $N$ the total size of the ensemble.
In the case of the cube array, we treat the repeating pattern of each simulation as an ensemble member while in the case of the realistic urban environment each ensemble member is a separate simulation.
When temporal averaging is combined with the ensemble averages to increase the sample size, we compute the ensemble mean using temporally averaged velocities:
\begin{equation}
  \langle u_i(t) \rangle_{e,T} = \frac{1}{N}\sum_{j=1}^N \frac{1}{T}
  \sum_{\substack{t-T/2 \le k \\ t+T/2 \ge k}} \frac{1}{\Delta t_k}
  u_{i,j,k},
\end{equation}
where $k$ refers to the discrete time and $\Delta t$ is the size of the time step.
Variance is calculated from the arithmetic means as $\sigma_i = \langle u_i^2 \rangle - \langle u_i \rangle^2$.
For this reason, we use the term ``averaging'' to refer to both the calculation of the mean and the variance.

\subsection{Statistical Measures}

We have utilized different statistical measures to characterize the ensemble and how it differs from the ensemble mean.
All members of the ensemble are equally valid representations of the flow, even though they may differ significantly.
By plotting the 25th and the 75th percentiles, we can visualize the region where half of the realizations are.
The separation between the two percentiles, the interquartile range, can be used as a measure of the variation within the ensemble \citep{harman_spatial_2016}.

We quantify the reliability of plain time averages or smaller ensembles by using various statistical performance measures.
The normalized standard deviation
\begin{equation} \label{nsd}
  \mathrm{NSD} = \frac{\sigma_p}{\sigma_o},
\end{equation}
the correlation coefficient
\begin{equation} \label{r}
  \mathrm{R} = \frac{\langle (C_o - \langle C_o \rangle )( C_p -
    \langle C_p \rangle \rangle}{\sigma_p \sigma_o},
\end{equation}
the root mean square error
\begin{equation} \label{rmse}
  \mathrm{RMSE} = \sqrt{\langle [ ( C_p - \langle C_p \rangle )
        - ( C_o - \langle C_o \rangle )]^2 \rangle },
\end{equation}
the normalized RMSE
\begin{equation} \label{nrmse}
  \mathrm{NRMSE} = \frac{\sqrt{\langle [ ( C_p - \langle C_p \rangle )
        - ( C_o - \langle C_o \rangle )]^2 \rangle } }{\sigma_o},
\end{equation}
and the fractional bias
\begin{equation} \label{fb}
  \mathrm{FB} = \frac{\langle C_o \rangle - \langle C_p
    \rangle}{\frac{1}{2} \left( \langle C_o \rangle + \langle C_p
    \rangle \right) }
\end{equation}
all compare a predicted value $C_p$  (with variance $\sigma_p$) to a reference value $C_o$  (with variance $\sigma_o$) \citep{chang_air_2004}.
A perfect correspondence to the reference result is indicated when NSD$=1.0$, R$=1.0$, NRMSE$=0.0$, and FB$=0.0$.
The NSD, NRMSE, and R measure only unsystematic errors while fractional bias measures only the systematic bias.
It is possible to display NSD, R, and NRMSE in a single plot, the Taylor diagram \citep{taylor_summarizing_2001,chang_air_2004}.
The Taylor diagram is a polar plot, where NSD is the radial coordinate, $\arccos R$ the azimuthal coordinate, and the NRMSE is the distance to the point (NSD$=1.0$, R$=1.0$). 

These error measures can be used in various different ways, depending on the nature of the data or the goals of the study.
In the present paper, we will use these measures pointwise in three dimensions.
This means that we calculate the measures \eqref{nsd}--\eqref{fb} at each grid point and then aggregate these data by spatial averaging for each simulation.
As reference data, we will use either ensemble-averaged data, calculated from instantaneous or temporally averaged values.
Depending on the purpose, we use RMSE with different normalizations then in the NRMSE given by \eqref{nrmse}.

\section{Results} \label{tulokset}

Simulation ensembles were created for both the staggered cube array and for the realistic urban environment using the approaches described earlier.
In both cases, we concentrate at the time in the middle of the simulation, at $t/T_\Omega = 0.526$ since the start of the turning of the pressure gradient.
All used time averages are centred around this time instant.

\subsection{Staggered Cube Array} \label{kuutiotulokset}

\begin{figure}[tp]
  a) \\ % Skripti R1_K_EAS1.py
  \includegraphics[width=\textwidth]{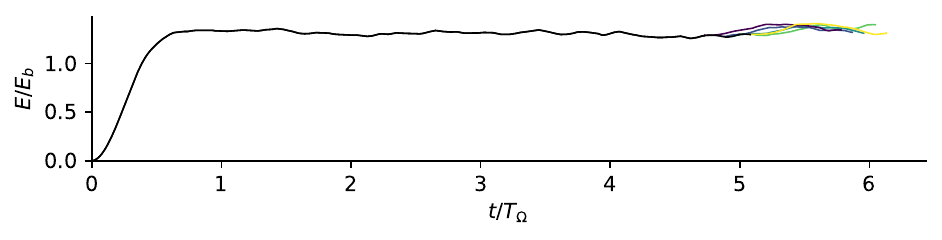} \\
  b) \\ % Skripti R1_K_EAS2.py
\includegraphics[width=\textwidth]{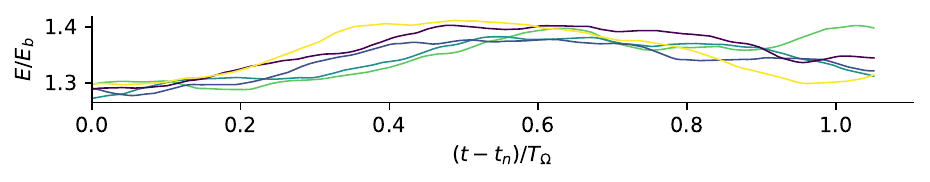}
\caption{\small Volume average of the resolved kinetic energy of the   simulations with the staggered cube array.
  a) The spinup simulation  (black dashed line) and the simulations with a turning pressure  gradient (coloured lines).
  The $n$th turning pressure gradient  simulation branches off from the spinup at time $t_n/T_\Omega = 4.73   + 0.0876(n-1)$.
  b) All turning pressure gradient simulations shown using time since start of pressure gradient turning}
\label{enaikas}       % Give a unique label
\end{figure}

Using the staggered cube array set-up, we have carried out a spinup simulation with a static pressure gradient and a duration of $t/T_\Omega = 4.99$.
The volume average of the total kinetic energy of the spinup
\begin{equation}
\frac{E}{E_b} = \frac{\langle \frac{1}{2}u_i u_i \rangle_V}{\frac{1}{2}U^2_b}
\end{equation}
is shown in Fig.\,\ref{enaikas} a) using a black dashed line.
The initial development from zero is quick and the flow reaches its final energy level already at approximately $0.6 T_\Omega$.
After this, the flow assumes relatively steady energy level with minor fluctuations.
The level of resolved TKE develops similarly as the total kinetic energy (not shown).
We monitored the convergence of the velocity profiles and allowed the turbulence to develop for further $t/T_\Omega = 3.94$.

Starting at $t/T_\Omega = 4.73$, five simulations with a turning pressure gradient are branched off from the spinup every $t/T_\Omega = 0.0876$ and simulated for $t/T_\Omega = 1.05$.
The introduction of the turning pressure gradient alters the kinetic energy content.
All runs show a local maximum after a nondimensional time of $0.35–0.7 T_\Omega$ after the start of pressure gradient turning, as seen in Fig.\,\ref{enaikas}~b).
Otherwise the kinetic energy content develops differently in all simulation although the level of fluctuation is still similar.
The spinup simulation required approximately 5\,000 core hours while each run consumed approximately 1\,000 core hours on average.

The simulated wind reacts to the turning of the pressure gradient with a delay.
the large-scale wind direction is not the same as the direction of the pressure gradient.
Similar beheviour between the changing forcing and the resulting flow was observed also in the case of the pulsatile flow by \citet{li_mean_2023}.
To quantify the difference between the directions of the pressure gradient and the flow, we have calculated the wind direction using the upper parts of the domain, above $0.665H$ (not shown).
At the beginning of the simulation, the large-scale wind direction is approximately \unit[241]{$^\circ$} and it follows the direction of the pressure gradient with a lag.
The wind turning happens in two parts: a fast acceleration during the first $t/T_\Omega = 0.175$ of the simulation up to a turning speed of approximately $0.533 \Omega$ and a somewhat slower acceleration during the rest of the simulation up to approximately $0.933 \Omega$.
The mean wind direction at middle of the simulation is approximately \unit[257]{$^\circ$} and \unit[282]{$^\circ$} at the end of the simulation.

The staggered cube array has 648 repeating units and hence the total ensemble size is 3\,240 when we utilize both simulations and the repeating units in the creation of the ensemble and in the calculation of ensemble statistics.
We have studied the statistical convergence of the ensemble as a function of ensemble size using the bootstrapping approach \citep{efron_introduction_1993}.
Starting from an ensemble size of five, the error in the velocity means decreases with an order of magnitude at the ensemble size of around 80 with the second order of magnitude requiring around 700 more members.
The total ensemble size of 3\,240 should hence be enough for proper statistical convergence.

\begin{figure}[tp]
  % Skripti R1_K_PK.py
  \center
  a) \hspace{0.4\textwidth} b) \\
  \includegraphics[width=0.37\textwidth]{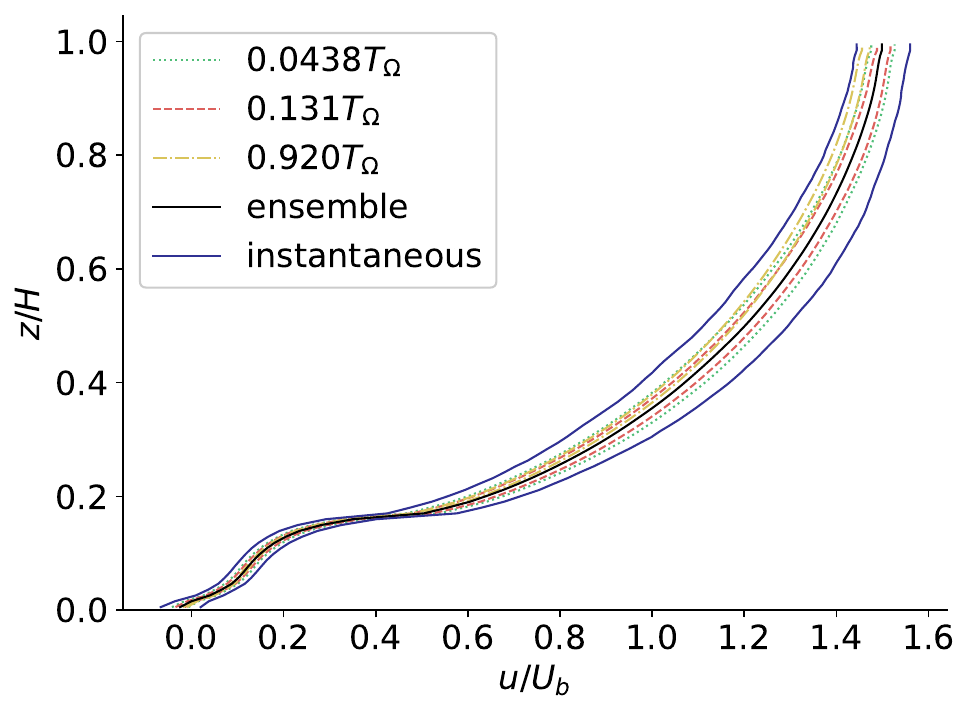}
  \includegraphics[width=0.37\textwidth]{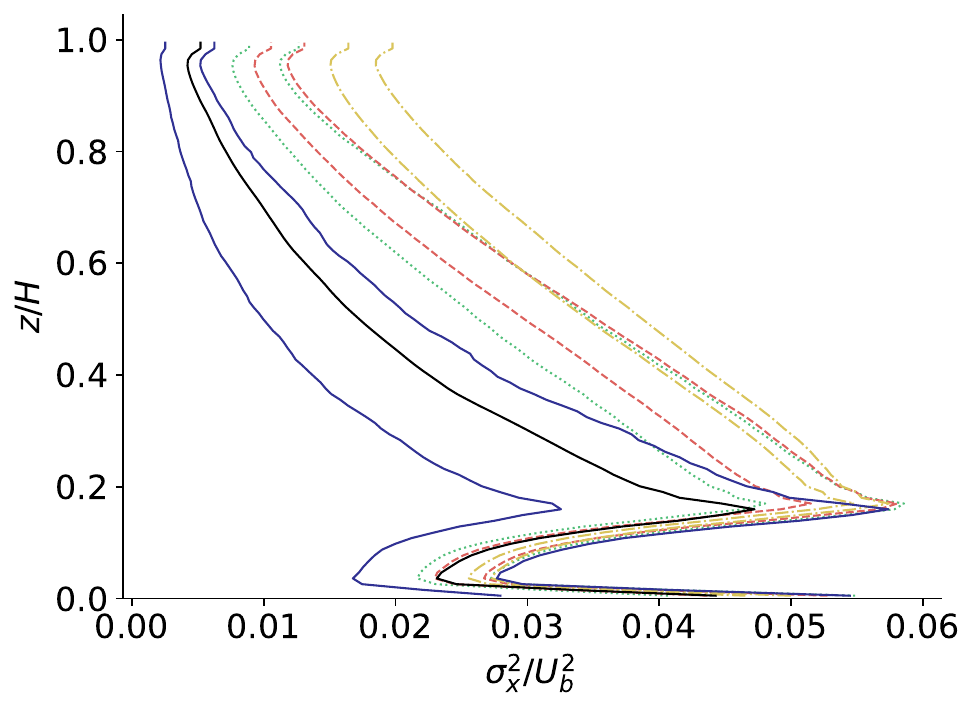} \\
  c) \hspace{0.4\textwidth} d) \\
  \includegraphics[width=0.37\textwidth]{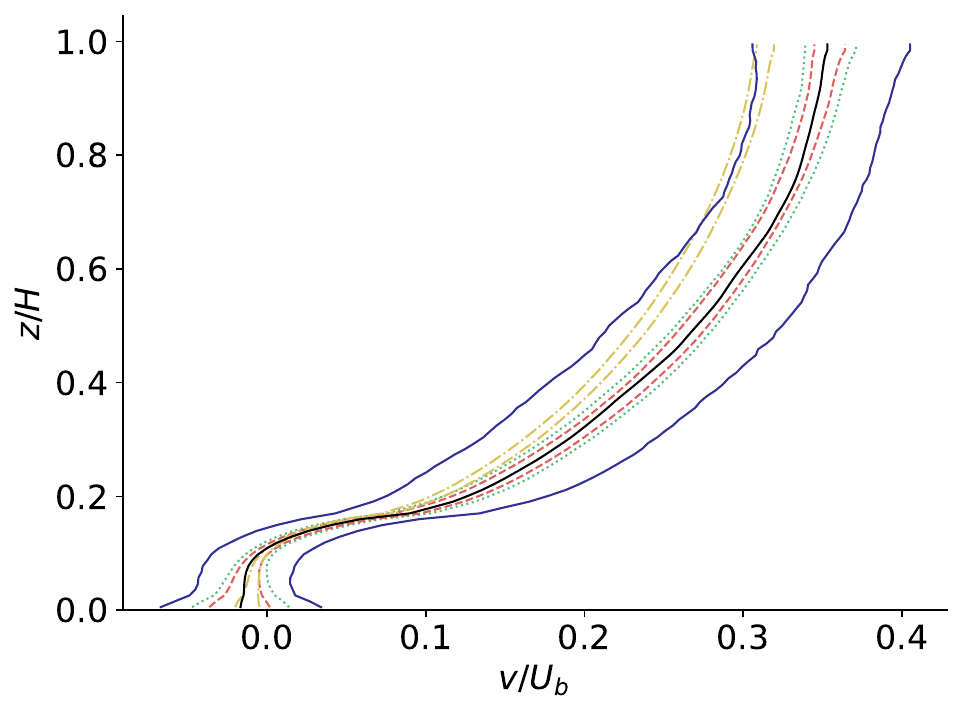}
  \includegraphics[width=0.37\textwidth]{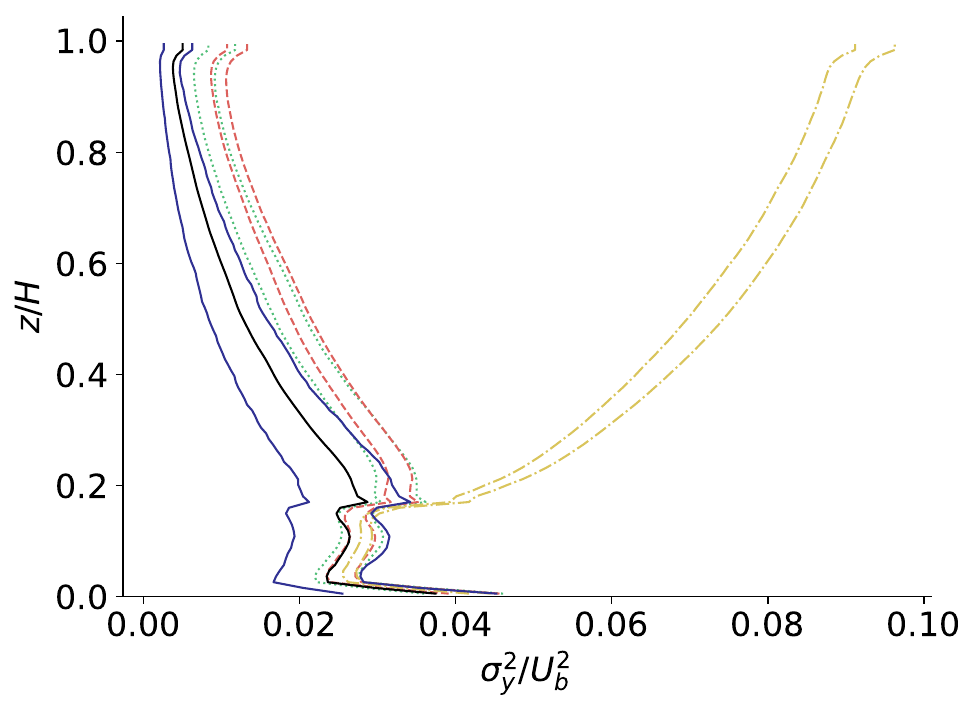} \\
  e) \hspace{0.4\textwidth} f) \\
  \includegraphics[width=0.37\textwidth]{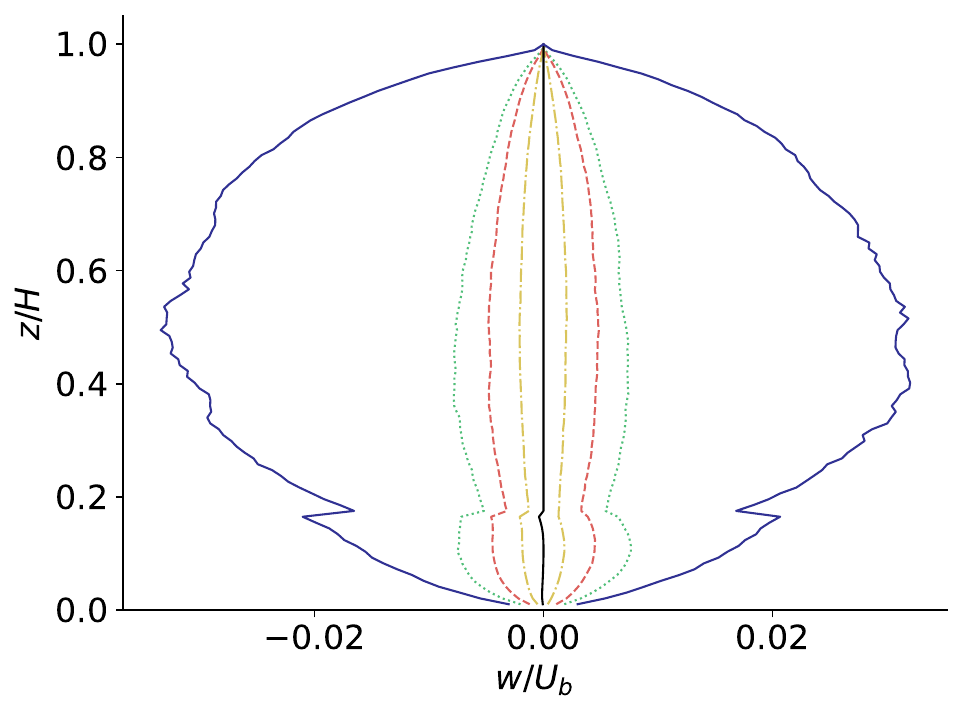}
  \includegraphics[width=0.37\textwidth]{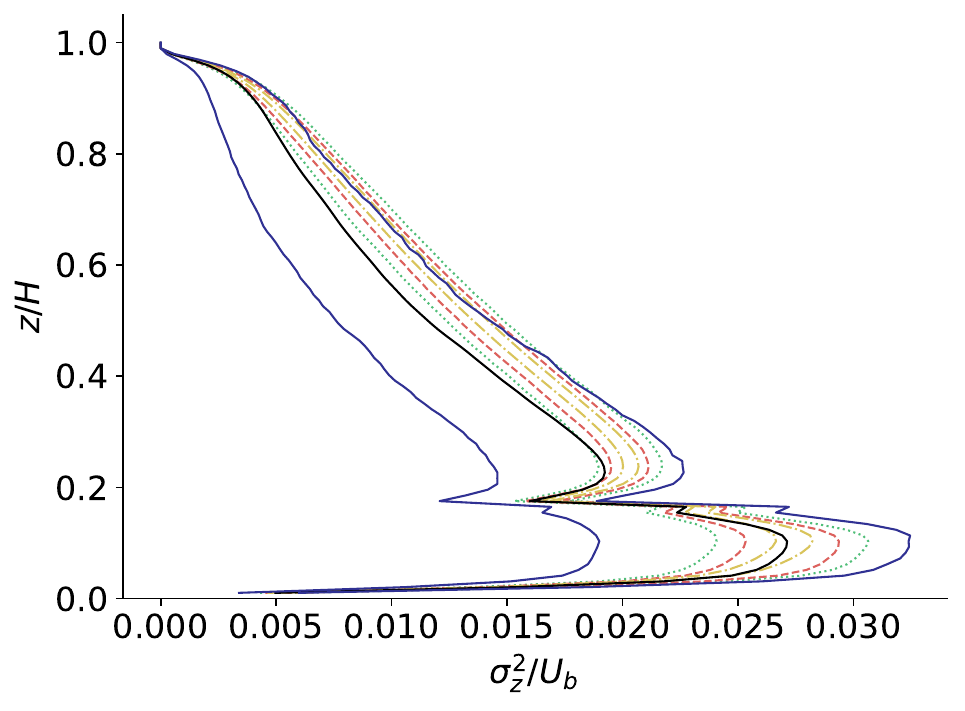} \\
  g) \hspace{0.4\textwidth} h) \\  
  \includegraphics[width=0.37\textwidth]{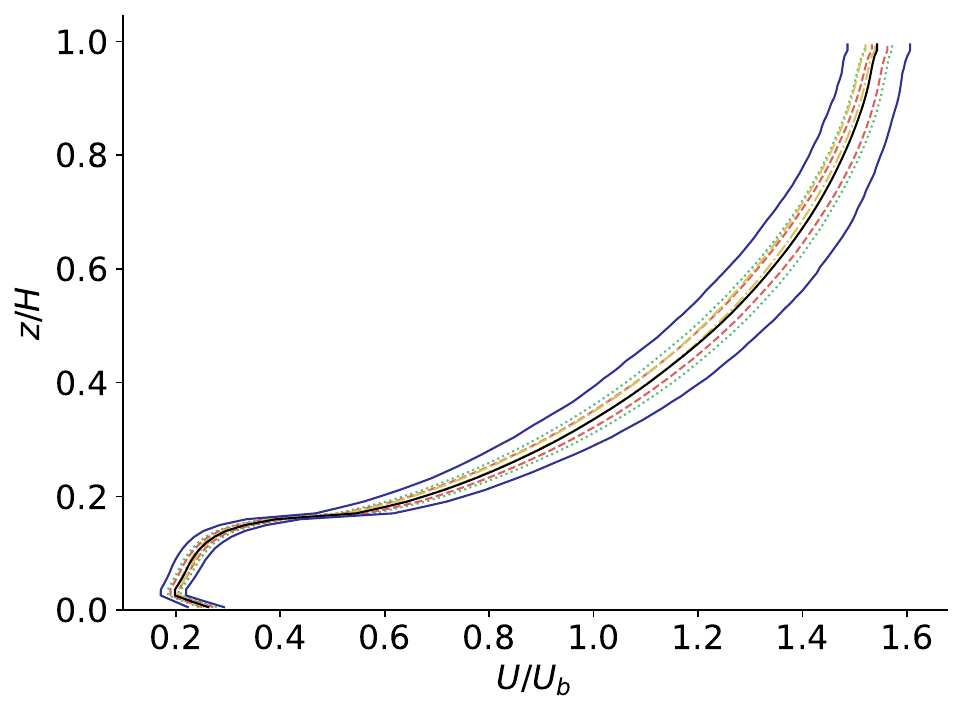}
  \includegraphics[width=0.37\textwidth]{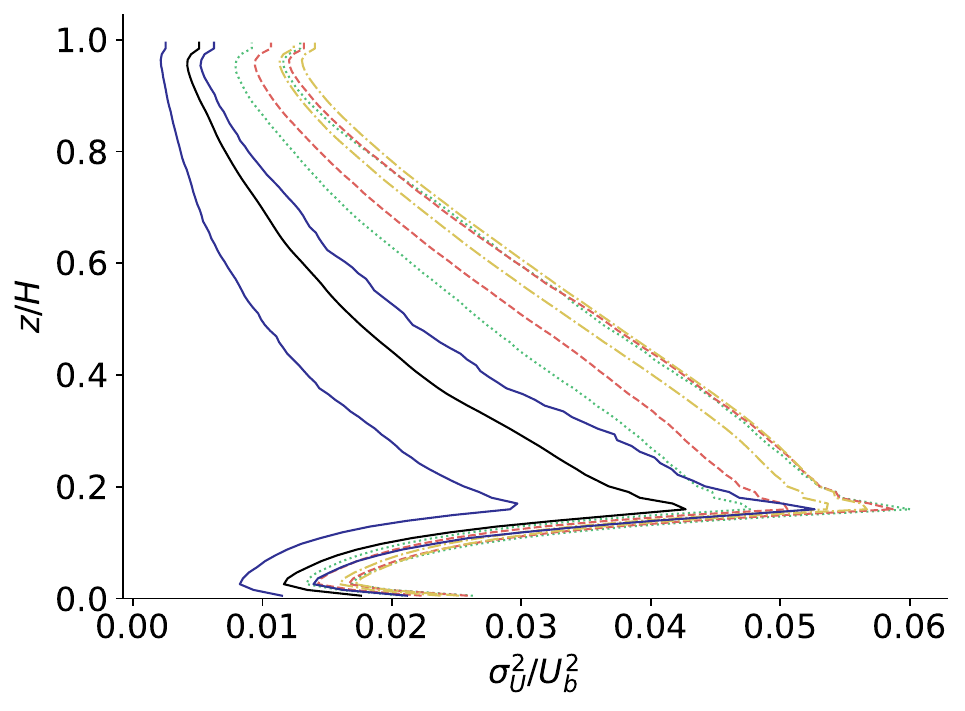}
  \caption{\small Vertical wind profiles in the case of the staggered cube array.
    The mean and variances for all quantities have been accumulated over the horizontal directions and the possible statistical accumulation over time is indicated with different lines.
    In time, all quantities are centred around the mid point of the simulation, at $ t/T_\Omega = 0.526 $.
    The black line indicates the ensemble mean.
    The coloured lines indicate the 25th and 75th percentiles of the ensemble where the specified time-averaging has been applied to each member.
    a) Mean velocity along $x$ axis ($u$), b) variance of $u$ ($\sigma^2_x$), c) mean velocity along $y$ axis ($v$), d) variance of $v$ ($\sigma^2_y$), e) mean velocity along $z$ axis ($w$), f) variance of $w$ ($\sigma^2_z$), g) mean horizontal velocity $U$, and h) variance of $U$ ($\sigma^2_U$)}
\label{kaprofiilit}       % Give a unique label
\end{figure}

Using the full ensemble together with averaging in the horizontal directions, we obtain vertical profiles for the ensemble mean velocities and variances, shown with black lines in Fig.\,\ref{kaprofiilit}.
The profiles indicate a typical flow over cubical obstacles with highest mean velocities at the top and a maxima for variances at the cube height (see, e.g. \citet{cheng_adjustment_2015,ahn_statistics_2013}).
In order to quantify the variation within the ensemble, we have calculated the vertical profiles for each ensemble member by averaging in the horizontal directions and then plotted the 25th and the 75th percentiles of the ensemble using blue lines in Fig.\,\ref{kaprofiilit}.
The separation between these percentiles is called the interquartile range and it quantifies the variation within an ensemble \citep{harman_spatial_2016}: Half of the 3\,240 ensemble members are within the interquartile range.

The use of time averaging in addition to horizontal averaging for the ensemble members allows one to obtain a better statistical convergence for the vertical profiles and hence a smaller interquartile range.
However, the changing wind direction can lead to departures from the ensemble average (errors) when long averaging times are utilized.
The 25th and 75th percentiles for mean velocity and variance profiles of the ensemble, calculated using horizontal and temporal averages, are shown Fig.\,\ref{kaprofiilit}.
Three representative averaging times are shown: a short time average of $0.0438 T_\omega$ with green dotted line, a medium time average of $0.131 T_\Omega$ with red dashed line, and long time average of $0.920 T_\Omega$ with yellow dot-dashed line.
The short time average is clearly smaller than the characteristic time scale of the turning pressure gradient $T_\Omega$ and hence the effects of the turning pressure gradient on the mean values are expected to be minor.
On the other hand, the long time average is almost as long as $T_\Omega$ and hence the turning pressure gradient is expected to influence the outcome.
These averaging times correspond approximately to \unit[0.92]{$^\circ$}, \unit[4.8]{$^\circ$}, and \unit[37]{$^\circ$} change in wind direction, respectively.

For the mean velocities in the coordinate directions $x$ in Fig.\,\ref{kaprofiilit} a), $y$ in c), and $z$ in e) the use of both the short and the medium averaging time narrows the interquartile range of the ensemble around the ensemble mean for all components, indicating smaller variation between the ensemble members compared to the instantaneous profiles.
However, the use of the long averaging time $0.920 T_\Omega$ results in a narrow interquartile range that's away from the ensemble mean for $u$ and, more strongly, in $v$.
This can be explained by the averaging time being at the same order as the characteristic time scale of the turning $T_\Omega$.
This translates into a large change in the overall wind direction during time averaging.
The narrowing of the interquartile range is hence not an indication of a correct solution but rather the agreement between the ensemble members.

In the case of variances, shown in Fig.\,\ref{kaprofiilit} b) for $\sigma^2_x$, d) b) for $\sigma^2_y$, and in f) for $\sigma^2_z$, the use of time averaging pushes the interquartile range away from the ensemble average for all heights above the roughness sublayer.
The most extreme case is seen for long time average in $\sigma^2_y$ where variance increases upwards while at the same time the ensemble value decreases.
Above the roughness sublayer the flow is homogeneous in the horizontal directions and so these dimensions can be used as a natural way of accumulating statistics.
Although a single simulation (ensemble member) is not likely to produce exactly the same profile as the ensemble would, variance profiles calculated using instantaneous values with only horizontal averaging are closer to the ensemble than when calculated using a combination of temporal and spatial averaging.
In all cases, the time averaged variances are larger than the true, ensemble averaged variances.
This indicates that a part of the change in mean wind ends up in the variances.

The use of horizontal wind velocity instead of velocities $u$ and $v$ of the Cartesian coordinate system could potentially be used to circumvent the errors induced especially to the weaker wind component $v$.
The profile of the horizontal mean velocity is shown in Fig.\,\ref{kaprofiilit} g) and its variance in h).
Both profiles are very similar to those of the of the $u$ velocity.
This is due to the main flow direction being directed mainly along the $x$ axis and hence the influence of $v$ to the horizontal velocity is minor.
There are some differences between the profiles of the horizontal velocity and $u$ in the roughness sublayer.
The mean horizontal wind has a small jump very close to the bottom of the domain and the ensemble variance is located outside of the interquartile range of even the shortest time averages. 

\begin{figure}[tp]
  % Skripti R1_K_NT.py
  % Jos tarvitsen kuville kirjainmerkinnät tyyliin a), b), c) voinen
  % käyttää overpic-pakettia.
  % Keskiarvot
  \center
  \begin{tabular}{ccc} 
  \includegraphics[width=0.21\textwidth]{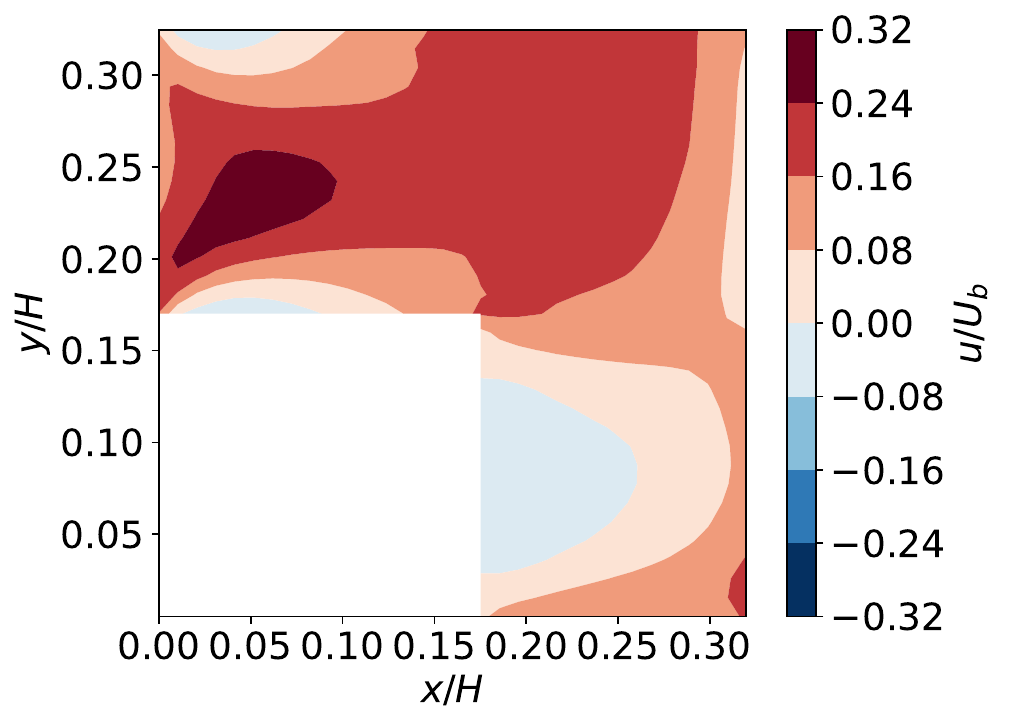} &
  \includegraphics[width=0.20\textwidth]{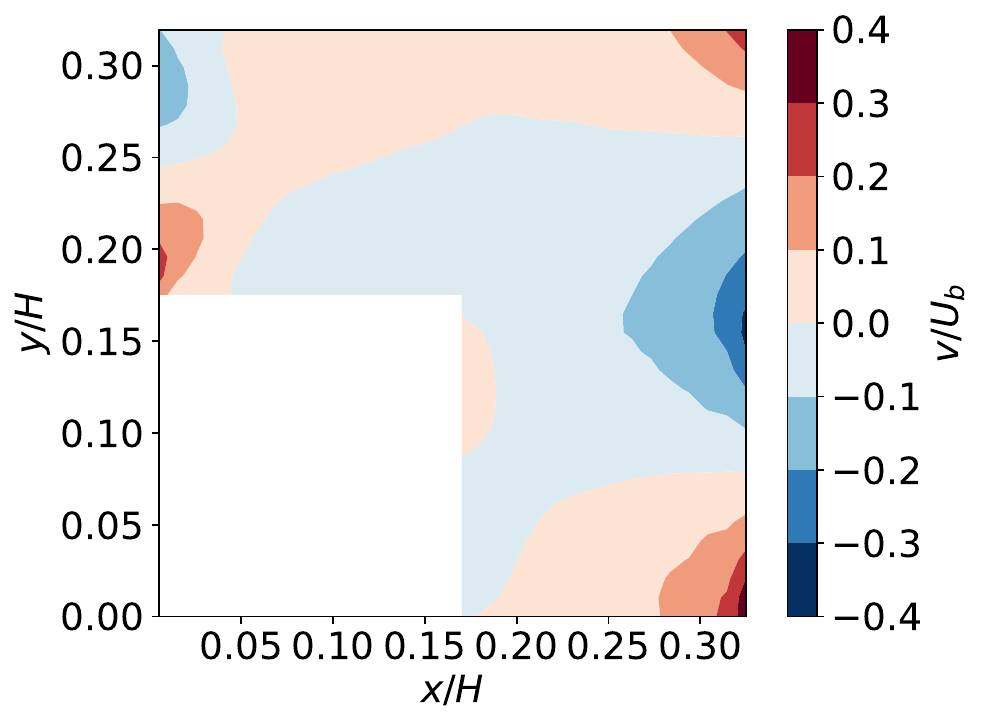} &
  \includegraphics[width=0.21\textwidth]{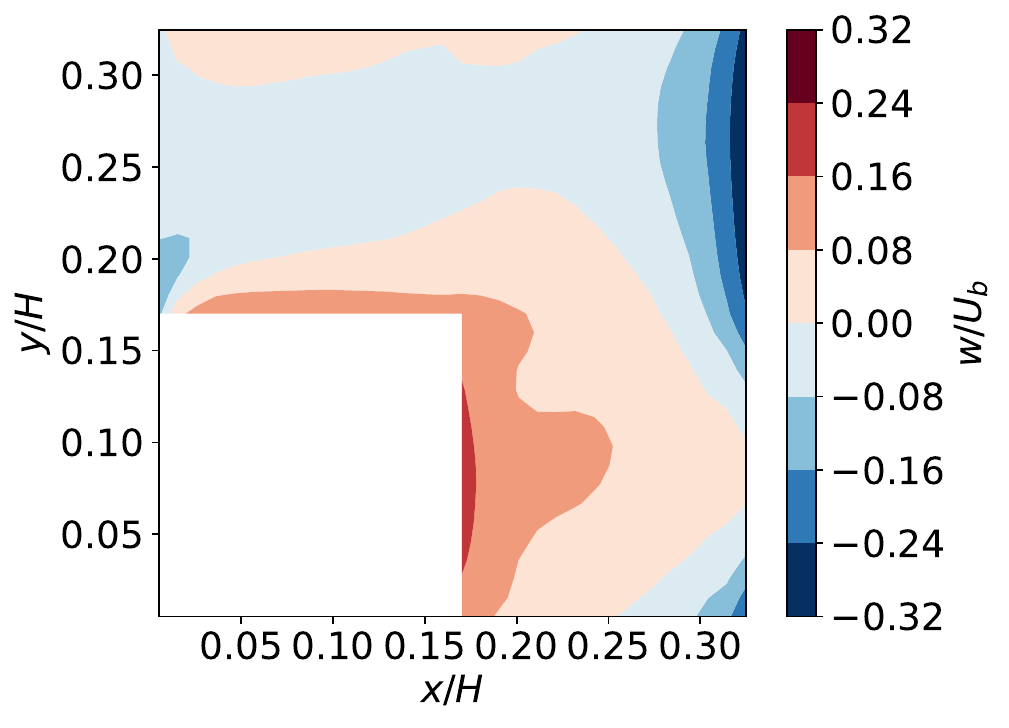} \\
  \includegraphics[width=0.21\textwidth]{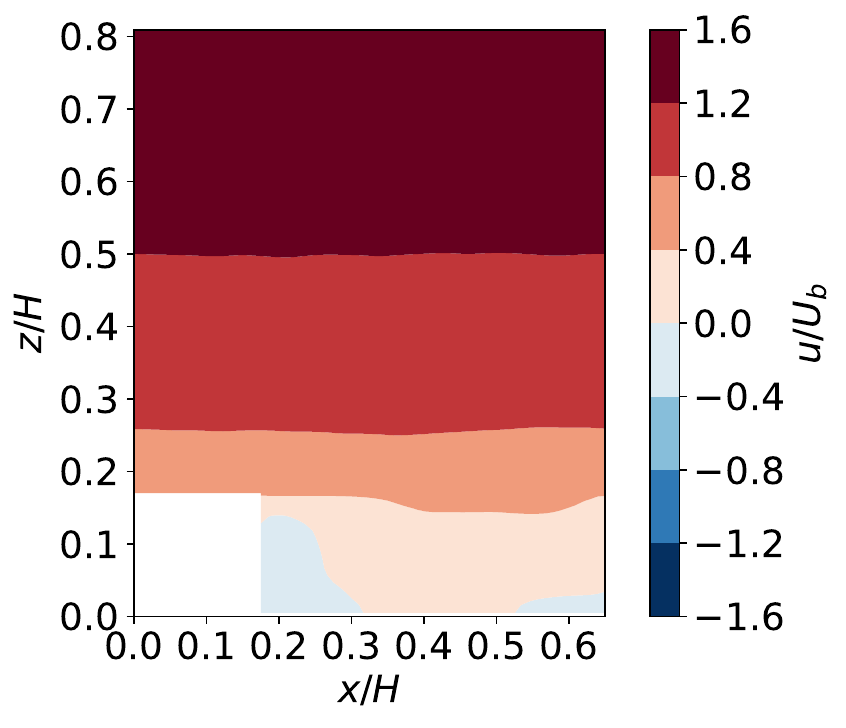}  &
  \includegraphics[width=0.21\textwidth]{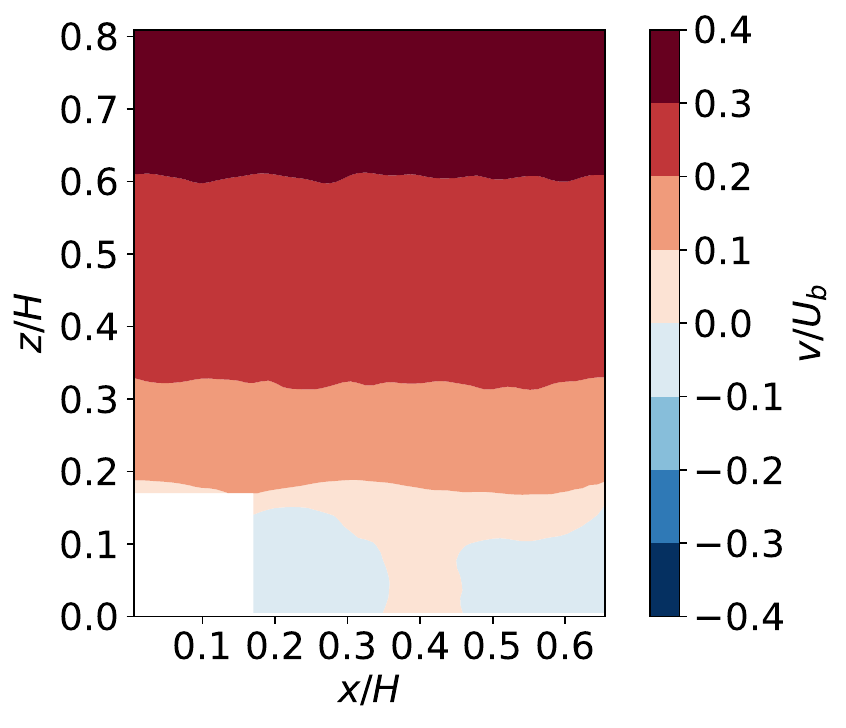} &
  \includegraphics[width=0.21\textwidth]{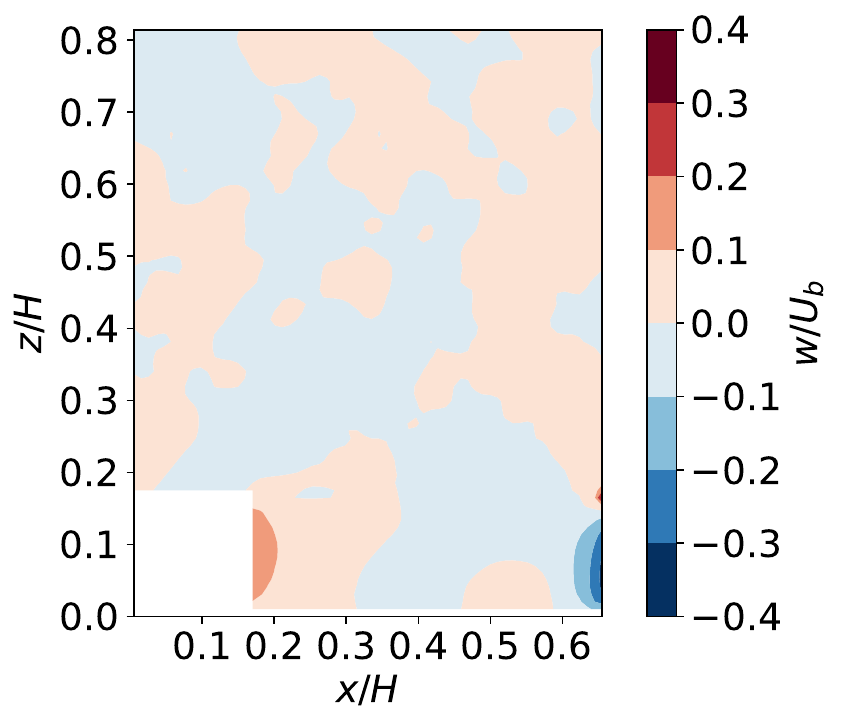}  \\
  % Varianssit
  \includegraphics[width=0.21\textwidth]{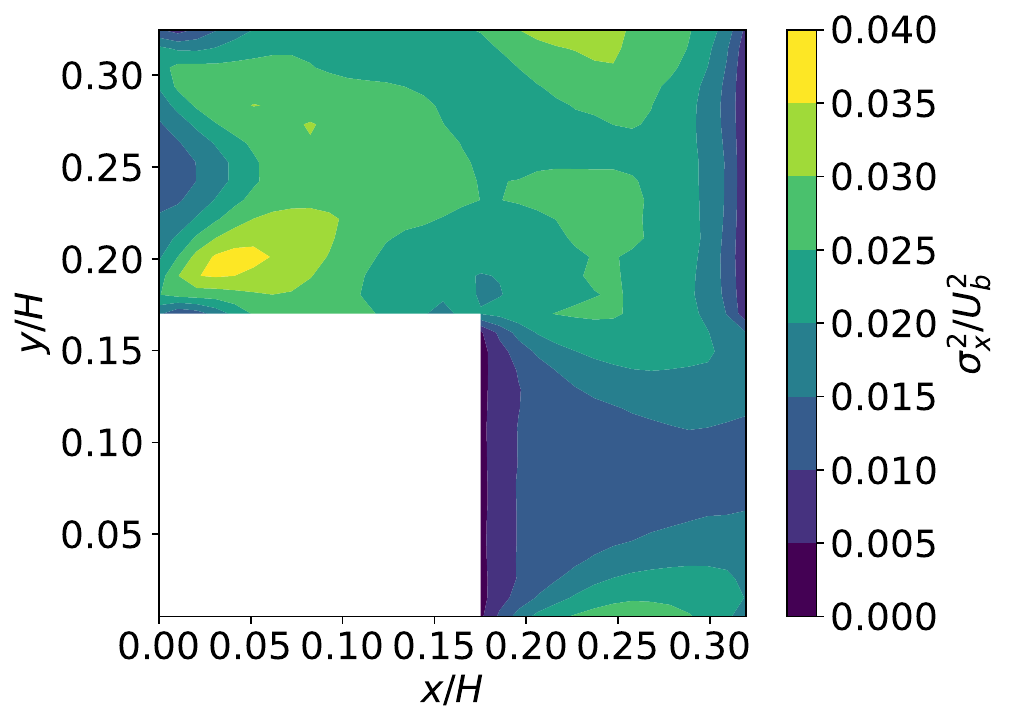} &
  \includegraphics[width=0.21\textwidth]{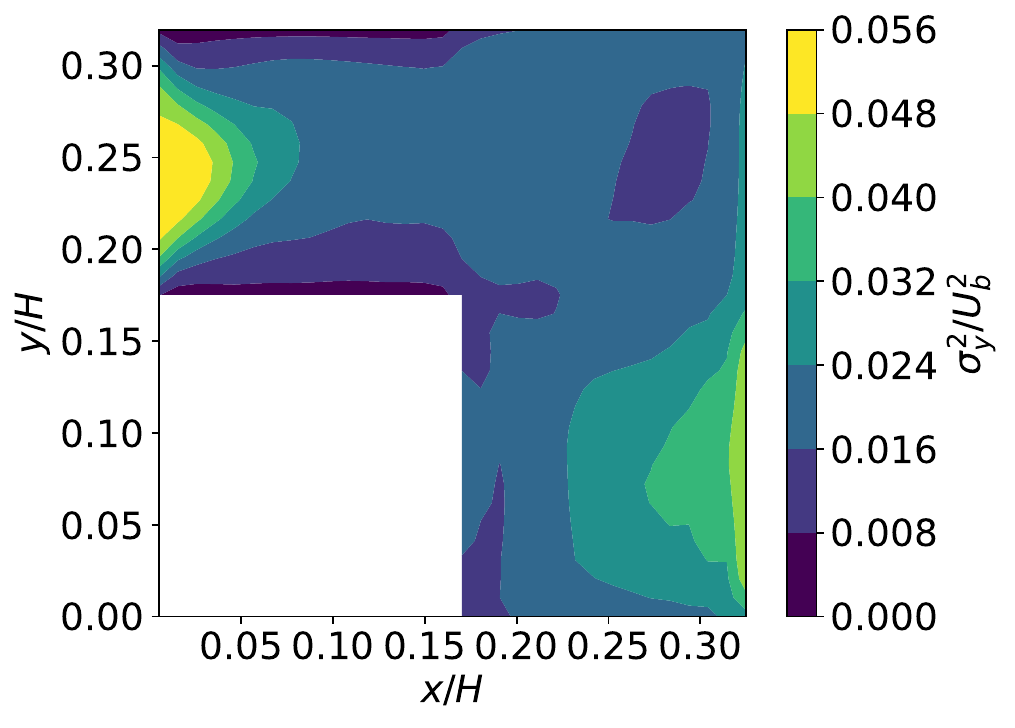} &
  \includegraphics[width=0.21\textwidth]{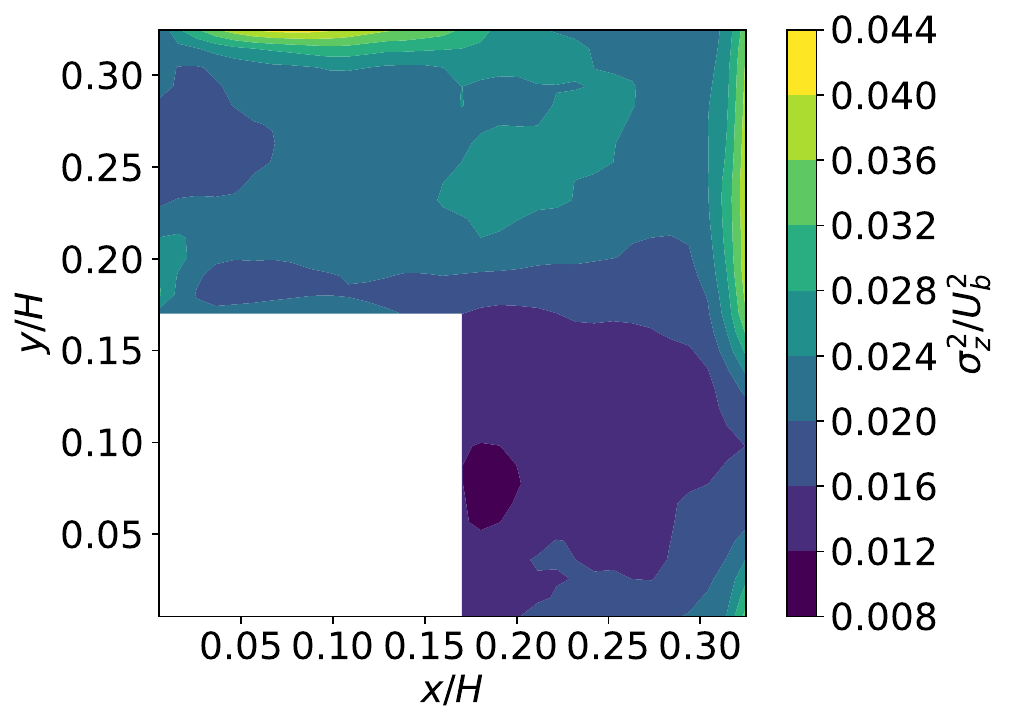} \\
  \includegraphics[width=0.20\textwidth]{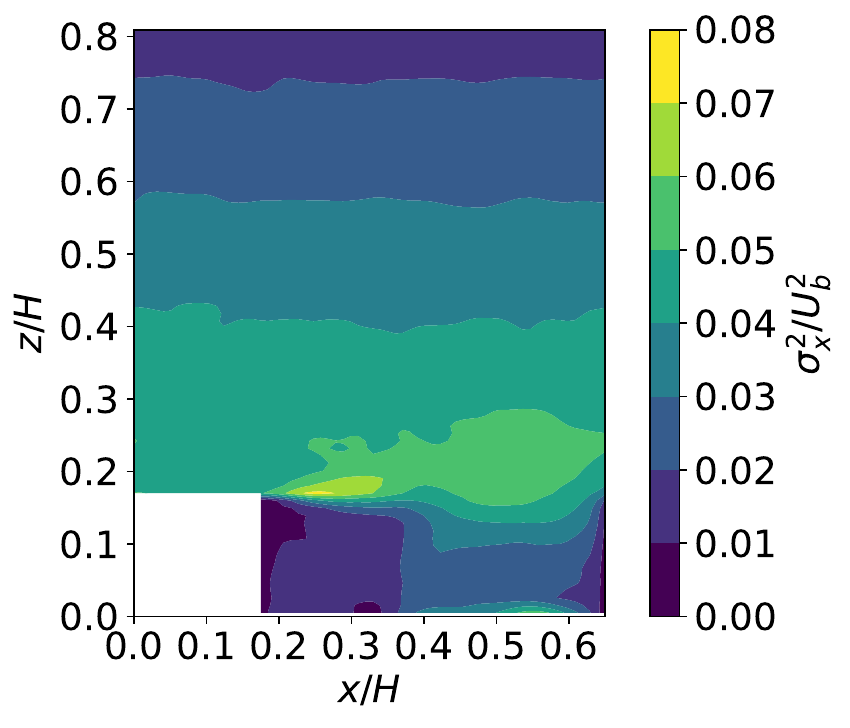}  &
  \includegraphics[width=0.21\textwidth]{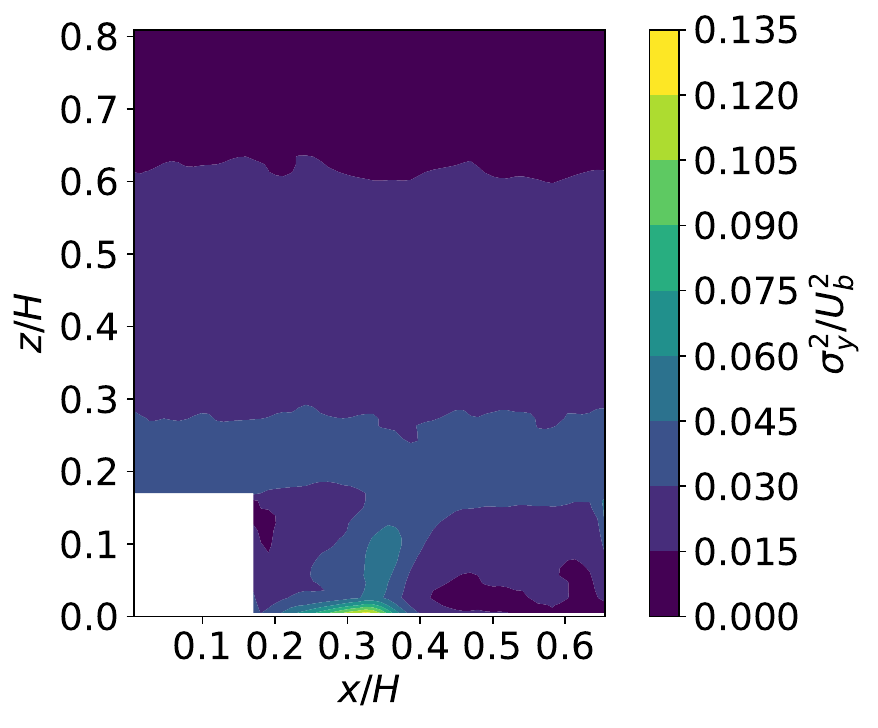} &
  \includegraphics[width=0.21\textwidth]{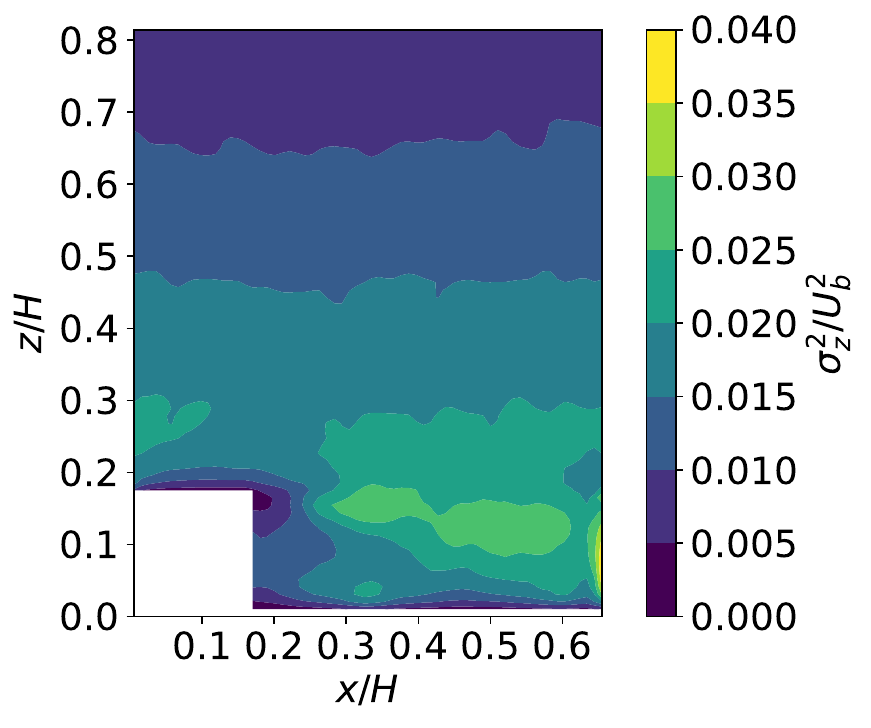} \\
    % Ylimääräiset
  \includegraphics[width=0.20\textwidth]{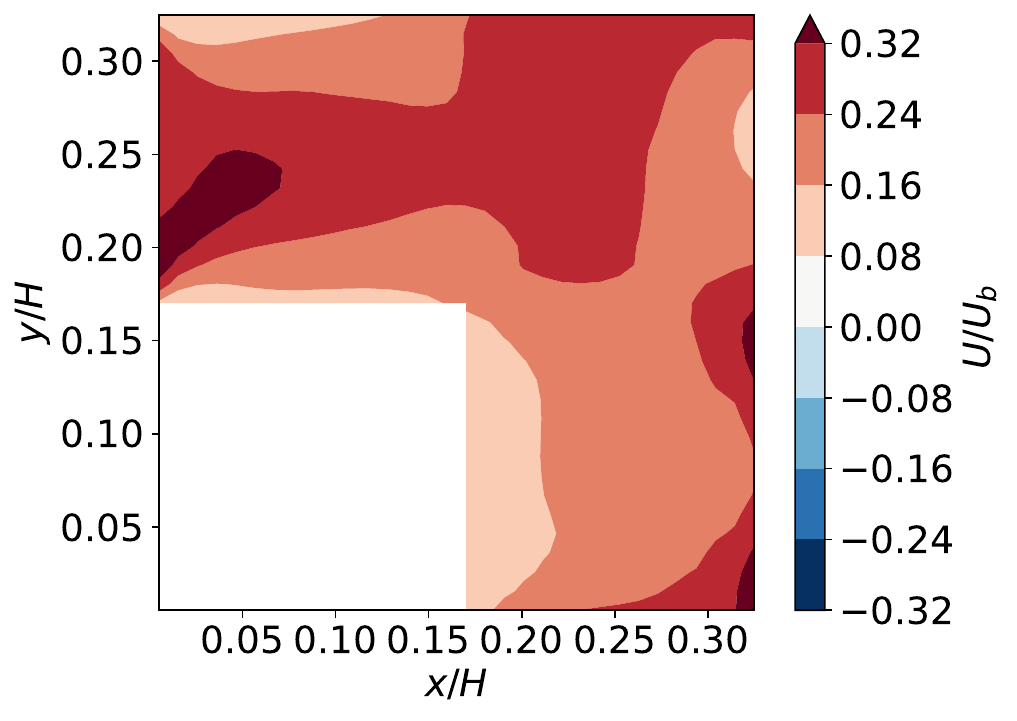} &
  \includegraphics[width=0.21\textwidth]{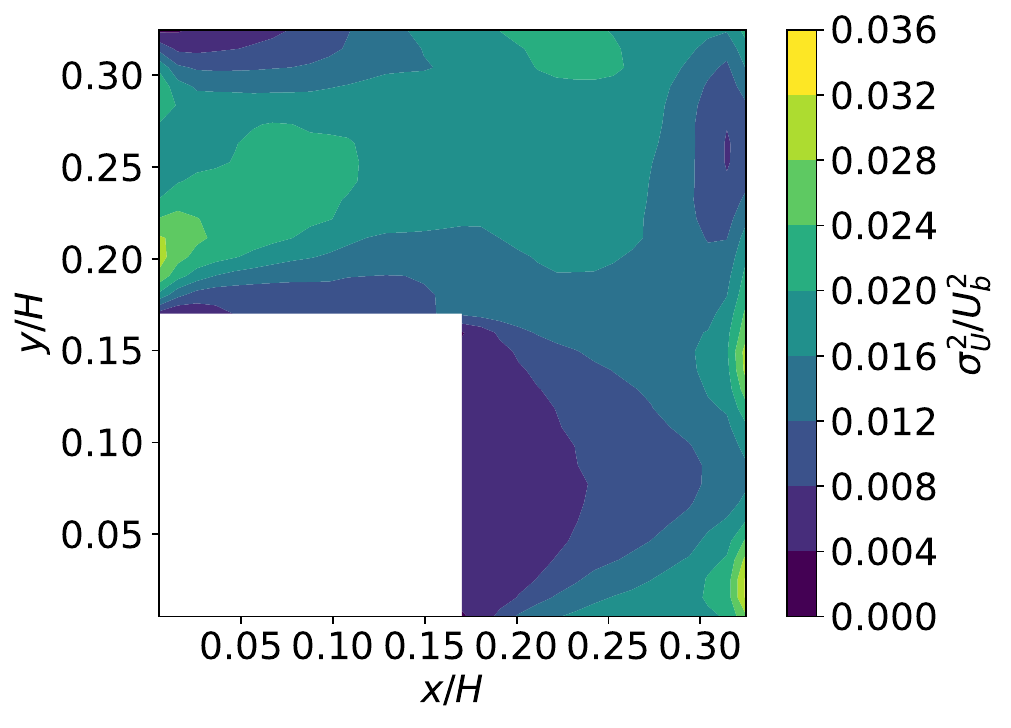}&
  \includegraphics[width=0.20\textwidth]{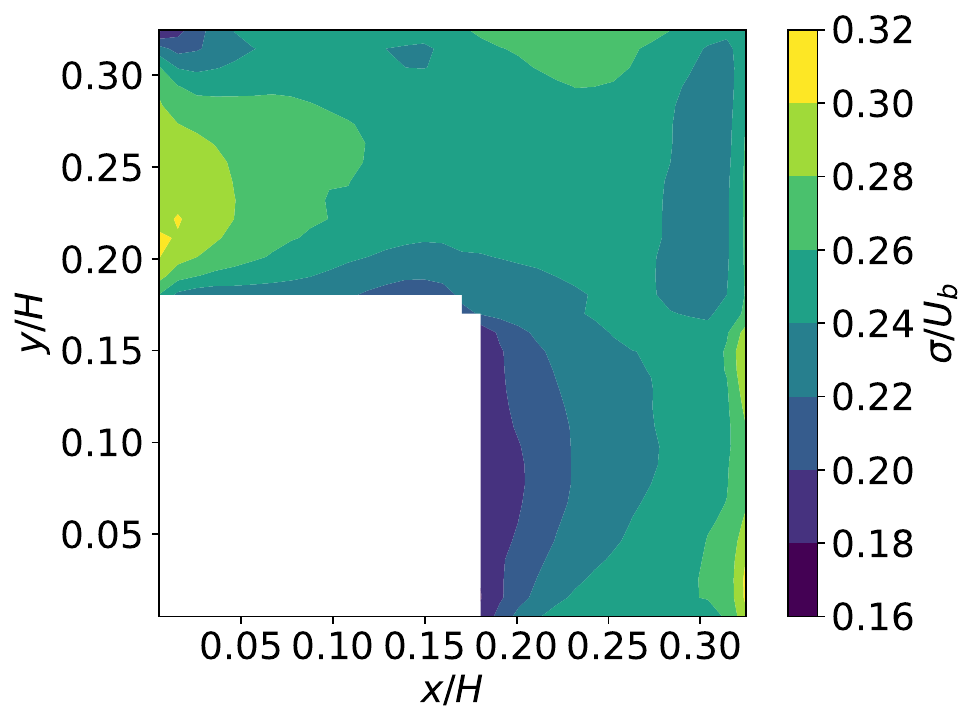} \\  
  \includegraphics[width=0.19\textwidth]{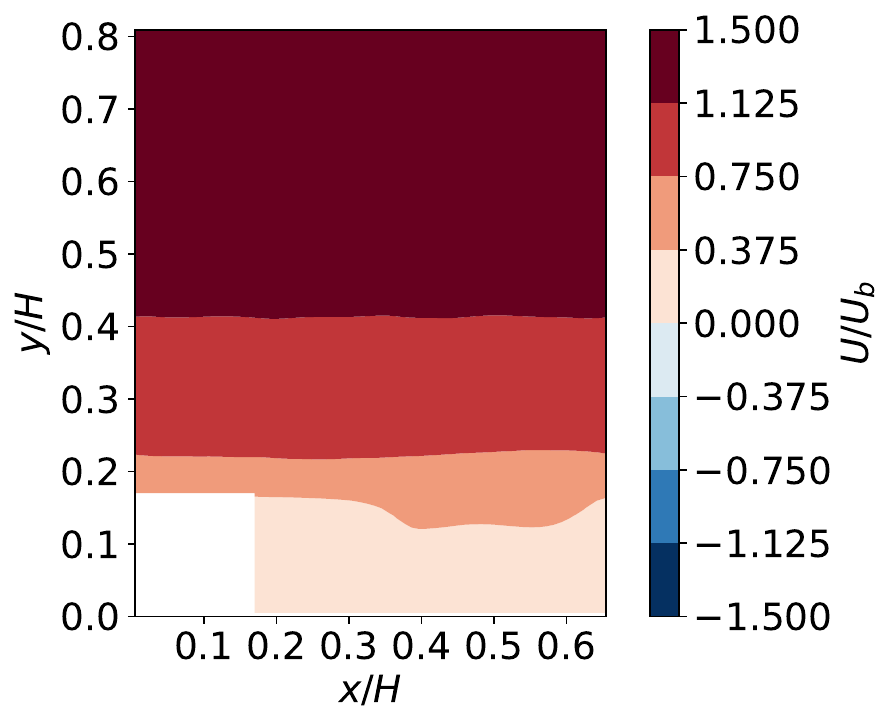}  &
  \includegraphics[width=0.21\textwidth]{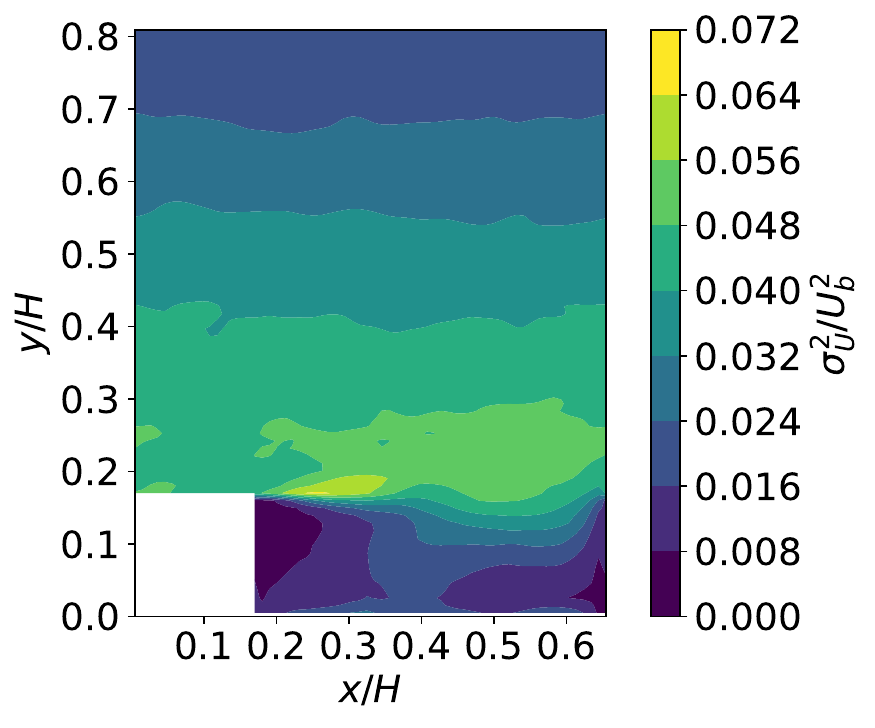}&
  \includegraphics[width=0.21\textwidth]{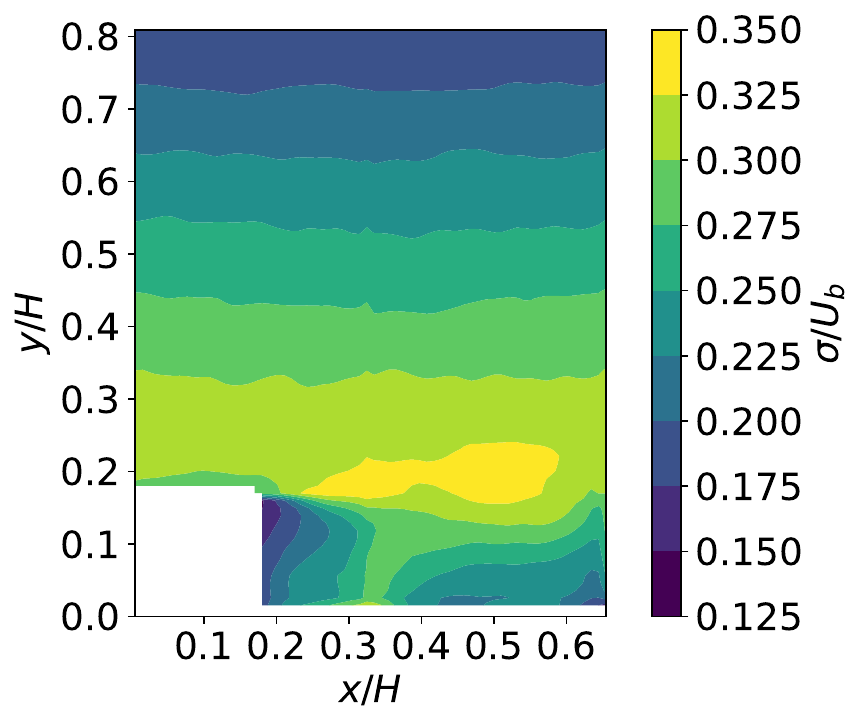} \\
  \end{tabular}
  \caption{\small Ensemble statistics in the case of the staggered cube array at time $0.526 T_\Omega$.
    Two different planes of a single repeating element are shown.
    One is horizontal through the roughness elements at $z/H=0.0825$ (panels on the first, third, and fifth rows from the top) and the other is vertical through the roughness elements at $y/H=0.0825$ for $0 \le x/H \le 0.330$ and at $y/H=0.247$ for $0.330 \le x/H \le 0.660$ (panels of the second, fourth, and sixth rows from the top).
    The coordinates refer to a single repeating element and the cube corner is at the origin.
    The simulations contained several repeating elements and hence the shown horizontal plane should be considered as a part of a staggered pattern in the $x$ direction and as an aligned patter in the $y$ (see, Fig.\,\ref{kuutiokuva}).
    The shown vertical plane wraps through the repeating element and should be considered to repeat in the $x$ direction only.
    The panels on the first two rows show the components of the mean velocity with $u/U_b$ on the first column, $v/U_b$ on the second columns, and $w/U_b$ on the third column.
    Variances $\sigma^2_x/U_b^2$, $\sigma^2_y/U_b^2$, and $\sigma^2_z/U_b^2$ are shown similarly on the panels in the third and fourth rows.
    The lowermost two rows show the horizontal mean wind speed $U/U_b$ on the first column, horizontal wind variance $\sigma^2_U/U_b^2$ on the second column, and turbulence intensity $\sigma/U_b$ on the third column}
    \label{poikki1}
\end{figure}

Further insight in to the flow field within in the close vicinity of the obstacles is obtained from Fig.\,\ref{poikki1} where instantaneous ensemble statistics on planes cutting the repeating element are shown.
Spatial and temporal dimensions have not been utilized to supplement the ensemble statistics for this figure.
The mean velocity components in panels on the two uppermost rows reveal that the main flow direction is toward positive $x$ for most part of plane.
Recirculation regions in the vertical direction can be seen on all vertical sides of the obstacle with the one on the leeward side being the largest.
The flow is directed towards negative $z$ only on the windward side.
Due to the staggered configuration, the flow encounters the obstacles as a narrow region, at $0 \le x/H \le 0.16$ and $0.16 \le y/H \le 0.32$, where it is accelerated and horizontal components are added to the recirculation regions on those sides of the obstacle.

The variances of the velocity components, shown in panels on the third and fourth rows of Fig.\,\ref{poikki1}, indicate higher levels of turbulence in three places.
Firstly, there is a shear layer at around the top of the obstacle.
Secondly, variances are larger on the vertical sides of the obstacle where the flow is directed upwards
Thirdly, the flow is accelerated by the constricted geometry in the region between two obstacles and then sheared by the horizontal recirculation regions.
At least a part of the turbulence generated at the shear layer at the top of the obstacle appears to be transported downwards by the recirculation region, in a similar manner as described by \citet{salizzoni_turbulent_2011}.
There is also a patch of high variance of the $v$ component of velocity near the bottom surface at around $x/H=0.309$.
It is unclear if this turbulence is transported down from the shear layer above by the vertical motions in the recirculation region or generated in-place by the wall through local shear.

The horizontal mean velocity in the first column of the two last rows in Fig.\,\ref{poikki1} shows that the horizontal velocity within the obstacles in not a direct copy of $u$ as it was for the profiles.
The regions where $v$ had significant non-zero values, such as the recirculation regions on the windward corners of the cube, are added to the mean horizontal velocity.
It is clear that the flow has a clear horizontal direction in only the narrow region between the cubes, at $0 \le x/H \le 0.16$ and $0.16 \le y/H \le 0.32$.
The horizontal variances in the second column of the last two rows of in Fig.\,\ref{poikki1} show a similar superposition as does the turbulence intensity $\sigma/U_b = \sqrt{\sigma^2_x + \sigma^2_y + \sigma^2_z}/U_b$ in the third column of the last two rows in Fig.\,\ref{poikki1} with an additional contribution from the vertical direction.

\begin{figure}[tp]
  % Skripti R1_K_NT.py
  \center
  \begin{tabular}{ccc}
  \includegraphics[width=0.21\textwidth]{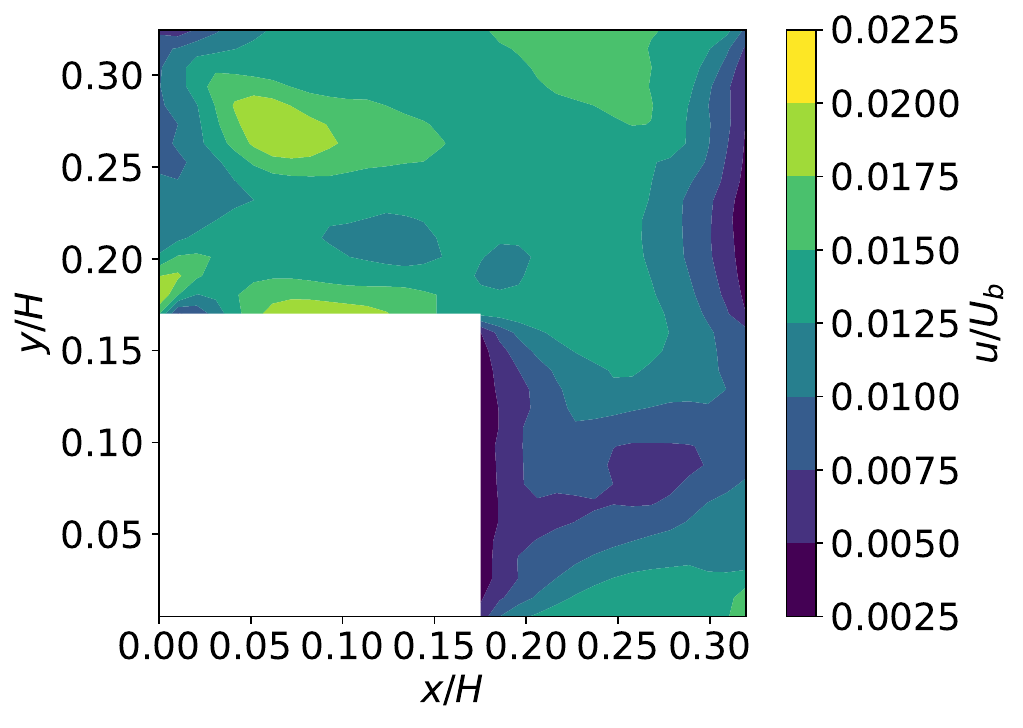} &
  \includegraphics[width=0.20\textwidth]{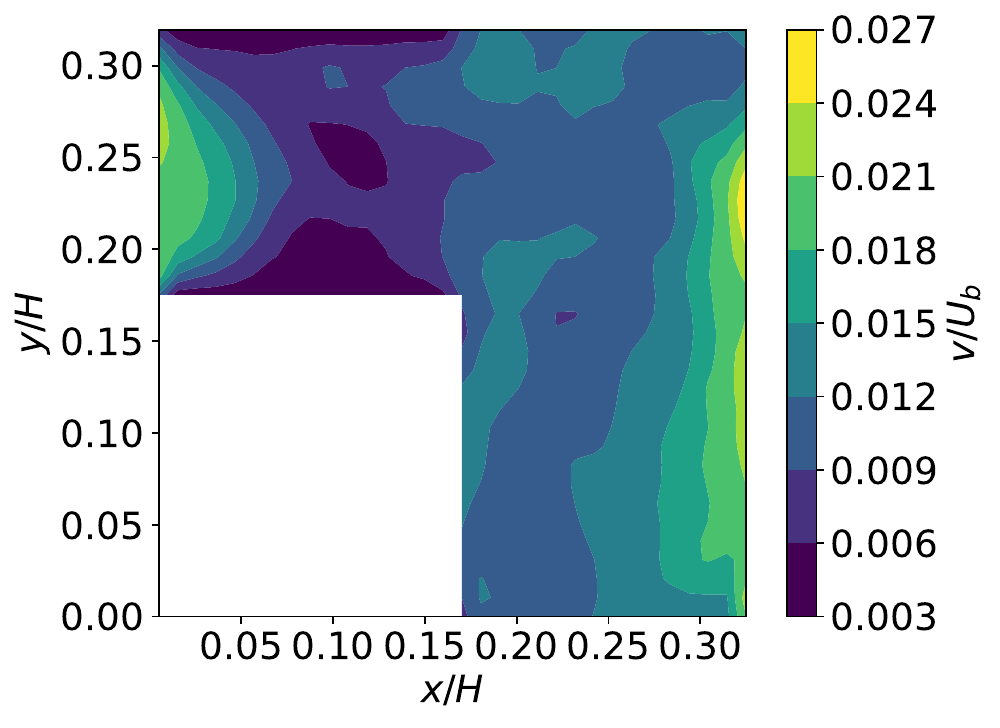} &
  \includegraphics[width=0.20\textwidth]{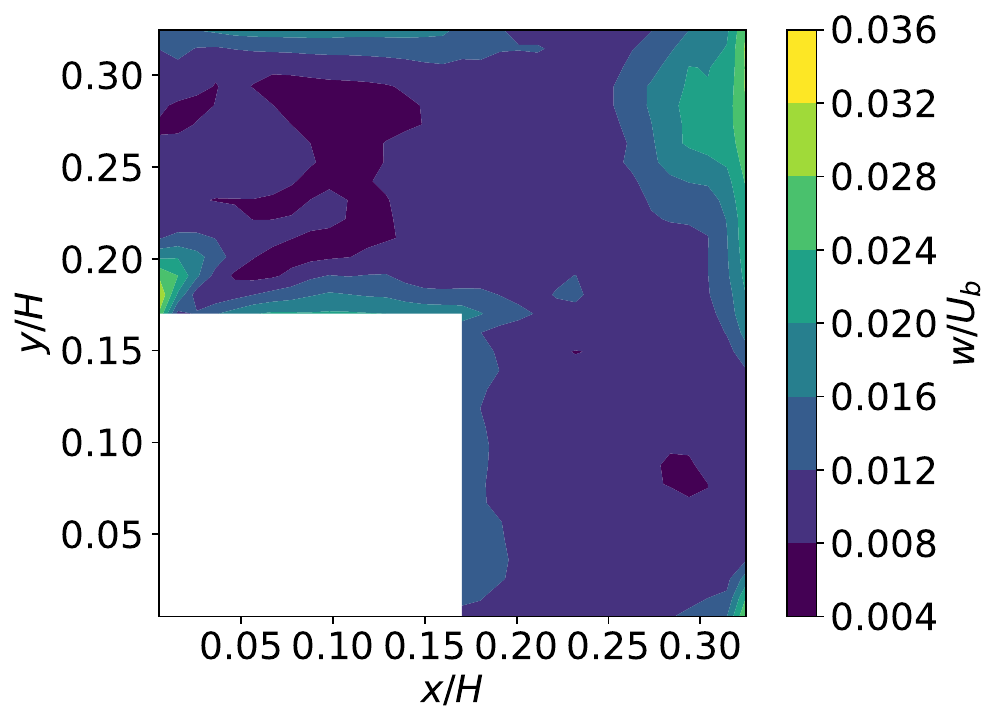} \\
  \includegraphics[width=0.21\textwidth]{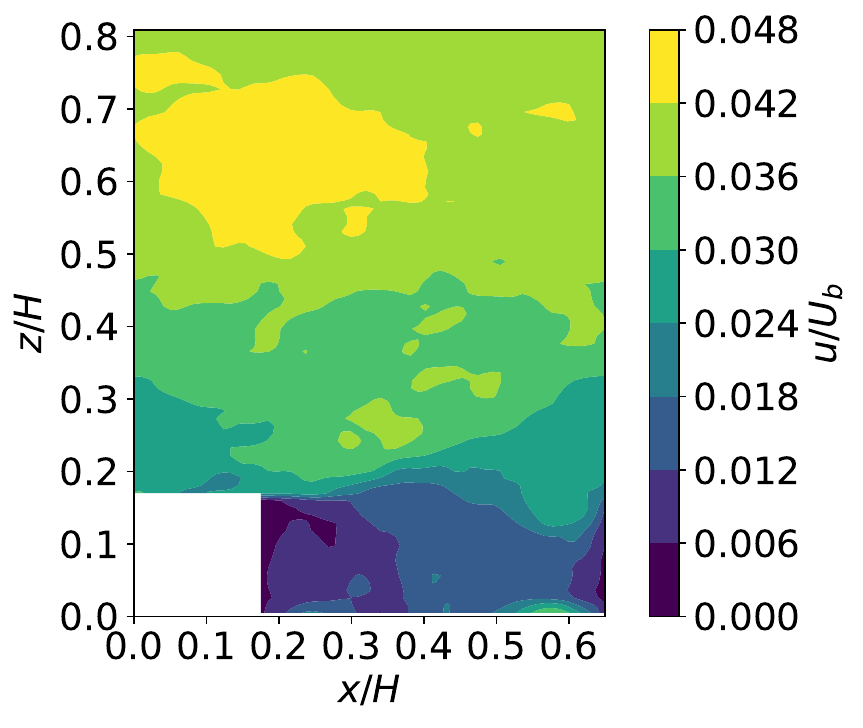} &
  \includegraphics[width=0.21\textwidth]{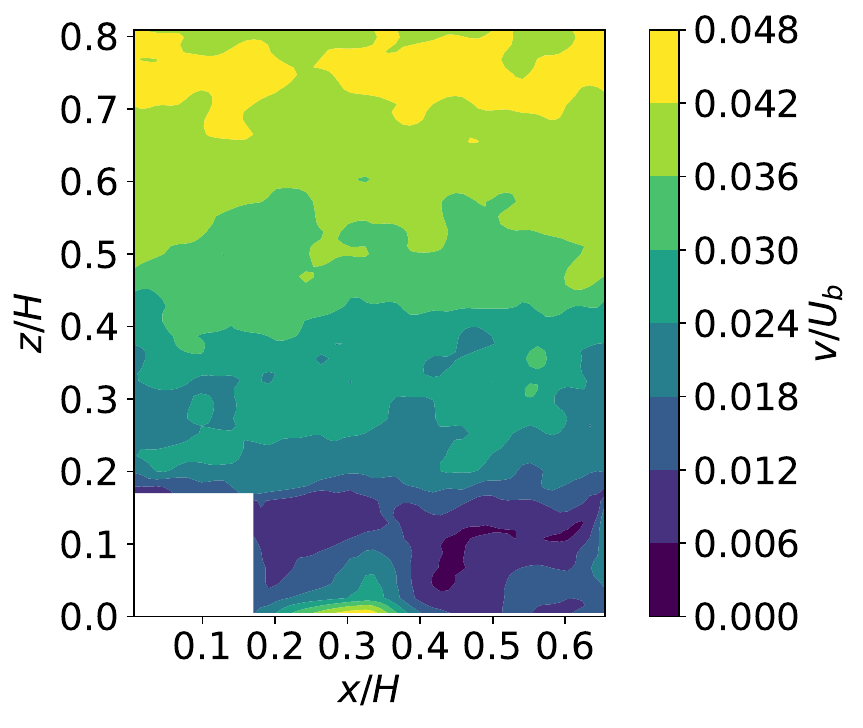} &
  \includegraphics[width=0.21\textwidth]{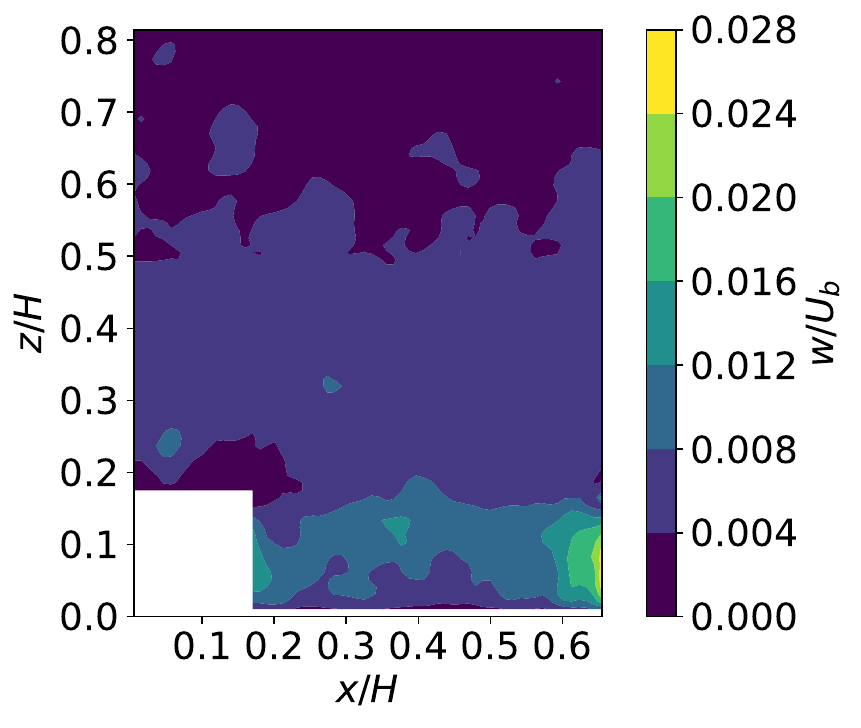} \\
  \includegraphics[width=0.21\textwidth]{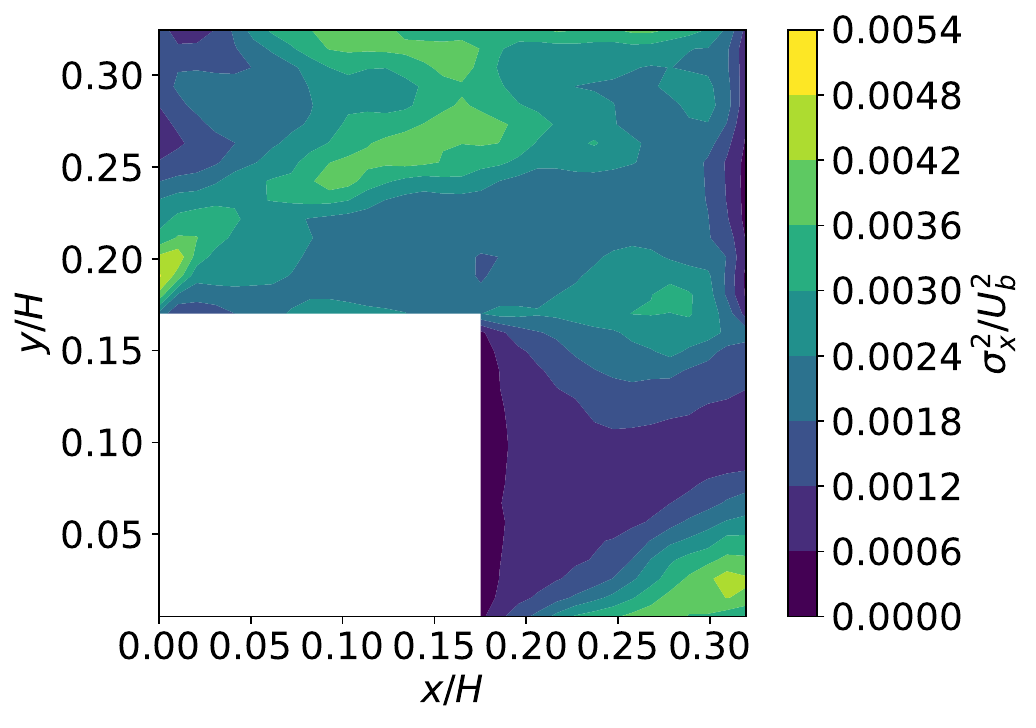} &
  \includegraphics[width=0.21\textwidth]{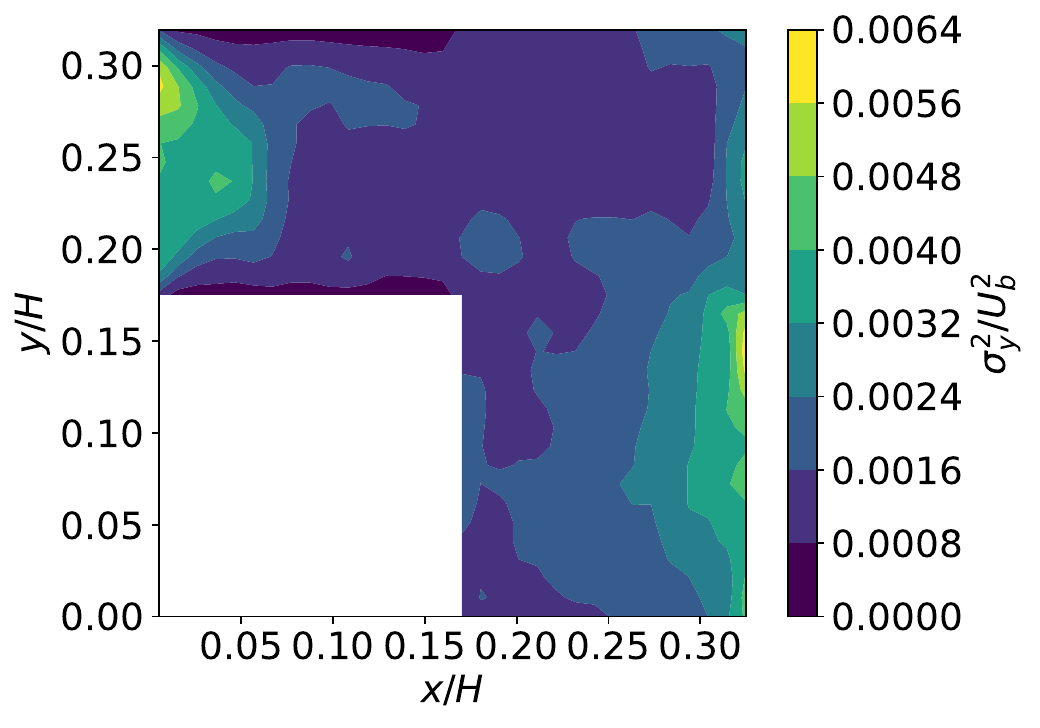} &
  \includegraphics[width=0.21\textwidth]{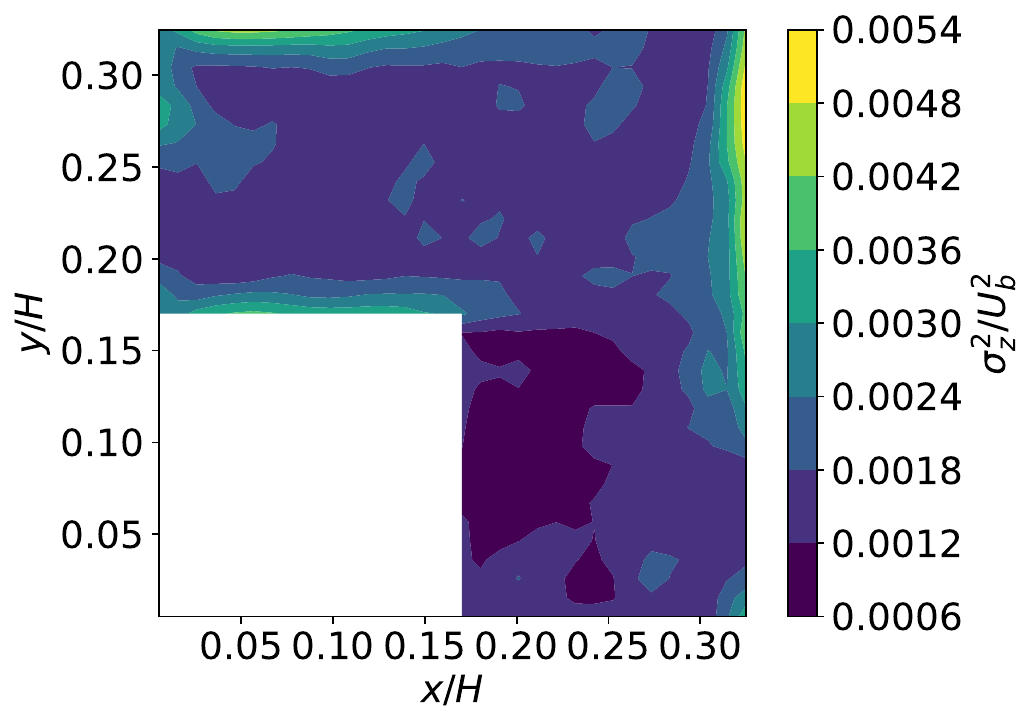} \\  
  \includegraphics[width=0.21\textwidth]{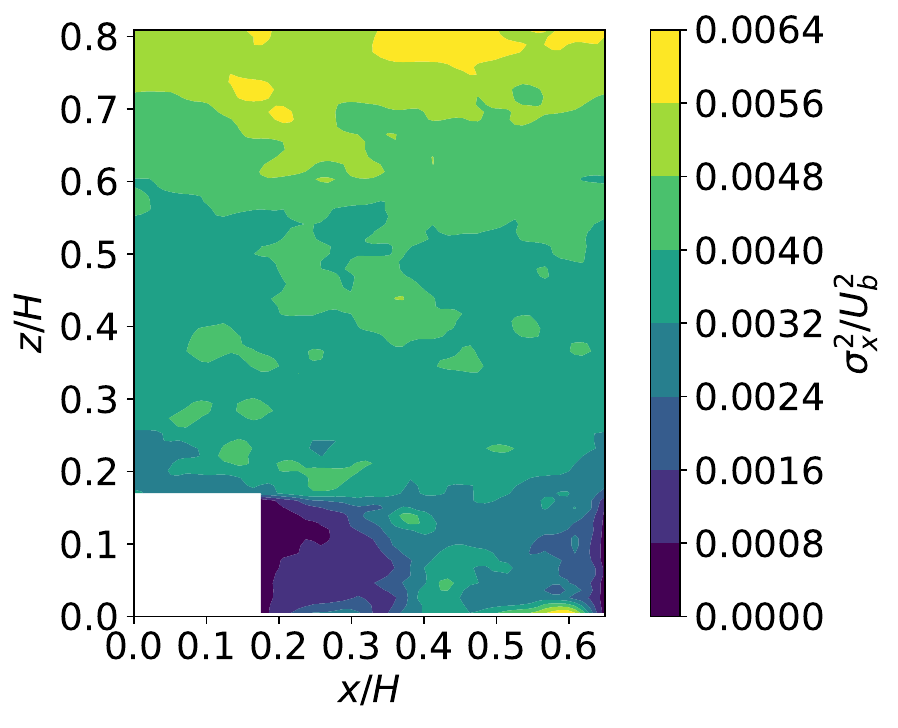} &
  \includegraphics[width=0.19\textwidth]{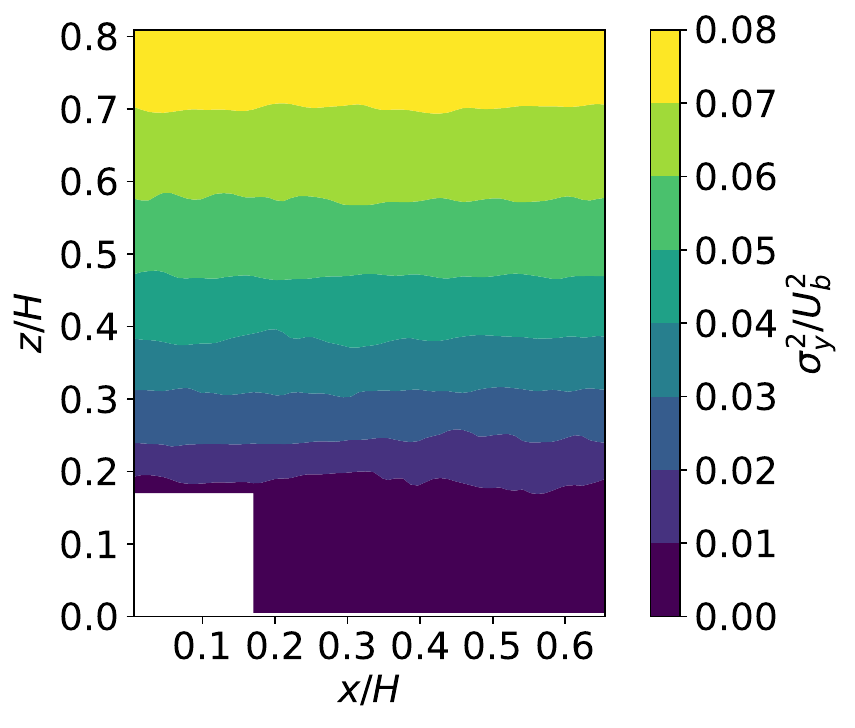} &
  \includegraphics[width=0.21\textwidth]{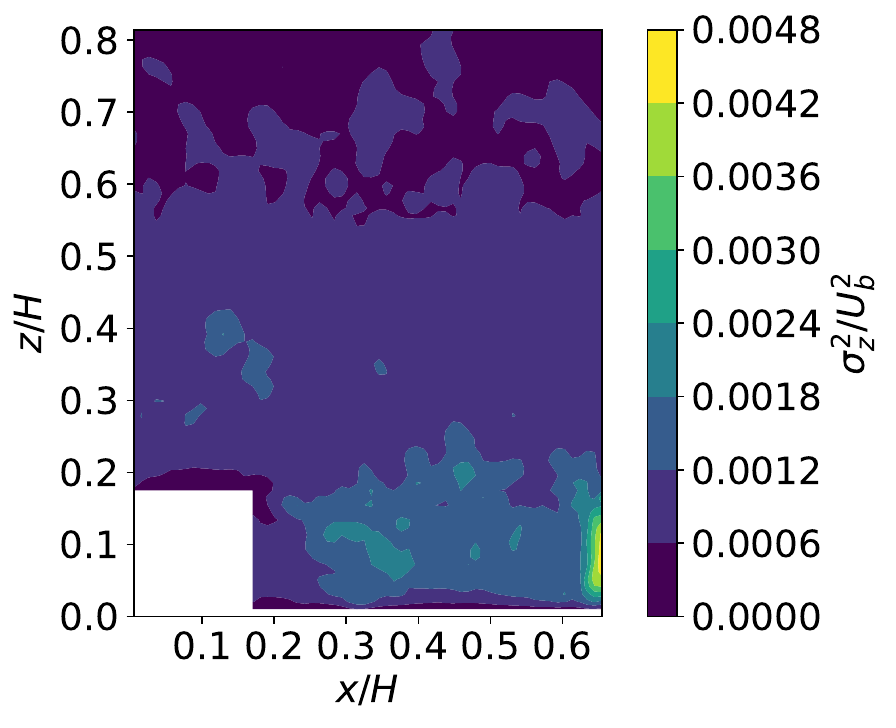} \\
  \includegraphics[width=0.21\textwidth]{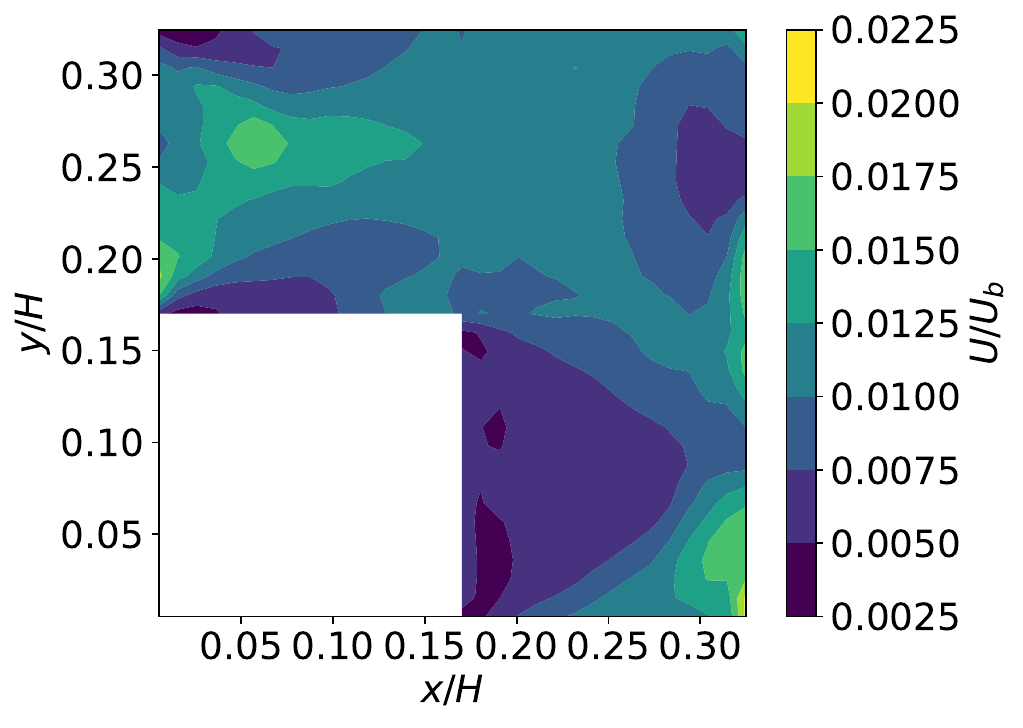} &
  \includegraphics[width=0.20\textwidth]{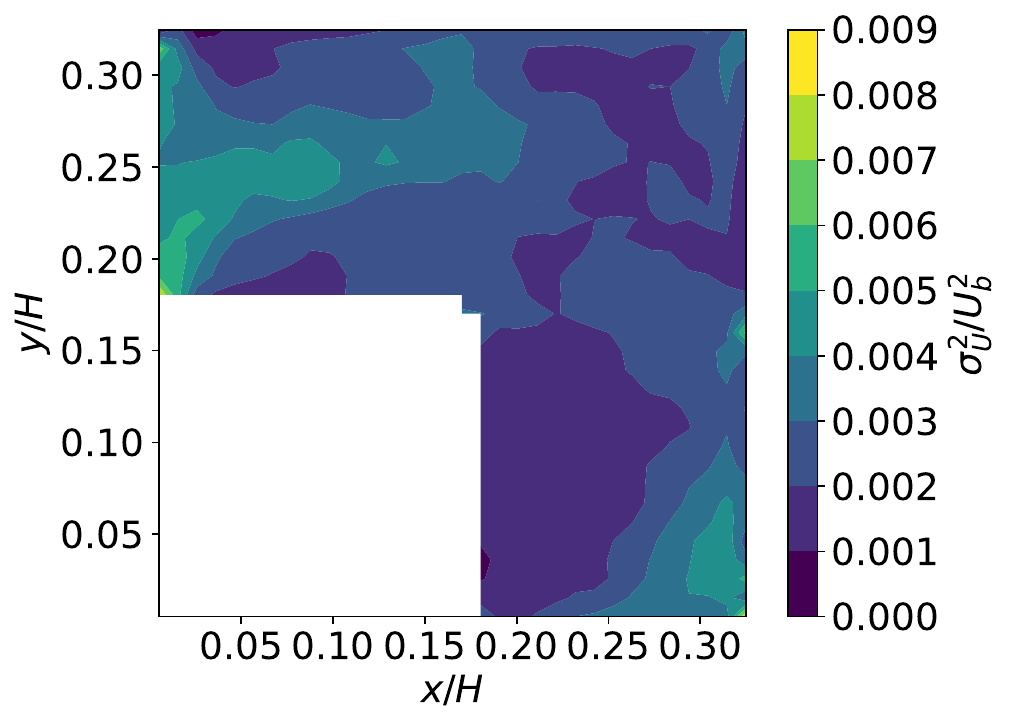} &
  \includegraphics[width=0.21\textwidth]{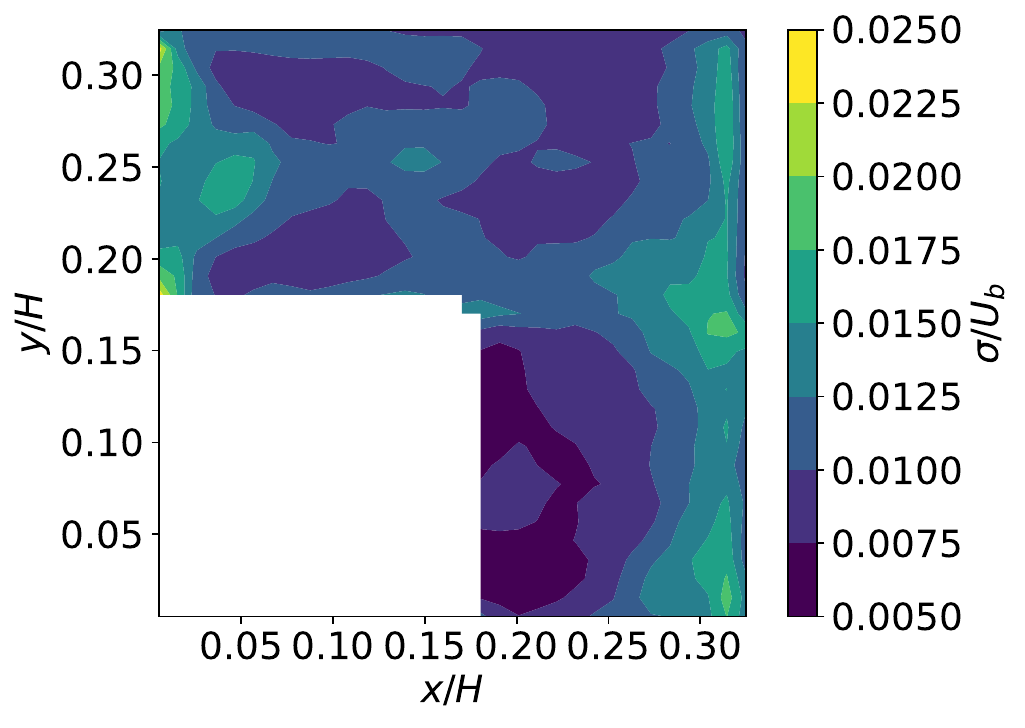} \\
  \includegraphics[width=0.21\textwidth]{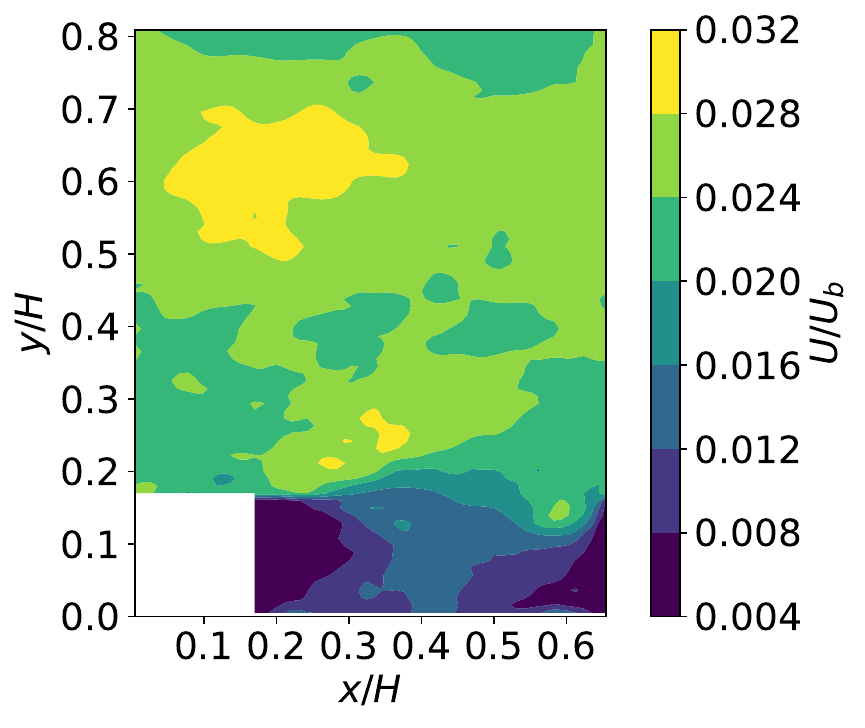} &
  \includegraphics[width=0.21\textwidth]{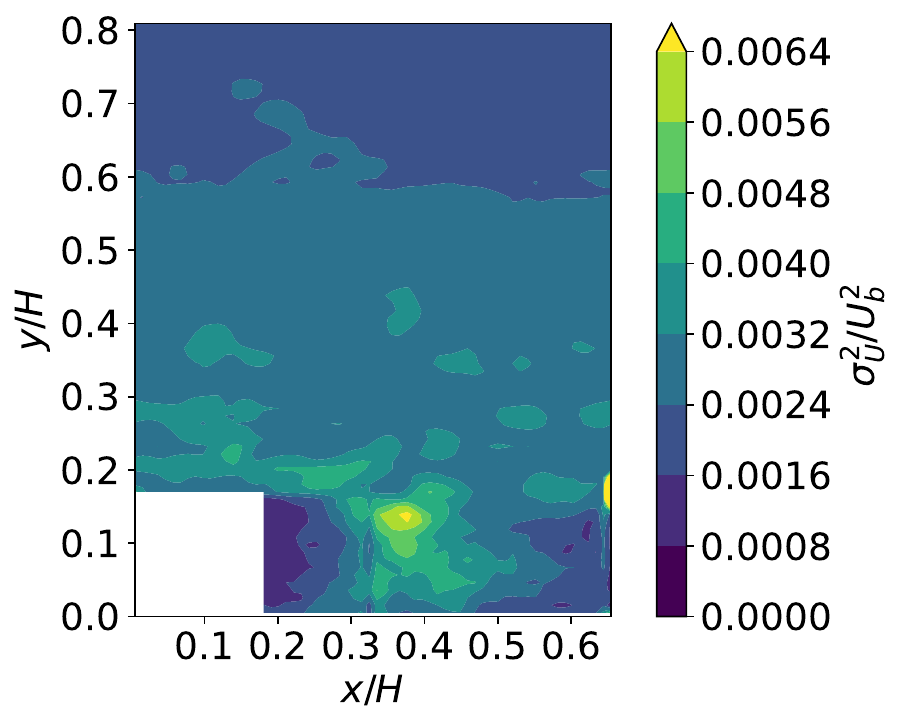} &
  \includegraphics[width=0.20\textwidth]{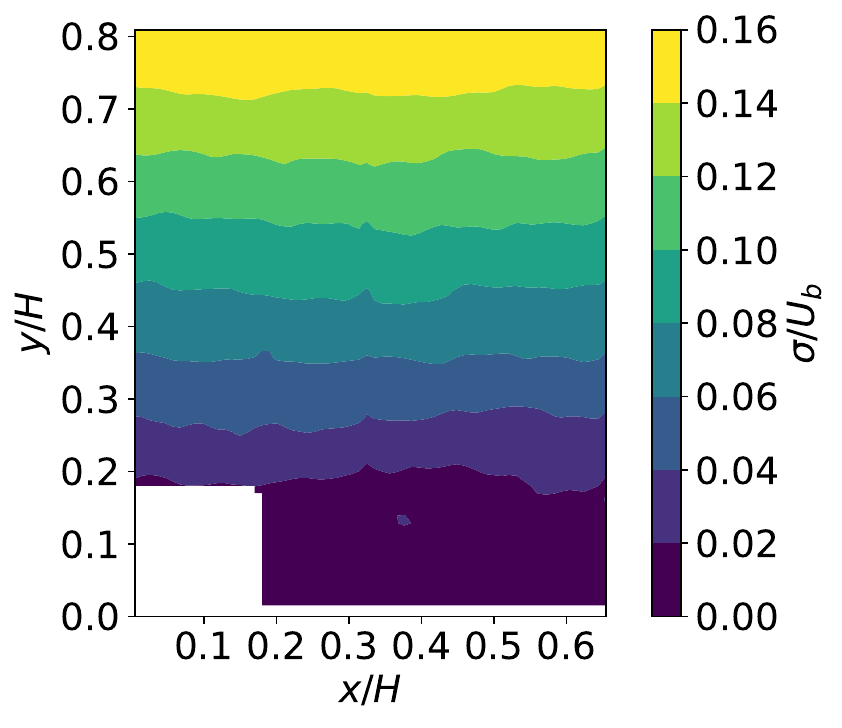} \\
  \end{tabular}
  \caption{\small RMSE between the ensemble velocity and the long ($t/T_\Omega = 0.920$) time average.
    Two different planes of a single repeating element are show, one horizontal through the roughness elements at $z/H=0.0825$ (panels on the first, third, and fifth rows from the top) and one vertical plane through the roughness elements at $y/H=0.0825$ for $0 \le x/H \le 0.330$ and at $y/H=0.247$ for $0.330 \le x/H \le 0.660$ (panels of the second, fourth, and sixth rows from the top).
    The coordinates refer to a single repeating element and the cube corner is at the origin.
    The simulations contained several repeating elements and hence the shown  horizontal plane should be considered as a part of a staggered pattern in the $x$ direction and as an aligned patter in the $y$ (see, Fig.\,\ref{kuutiokuva}).
    The shown vertical plane wraps through the repeating element and should be considered to repeat
    in the $x$ direction only.
    The panels on the first two rows show the components of the mean velocity with $u/U_b$ on the first column, $v/U_b$ on the second columns, and $w/U_b$ on the third column.
    Variances $\sigma^2_x/U_b^2$, $\sigma^2_y/U_b^2$, and $\sigma^2_z/U_b^2$ are shown similarly on the panels in the third and fourth rows.
    The lowermost two rows show the horizontal mean wind speed $U/U_b$ on the first column, horizontal wind variance $\sigma^2_U/U_b^2$ on the second column, and turbulence intensity $\sigma/U_b$ on the third column}
  \label{poikki2}
\end{figure}

To obtain a view on the spatial distribution of the differences (errors) between ensemble-averaged and time-averaged mean and variance, we utilize the RMSE.
It is calculated for each time-averaged ensemble member using the ensemble mean as the reference.
The average RMSE for the the long time average ($t/T_\Omega = 0.920$) with the quantities and planes shown in Fig.\,\ref{poikki1} are displayed in Fig.\,\ref{poikki2}.
The largest errors, seen most clearly on the mean and variance of $u$ and $v$ as well as for the mean horizontal velocity, are above the obstacles.
The large scale mean wind and the wind turning have their strongest effect there and a part of the reason for the large absolute errors is the relatively large wind speed.
As discussed earlier, these errors can be mitigated by the use of spatial averaging in the horizontal directions when this is done sufficiently high above the roughness elements.

Within the obstacles, a somewhat different image of the errors is seen.
The overall distribution of the errors is similar for all components and for both mean and variance.
The biggest errors are located in the narrow area between two obstacles at at $0 \le x/H \le 0.16$ and $0.16 \le y/H \le 0.32$, in the in the vicinity of the windward face of the obstacle, and at the at bottom boundary in front of the obstacle.
In the narrow area between two obstacles, the flow is constrained and accelerated.
This is expected to create a stronger fluctuations that are probably prone to change with the changing wind direction.
In the case of pulsatile flow studied by \citet{li_mean_2023}, the spatial locations of the differences between the time-averaged and ensemble-averaged flow is different as the largest variations between the different phases are located in the recirculation regions behind and on the sides of the obstacle.
The pulsatile forcing and the turning pressure gradient thus produce different flows with different nonstationary characteristics.

The errors close to the windward face and on the bottom boundary in front of it are most likely related.
Based on the ensemble averages in Fig.\,\ref{poikki1}, there is a recirculation system between the obstacles with downward flow on the windward face.
Turbulence is created in the shear layer at the top of the obstacles, indicated by elevated variances downwind from the obstacle at the obstacle height, and then transported towards the next obstacle by the mean wind.
This turbulence is then transported down towards the bottom of the domain by the recirculation region.

The mean and variance of the horizontal wind speed, shown in the two first columns of the last two rows in Fig.\,\ref{poikki2}, indicate mostly a similar distribution of errors as do the Cartesian wind components.
However, the errors in the upper parts of the domain are gone for the variance.
This confirms our earlier observation on the suitability of the horizontal wind instead of Cartesian components for the study of the region above the roughness sublayer.

\begin{figure}[tp]
  % Skripti R1_K_TD.py
  a) \hspace{0.32\textwidth} b) \hspace{0.32\textwidth} c) \\
  \includegraphics[width=0.32\textwidth]{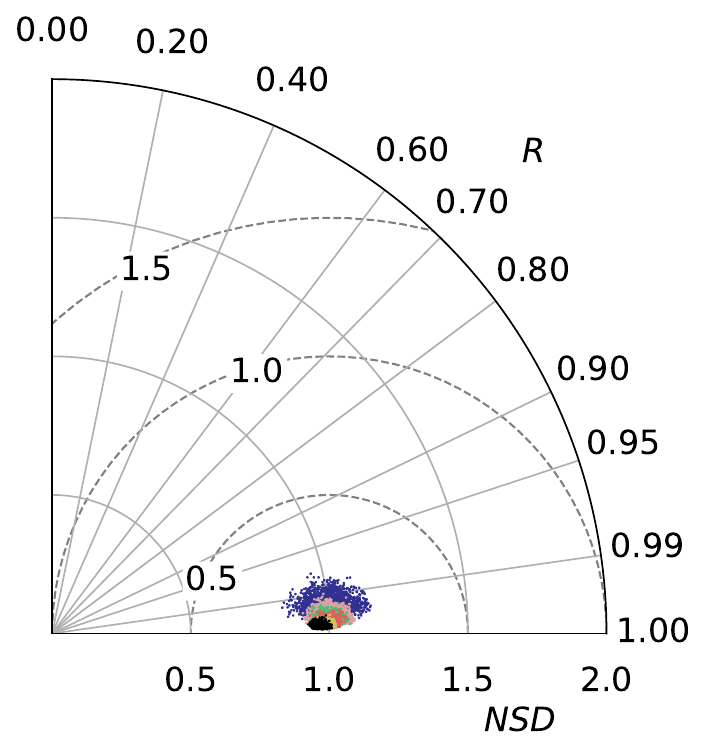}
  \includegraphics[width=0.32\textwidth]{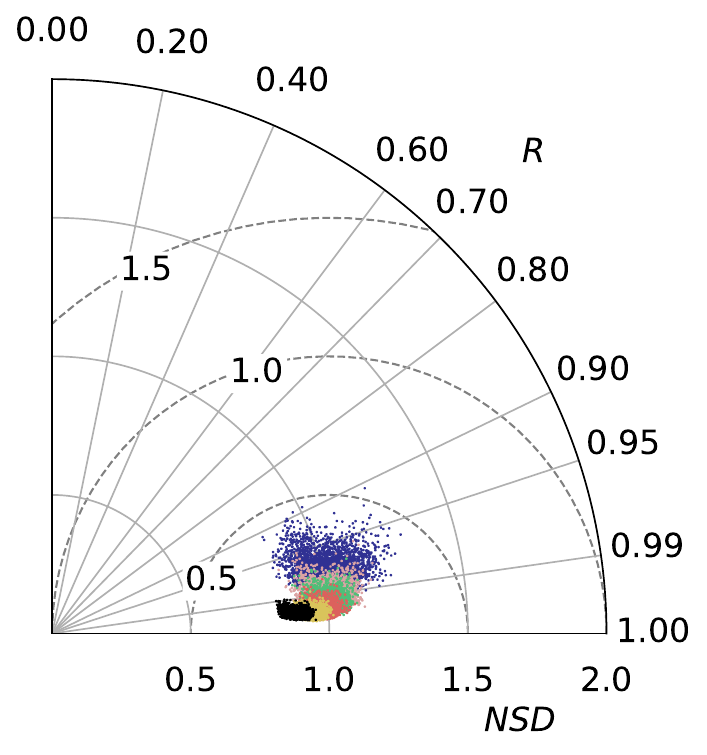}
  \includegraphics[width=0.32\textwidth]{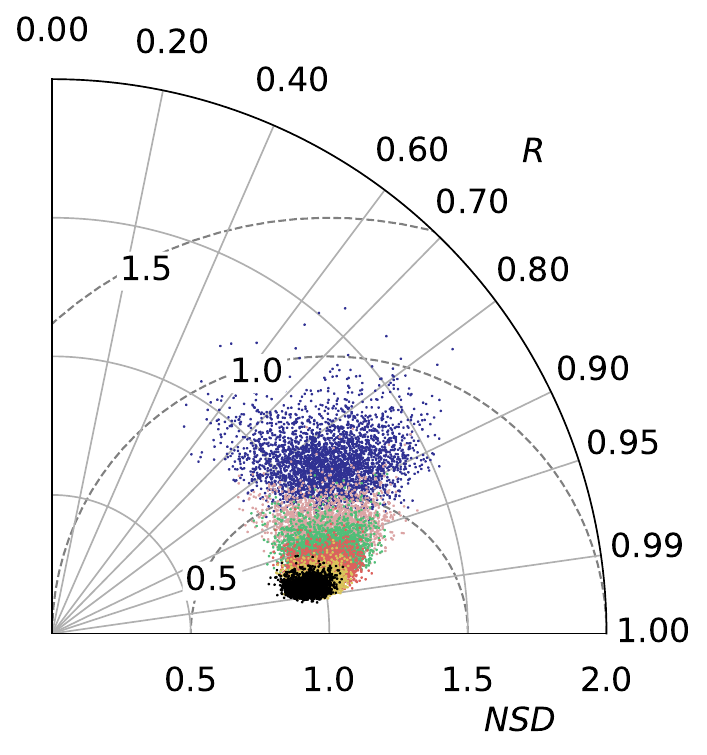} \\
  d) \hspace{0.32\textwidth} e) \hspace{0.32\textwidth} f) \\
  \includegraphics[width=0.32\textwidth]{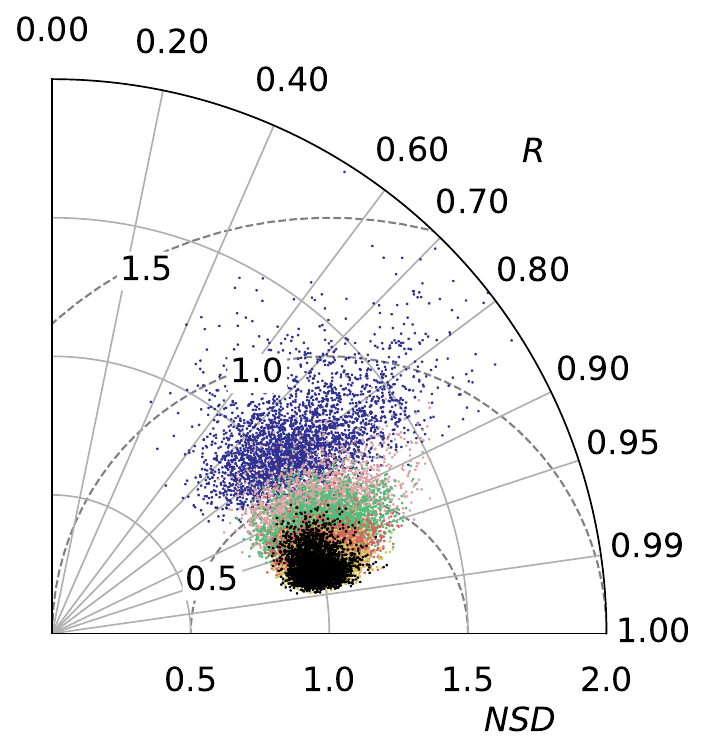}
  \includegraphics[width=0.32\textwidth,clip]{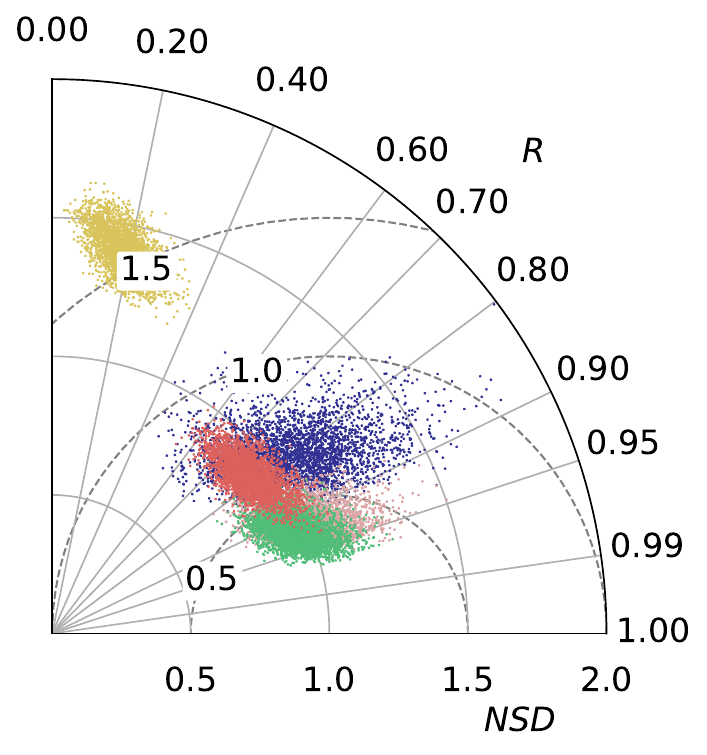}
 \includegraphics[width=0.32\textwidth]{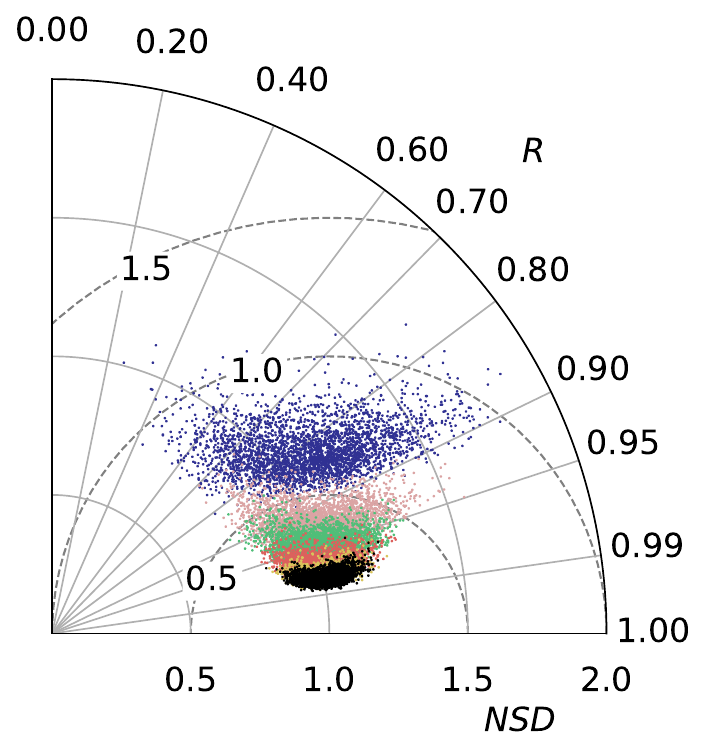}   \\
  g) \hspace{0.32\textwidth} h) \hspace{0.32\textwidth}  \\
  \includegraphics[width=0.32\textwidth]{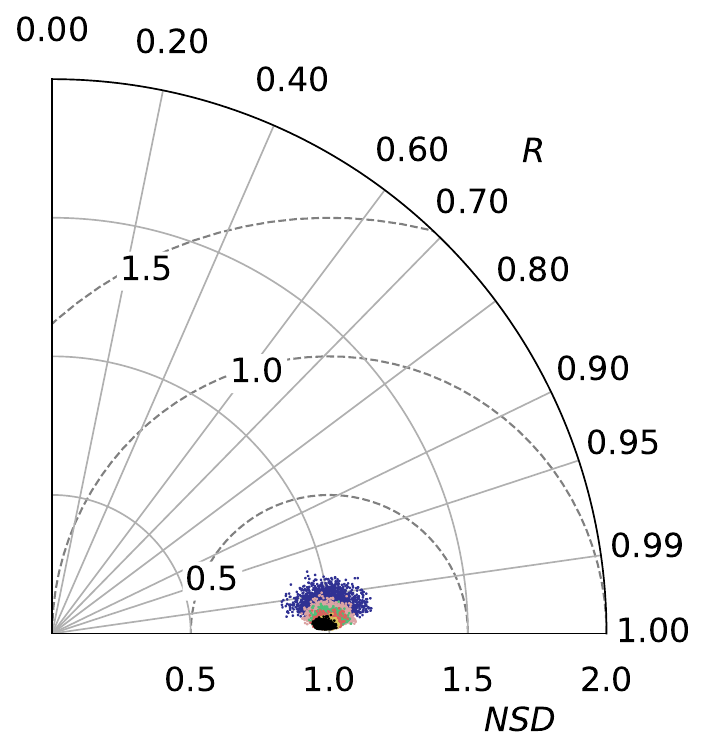} 
  \includegraphics[width=0.32\textwidth]{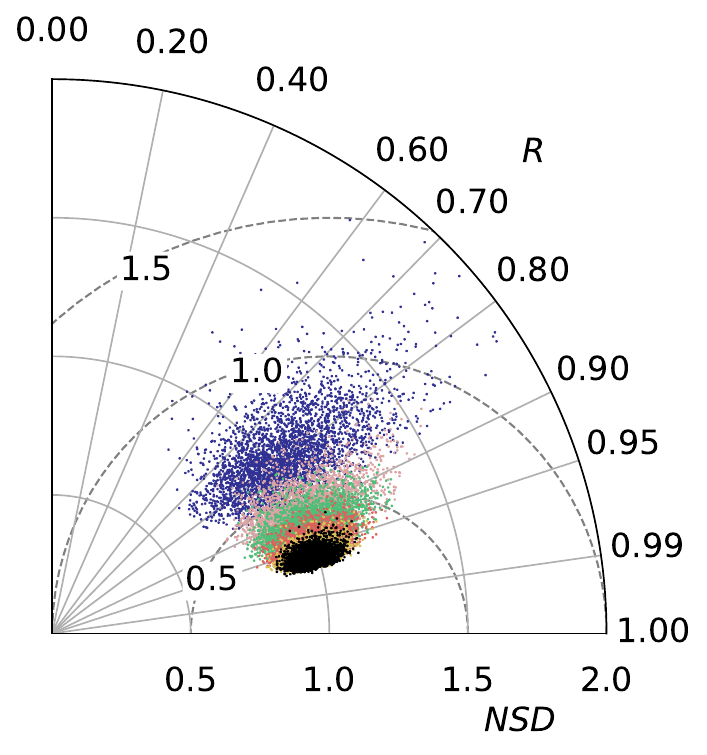}  \\
  \caption{\small 
    Taylor diagrams for the mean (a--c, g) and variance (d--f, h) of $u$, $v$, $w$, and $U$ components of velocity with different averaging intervals in roughness sublayer ($z/H < 0.660$) for the case of the staggered cube array.
    The ensemble mean calculated with instantaneous values is used as the reference.
    The circles drawn using a dashed lines indicate NRMSE with a perfect model at point (1,1).
    Each dot indicates a time-averaged ensemble member and the colours indicate different averaging intervals for $0.0438 T_\Omega$ (blue), $0.131 T_\Omega$ (pink), $0.219 T_\Omega$ (green), $0.394 T_\Omega$ (red), $0.657 T_\Omega$ (yellow), and $0.920 T_\Omega$ (black).
    Note that in e) the $0.657 T_\Omega$ and $0.920 T_\Omega$ values are outside the plot area and not shown for this reason}
  \label{TD1}
\end{figure}

To further quantify the errors, we utilize Taylor diagrams \citep{taylor_summarizing_2001,chang_air_2004}.
We evaluate the required error measures for different time averages in each cell in the roughness sublayer i.e. the flow below the height of $4h=0.660H$ using the ensemble average as the reference value.
Finally, the cell-values of the error measures are spatially averaged to produce a single value per ensemble member which are shown in Fig.\,\ref{TD1}.
Above the roughness sublayer the flow field is not sensitive to the details of individual roughness elements, the flow is horizontally homogeneous, and hence spatial averaging in the horizontal directions can be utilized instead of ensemble or time averaging.
One has to keep in mind, however, that the use of spatial averaging requires that there are no deviations from homogeneity in the roughness pattern on scales much larger than the representative size of the roughness elements. 

The effect of averaging time on the error as compared to the ensemble mean is shown in Fig.\,\ref{TD1}~a)–c) for the Cartesian velocity components $u$, $v$, and $w$.
Each dot represents an ensemble member and the colours indicate different averaging periods.
For all components, a longer averaging time results in a smaller spread within the ensemble.
However, in the case of $v$, the $0.657 T_\Omega$ and $0.920 T_\Omega$ averages show a larger error than the $0.394 T_\Omega$ average due to decrease in NSD.
Nevertheless, time averaging appears to improve the accuracy of the mean velocity components for averaging times up to $0.394 T_\Omega$ and even after that the error is not very large.
Averaging over the whole duration of the simulation with a changing wind direction hence does not have a major effect on the mean velocity components in the roughness sublayer.
The horizontal mean velocity $U$ in Fig.\,\ref{TD1} g) behaves in a manner very similar to that of $u$.

The variances, shown in the same manner in Fig.\,\ref{TD1}~d)–f), are different to the means.
Firstly, the spread between the ensemble members is much larger than with the averages.
Secondly, both the $u$ and the $v$ component of variance show deteriorating accuracy for longer time
statistics beyond $0.657 T_\Omega$ and $0.219 T_\Omega$, respectively. Furthermore, the accuracy of the $v$ variance decreases catastrophically for averaging periods longer than $0.394 T_\Omega$.
The errors for $0.657 T_\Omega$ and $0.920 T_\Omega$ $v$ variances are so large that they are outside the plot shown in panel e).
The variance of $w$, however, appears well-behaved and averaging over any length of time improves the statistic.
The variance of horizontal wind in Fig.\,\ref{TD1} h) is otherwise very similar to $\sigma_x^2$ but it has a slightly worse performance in $R$.

The error measures included in the Taylor diagram (NSD, R, and NRMSE) are indicators of unsystematic errors only \citep{chang_air_2004} and hence a Taylor diagram might give a false impression of the performance of the simulations.
To explore the systematic errors, we have calculated the fractional bias \citep{chang_air_2004} in the same manner and for the same time averaging intervals as in the Taylor diagram in Fig.\,\ref{TD1}.
The resulting interquartile ranges are shown in the upper half of Table\,\ref{FB1}.
Both $u$ and $v$ display a similar but not exactly the same kind of performance measured with fractional bias as they did in the Taylor diagram.
The fractional bias decreases for both components with increasing averaging time until $0.394 T_\Omega$.
After that, the fractional bias becomes worse for both but more so for $v$.
The interquartile range of the vertical component $w$, however, does not tell much about the bias but more about the failure of fractional bias for this case.
The fractional bias is calculated using a fraction and because the mean of $w$ is zero for large parts of the domain, division by very small numbers is heavily influencing this case.
The fractional bias for the horizontal velocity is similar to that of $u$, especially up to $0.394 T_\Omega$ after which the horizontal velocity shows a slightly better performance than $u$.

Based on the interquartile ranges of fractional bias, the systematic errors in variances behave in a similar manner to the unsystematic errors as
indicated by the Taylor diagram in Fig.\,\ref{TD1}.
The systematic biases in the variance of $u$ decrease until $0.657 T_\Omega$ and are relatively small even for the $0.920 T_\Omega$ averaging time.
In the case of $w$ variance, the smallest interquartile range of fractional bias is reached at $0.920 T_\Omega$.
The variance of $v$ shows improvement in fractional bias until $0.131 T_\Omega$ averaging time and deterioration after that.
As with the Taylor diagrams, $0.657 T_\Omega$ and $0.920 T_\Omega$ time averages show a very poor performance for the variance $\sigma_y^2$.
Unlike in the case of the mean velocities, the horizontal variance has fractional bias at similar levels as $\sigma_x$ for all averaging times.

\begin{table}
  % Skriptit R1_T_V.py ja R1_T_V2.py
  \caption{\small The interquartile ranges of the fractional bias for different averaging times (upper half) and for different ensemble sizes (lower half).
    All values have been calculated within the roughness sublayer ($z < 4h = 0.660H$) of the staggered cube-array.
    In the upper half, fractional bias has been calculated in each cell for different averaging times using the full ensemble average as the reference.
    The cell values of fractional bias are then spatially averaged to produce a single value per ensemble member which are then used to calculate the shown interquartile range.
    In the lower half, new ensembles with different sizes are created using $0.131 T_\Omega$ time averages by randomly sampling the original ensemble.
    As with different averaging times, the fractional bias are first calculated for each cell using the originalm instantaneous ensemble mean as the reference and then averaged to create a single value per new ensemble.
    For each ensemble size, 1\,000 new ensembles are created and the interquartile range is calculated using this ensemble of new ensembles}
  \begin{center} \small
  \begin{tabular}{llcccc}
    \hline 
        &    & $u$ & $v$ & $w$ & $U$ \\ \hline
   $\langle \cdot \rangle_t$ & 0.0438 $T_\Omega$  & [-0.036,\,0.040] & [-0.045,\,0.048] & [-2.012,\,-1.988] & [-0.035,\,0.039] \\ 
    & 0.131 $T_\Omega$  & [-0.019,\,0.025] & [-0.027,\,0.028] & [-2.019,\,-1.982] & [-0.019,\,0.023] \\
    & 0.219 $T_\Omega$  & [-0.014,\,0.021] & [-0.014,\,0.027] & [-2.023,\,-1.972] & [-0.014,\,0.020] \\
    & 0.394 $T_\Omega$  & [-0.005,\,0.021] & [-0.007,\,0.038] & [-2.032,\,-1.966] & [-0.007,\,0.017] \\
    & 0.657 $T_\Omega$  & [0.007,\,0.027] & [0.051,\,0.079] & [-2.042,\,-1.956] & [-0.001,\,0.019] \\
    & 0.920 $T_\Omega$ & [0.017,\,0.039] & [0.111,\,0.152] & [-2.045,\,-1.955] & [0.003,\,0.024] \\ \hline
  $\sigma$  & 0.0438 $T_\Omega$  & [0.048,\,0.230] & [-0.022,\,0.093] & [-0.015,\,0.086] & [0.010,\,0.204] \\
   & 0.131 $T_\Omega$  & [0.006,\,0.119] & [-0.055,\,0.016] & [-0.013,\,0.046] & [-0.025,\,0.090] \\
   & 0.219 $T_\Omega$  & [-0.013,\,0.074] & [-0.104,\,-0.047] & [-0.011,\,0.036 ] & [-0.042,\,0.049] \\
   & 0.394 $T_\Omega$ & [-0.024,\,0.042] & [-0.261,\,-0.218] & [-0.011,\,0.023 ] & [-0.044,\,0.019] \\
   & 0.657 $T_\Omega$ & [-0.045,\,0.013] & [-0.560,\,-0.523] & [-0.010,\,0.016] & [-0.048,\,0.007]\\
   &  0.920 $T_\Omega$  & [-0.070,\,-0.009] & [-0.841,\,-0.795] & [-0.012,\,0.011] & [-0.054,\,-0.008] \\ \hline 
   $\langle \cdot \rangle_t$ & 2 memb & [-0.013,\,0.022] & [-0.018,\,0.021] & [-2.025,\,-1.973] & [-0.015,\,0.016] \\
    & 5 memb      & [-0.008,\,0.012] & [-0.011,\,0.013] & [-2.036,\,-1.953] & [-0.008,\,0.012] \\
    & 10 memb     & [-0.006,\,0.009] & [-0.010,\,0.009] & [-2.062,\,-1.937] & [-0.006,\,0.009] \\
    & 25 memb     & [-0.003,\,0.006] & [-0.006,\,0.006] & [-2.099,\,-1.904] & [-0.003,\,0.006] \\
    & 50 memb     & [-0.002,\,0.005] & [-0.004,\,0.004] & [-2.131,\,-1.853] & [-0.002,\,0.005] \\
    & 100 memb    & [-0.001,\,0.004] & [-0.003,\,0.003] & [-2.193,\,-1.820] & [-0.001,\,0.003] \\ \hline
   $\sigma$ & 2 memb & [-0.008,\,0.082] & [-0.052,\,-0.002] & [-0.013,\,0.030] & [-0.041,\,0.046] \\
    & 5 memb      & [-0.008,\,0.045] & [-0.047,\,-0.016] & [-0.006,\,0.021] & [-0.040,\,0.017] \\
    & 10 memb     & [-0.003,\,0.034] & [-0.045,\,-0.023] & [-0.005,\,0.014] & [-0.037,\,0.005] \\
    & 25 memb     & [-0.002,\,0.021] & [-0.040,\,-0.027] & [-0.001,\,0.011] & [-0.033\,\-0.008] \\
    & 50 memb     & [0.000,\,0.018] & [-0.039,\,-0.030] & [ 0.001,\,0.009] & [-0.031,\,-0.01] \\
    & 100 memb    & [0.002,\,0.014] & [-0.038,\,-0.034] & [0.002,\,0.008] & [-0.028,\,-0.016] \\ \hline   
  \end{tabular}
  \end{center}
  \label{FB1}
\end{table}

The statistical convergence of an ensemble can be improved by utilizing time averaged values instead of instantaneous values.
In an earlier study concerning a forest edge flow, \citet{kanani_what_2014} showed that a ten-member ensemble is sufficient for their case when 15-minute time averages are used as ensemble members.
The compromise between the deterioration of accuracy due to time averaging and the cost of producing a large ensemble can be expected to be different in different flow cases. 

Based on the Taylor diagrams in Fig.\,\ref{TD1} and the fractional bias in Table \ref{FB1}, loss of solution quality is observed in $\sigma_y^2$ starting at the averaging time $0.219T_\Omega$ and culminating at catastrophic level of inaccuracy at $0.920 T_\Omega$.
Some of the other quantities are also affected but not as badly.
In order to avoid deteriorating effects from plain time averaging, the averaging time needs to be much smaller than the characteristic time scale of the turning $T_\Omega$.
In this paper we make a conservative choice and calculate ensemble statistic utilizing the averaging interval of $0.131 T_\Omega$.
This averaging time should not have a negative impact on the solution quality but, due to the time averaging, should increase the sample size and hence allow for smaller ensemble size.

\begin{figure}[tp]
  % Skripti R1_K_TD2.py
  a) \hspace{0.32\textwidth} b) \hspace{0.32\textwidth} c) \\
  \includegraphics[width=0.32\textwidth]{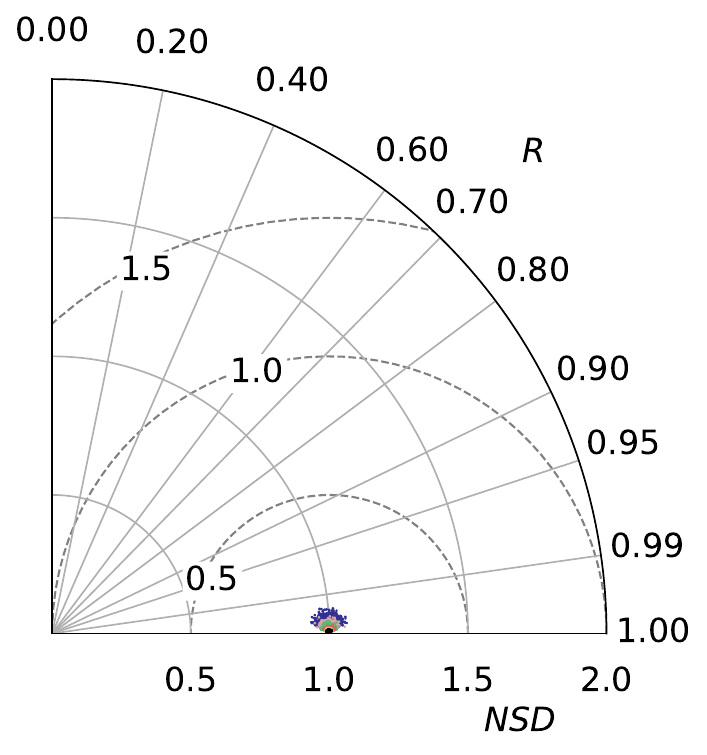}
  \includegraphics[width=0.32\textwidth]{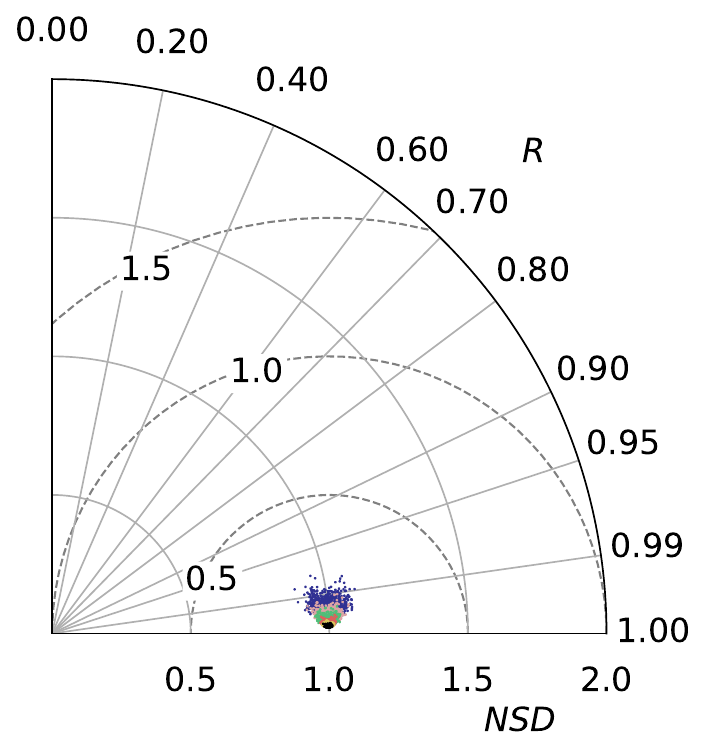}
  \includegraphics[width=0.32\textwidth]{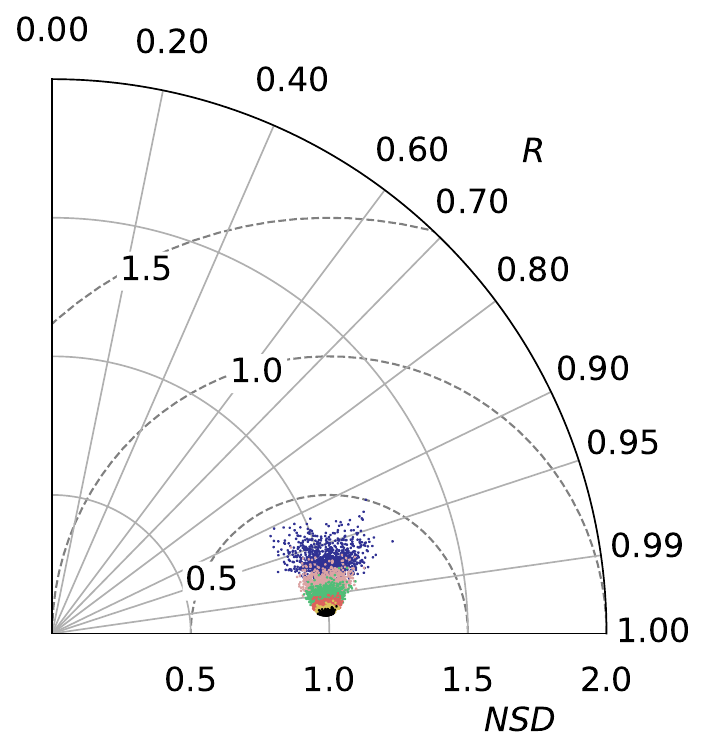} \\
  d) \hspace{0.32\textwidth} e) \hspace{0.32\textwidth} f)\\
  \includegraphics[width=0.32\textwidth]{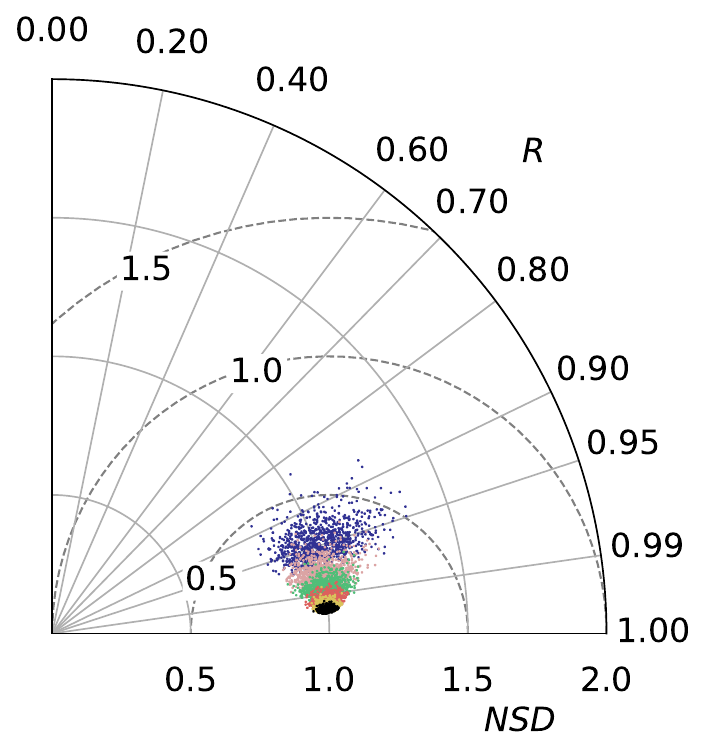}
  \includegraphics[width=0.32\textwidth,clip]{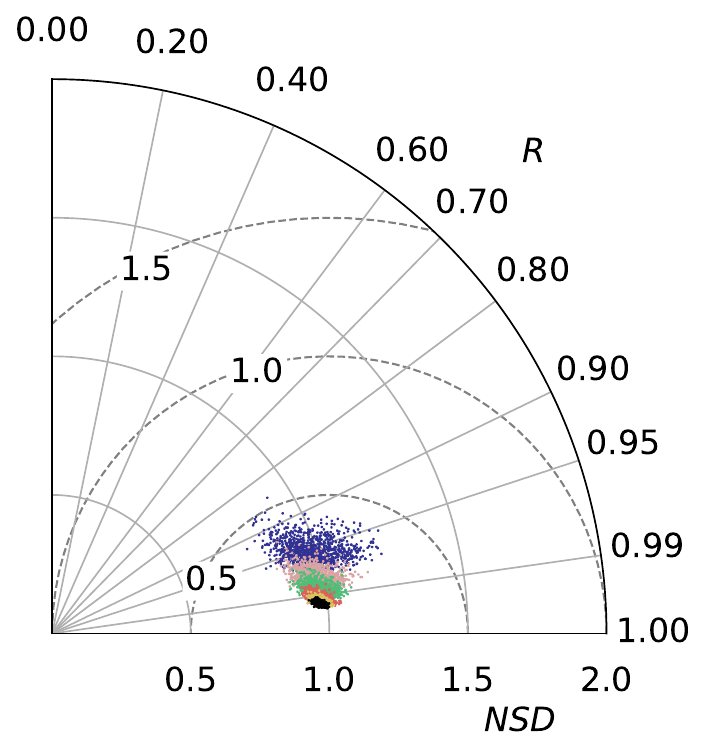} 
  \includegraphics[width=0.32\textwidth]{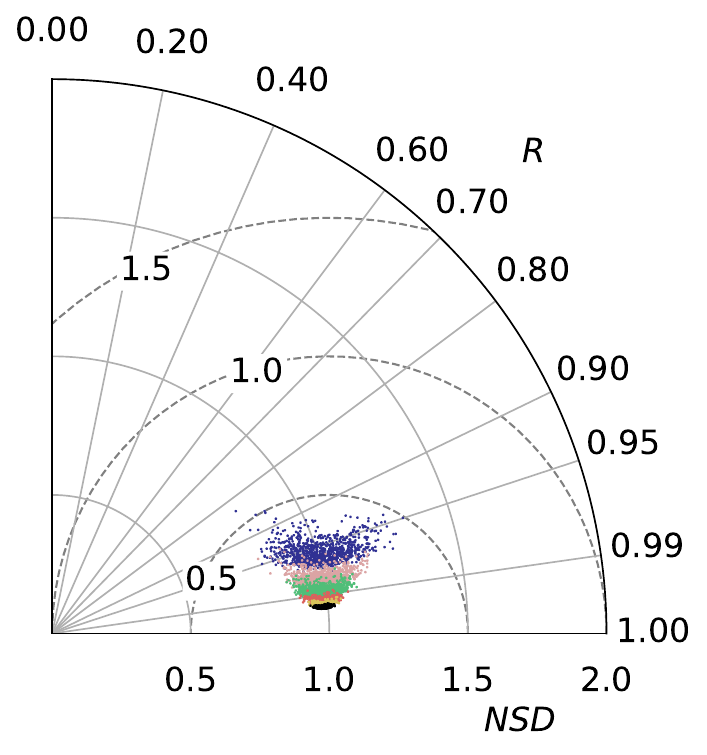}  \\
  g) \hspace{0.32\textwidth} h) \hspace{0.32\textwidth} \\
  \includegraphics[width=0.32\textwidth]{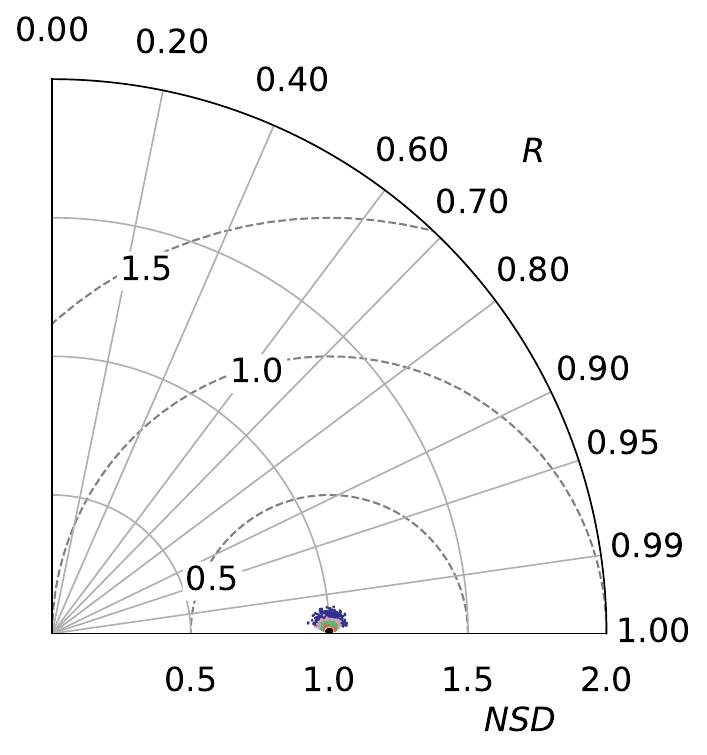} 
  \includegraphics[width=0.32\textwidth]{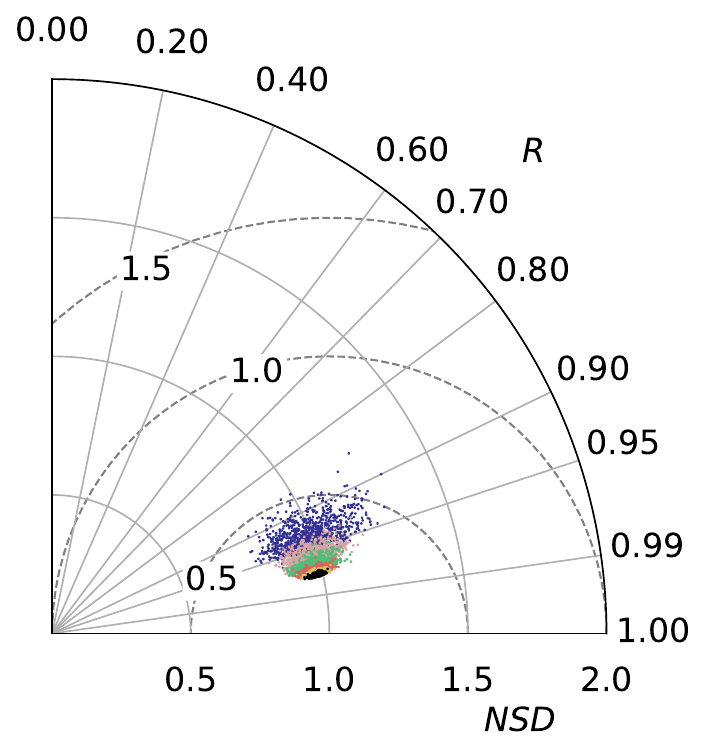}  \\
  \caption{\small 
    Taylor diagrams for ensembles with different sizes calculated using $0.131 T_\Omega$ (a--c, g) and variance (d--f, h) of $u$, $v$, $w$, and $U$ components of velocity for the case of staggered cube-array.
    The ensemble mean calculated with instantaneous values is used as the reference.
    The circles drawn using a dashed lines indicate NRMSE with a perfect model at point (1,1).
    Each dot indicates an ensemble and the colours indicate different ensemble sizes: 2 members (blue), 5 members (pink), 10 members (green), 25 members (red), 50 members (yellow), and 100 members (black)}
  \label{TD2}
\end{figure}

Counting both different simulations and the repeating elements, we have a total of 3\,240 ensemble members available for the analysis, each with temporally accumulated statistics over $0.131 T_\Omega$.
By selecting members from this full, temporally averaged ensemble, we have created smaller ensembles with sizes ranging from two to 100 members.
As each ensemble member is an equally correct representation of the flow, the sampling of the full ensemble is done randomly with equal probability.
For each considered ensemble size, 1\,000 different new ensembles are created.
We then calculate ensemble means and variances from the new ensembles for all cells in the roughness sublayer, hence combining the statistics accumulated temporally over $0.131 T_\Omega$ in all members of the new ensemble.
As earlier, the error measures are then calculated for each cell by using the instantaneous ensemble value as the reference and which are then spatially averaged to produce a single value for each new ensemble.
Each dot in the Taylor diagram in Fig.\,\ref{TD2} represent one new ensemble and the interquartile ranges of fractional bias in the lower half of Table\,\ref{FB1} represent the variability within the ensemble of the new ensembles.

For all components of both the mean and variance, a steady convergence towards a smaller error can be seen in the Taylor diagram in Fig.\,\ref{TD2}.
As seen from the scatter of the points, the mean variables show less spread at all averaging times compared to the variances.
The $w$ component of mean velocity shows the largest spreading at small ensemble sizes and hence benefits the most from increasing ensemble size.
A similar behaviour can be observed for all components of variance.
Disregarding the mean of $w$, the interquartile ranges of fractioal bias in Table \ref{FB1} agree with this observation.
There are very few changes in the metrics between 50 and 100 members and not very much between 10 and 50 members.
An ensemble size of 10--50 members could thus be a good compromise between computational burden and accuracy when used together with $0.131 T_\Omega$ time averaging.

\subsection{Realistic Urban Environment}

\begin{figure}[tp]
  a) \\ % Skripti R1_KT_EAS1.py
  \includegraphics[width=\textwidth]{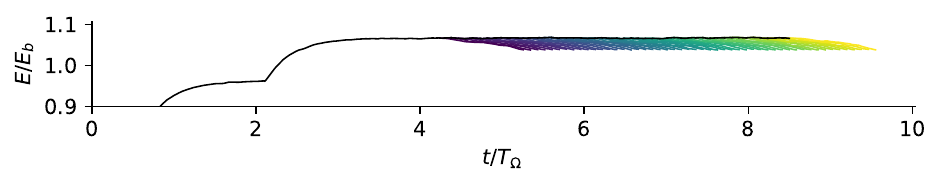} \\ b)
  \\ \includegraphics[width=\textwidth]{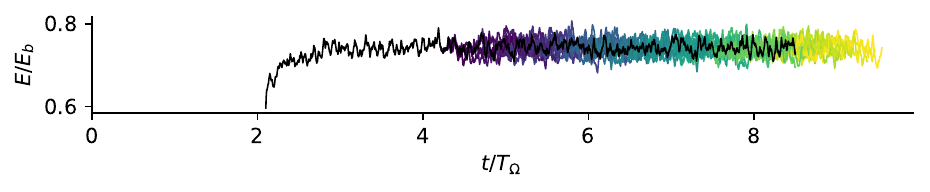} \\ c) \\ %
  %Skripti R1_KT_EAS2.py
  \includegraphics[width=\textwidth]{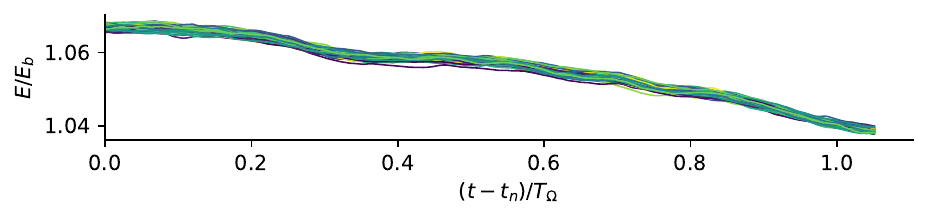} \\ d)
  \\ \includegraphics[width=\textwidth]{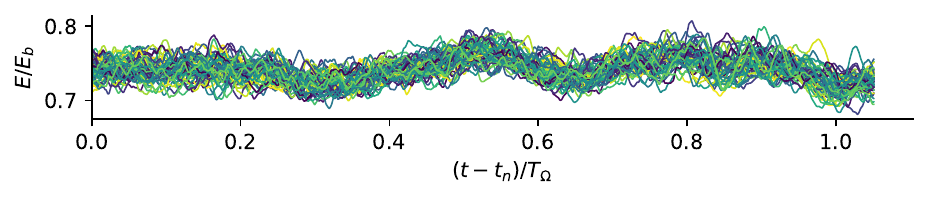}
  \caption{\small The volume-averaged, resolved kinetic energy of the realistic urban environment simulations.
    The $n$th turning pressure gradient simulation branches off from the spinup at time $t_n/T_\Omega = 4.20 + 0.0876(n-1)$.
    a) The spinup simulation (black) and the simulations with a turning pressure gradient (colours) for the largest simulation domain.
    b) The spinup simulation (black) and the simulations with a turning pressure gradient (colours) for the innermost simulation domain.
    c) All turning pressure gradient simulations for the largest domain shown using time since start of pressure gradient turning.
    d) All turning pressure gradient simulations for the innermost domain shown using time since start of pressure gradient turning}
\label{Turkuenaikas}       % Give a unique label
\end{figure}

In the case of the realistic urban environment, we have carried out a $6.39 T_\Omega$ long spinup simulation with constant pressure gradient directed at \unit[240]{$^\circ$}.
The simulation was started using only the largest domain in order to speed up to spinup process.
The two-way coupled, three-domain set-up was included after $2.10 T_\Omega$ single-domain spinup.
The volume-averaged, resolved mean kinetic energy within the largest domain, shown in Fig.\,\ref{Turkuenaikas} a) using the black line, develops quickly for the first $1.31 T_\Omega$ and then levels off.
We interpret this as the flow reaching a fully developed state at the largest domain and switch on the two-way coupled intermediate and small domains after $2.10 T_\Omega$.
This results in a jump in the kinetic energy content on the largest domain.
The kinetic energy content stabilizes at approximately $t/T_\Omega=2.63$ although there are clear small-scale, turbulent variations visible in the smallest domain, shown in Fig.\,\ref{Turkuenaikas} b).
The overall development of the resolved TKE (not shown) is similar to the total kinetic energy.

Two-way coupled, three-domain simulations with a turning pressure gradient are branched off every $0.0876 T_\Omega$ from the spinup starting at $t/T_\omega = 4.20$.
In total, an ensemble of 50 simulations with a turning pressure gradient are carried out.
As there are no repeating units in the realistic urban environment, the total ensemble size is the same as the number of simulations.
Based on the results of the staggered cube array in Sect. \ref{kuutiotulokset}, 50 ensemble members and short temporal averaging should provide sufficiently well-converged ensemble statistics.
The spinup simulation consumed approximately 61\,000 processing element hours in total while each of the turning pressure gradient simulations required approximately 9\,800 core hours on average.

The direction of the pressure gradient was at \unit[240]{$^\circ$} at the beginning of the simulation and \unit[300]{$^\circ$} at the end of the simulations.
As in the case of the staggered cube-array, we have calculated (not shown) the large-scale wind direction using only values in upper parts of the domain, above \unit[355]{m}.
Initially the large-scale wind direction was \unit[239]{$^\circ$} and it followed the pressure gradient with a lag in a similar manner to the cube array case.
There was a period of faster acceleration at the beginning of the simulation, until $0.262 T_\Omega$ when the approximate turning speed of $0.533 \Omega$ was reached.
For the rest of the simulation, there was a slower acceleration for the turning speed of the wind so that at the end of the simulation the turning speed was approximately $0.800 \Omega$.
At the middle of the simulation the wind direction was \unit[251]{$^\circ$} and \unit[271]{$^\circ$} at the end of the simulation.

The volume-averaged, resolved kinetic energy for the simulations with a turning pressure gradient are shown branching off from the spinup in Fig.\,\ref{Turkuenaikas} in panel a) for the largest domain and in panel b) for the innermost domain, using coloured lines.
Panels c) and d) show the volume-averaged, resolved kinetic energy of these simulations for time since the start of turning for the largest domain and the innermost domain, respectively.
For all simulations, the volume-averaged, resolved kinetic energy shows an overall decrease during the simulations in the outermost domain.
We interpret this to reflect the overall topography of the simulated domain: the river valley at the centre of the domain as well as the bays within the archipelago are all aligned in approximately northeast–southwest direction as a result of the latest ice age, as can be seen from the topography in Fig.\,\ref{kartta}.
The flow is expected to encounter less form drag due to large-scale channeling effects when it is approximately aligned with these large-scale structures of the simulated terrain.

The variation between the ensemble members appears stronger in the innermost domain than in the largest domain.
We expect the stronger variation to be due to a larger share of resolved turbulence, created by the roughness elements, in the innermost domain.
To quantify this variation, we normalized the kinetic energy content using time and ensemble-averaged kinetic energy and calculated the ensemble variance for each time instant.
With this, we observed (not shown) that the ensemble has three orders of magnitude less variation in the largest domain than in the innermost domain.
A comparable measure can be extracted from the staggered cube array ensemble by sampling the simulations and the repeating units.
Using 64 randomly selected repeating units from all simulations and a thousand 50 member ensembles, we observe the normalized mean variance to be of the same order as in the case of the innermost domain in the realistic flow.
For the remainder of the article, we focus on the innermost domain only.

\begin{figure}[tp]
  % Skripti R1_KT_PK.py
  \center
  a) \hspace{0.4\textwidth} b) \\
  \includegraphics[width=0.37\textwidth]{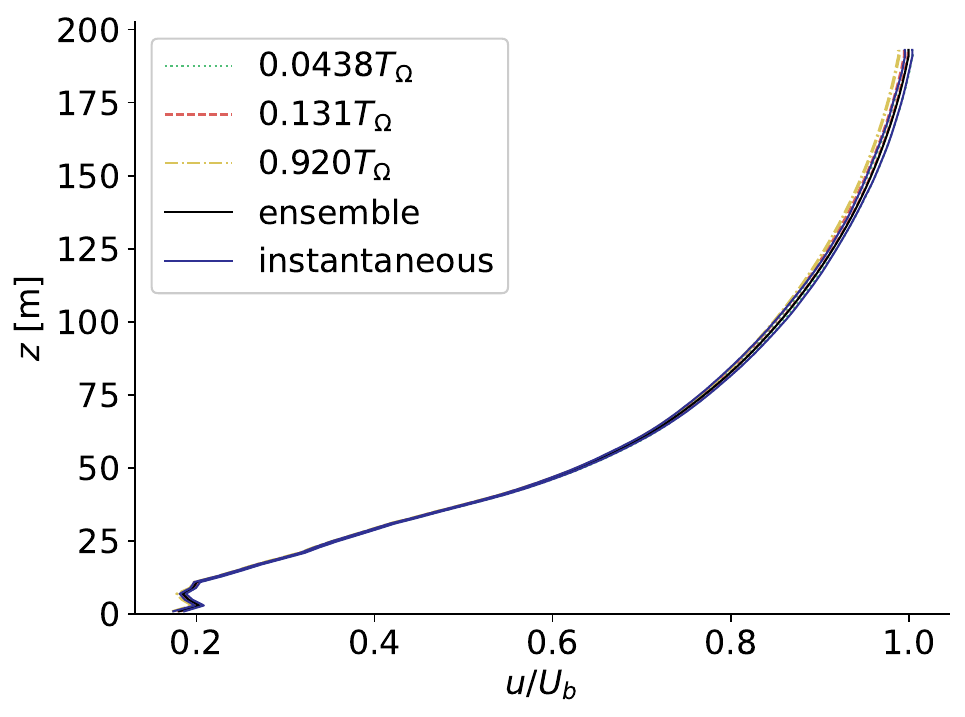}
  \includegraphics[width=0.37\textwidth]{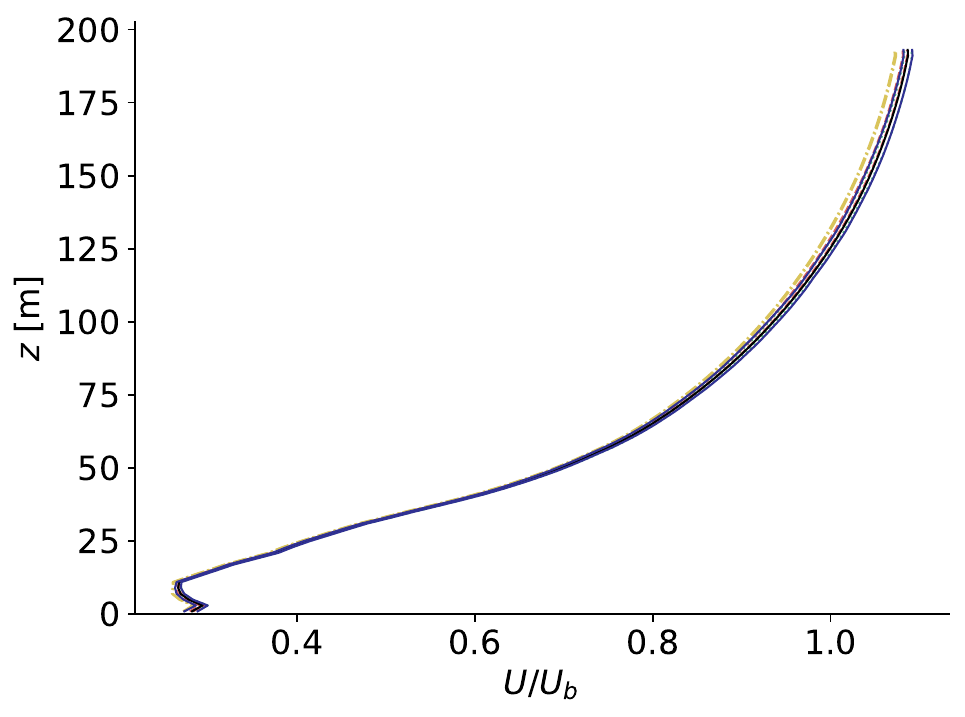} \\
  c) \hspace{0.4\textwidth} d) \\
  \includegraphics[width=0.37\textwidth]{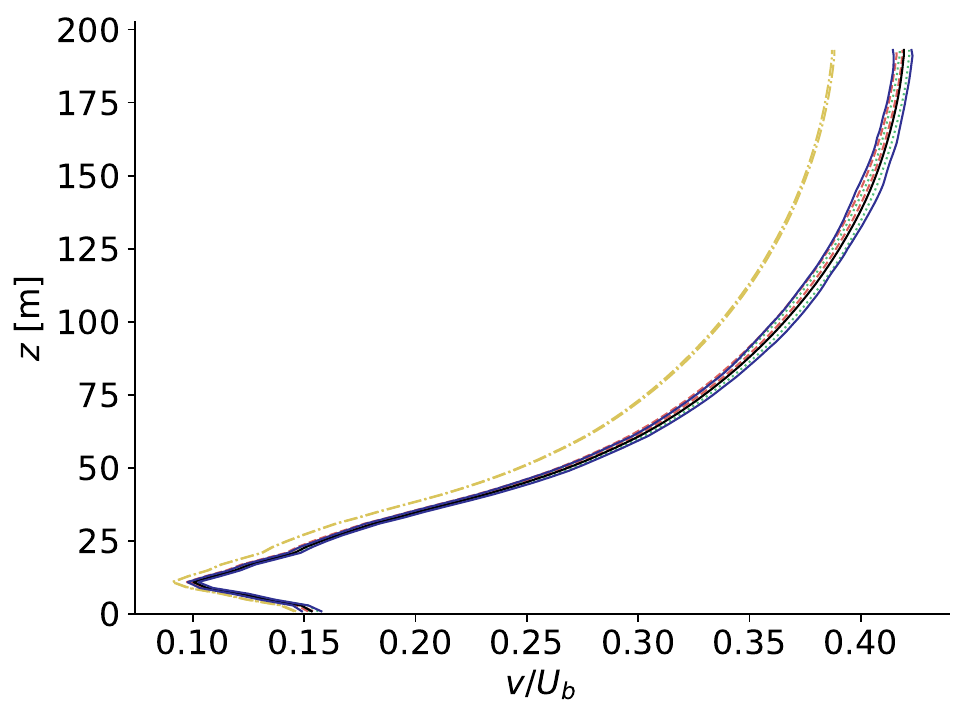}
  \includegraphics[width=0.37\textwidth]{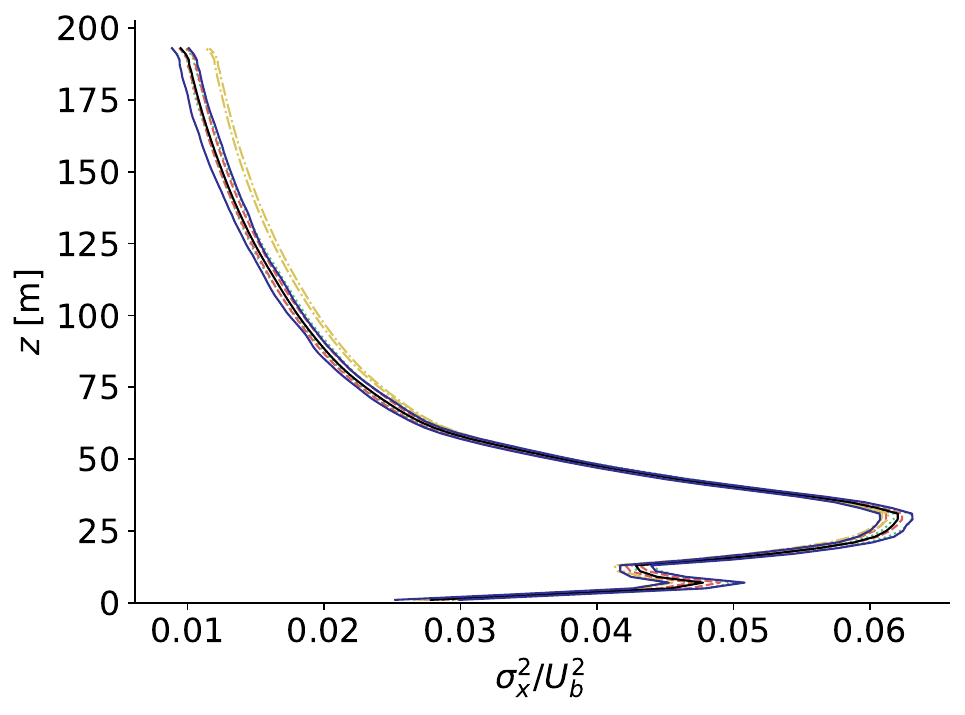} \\
  e) \hspace{0.4\textwidth} f) \\  
  \includegraphics[width=0.37\textwidth]{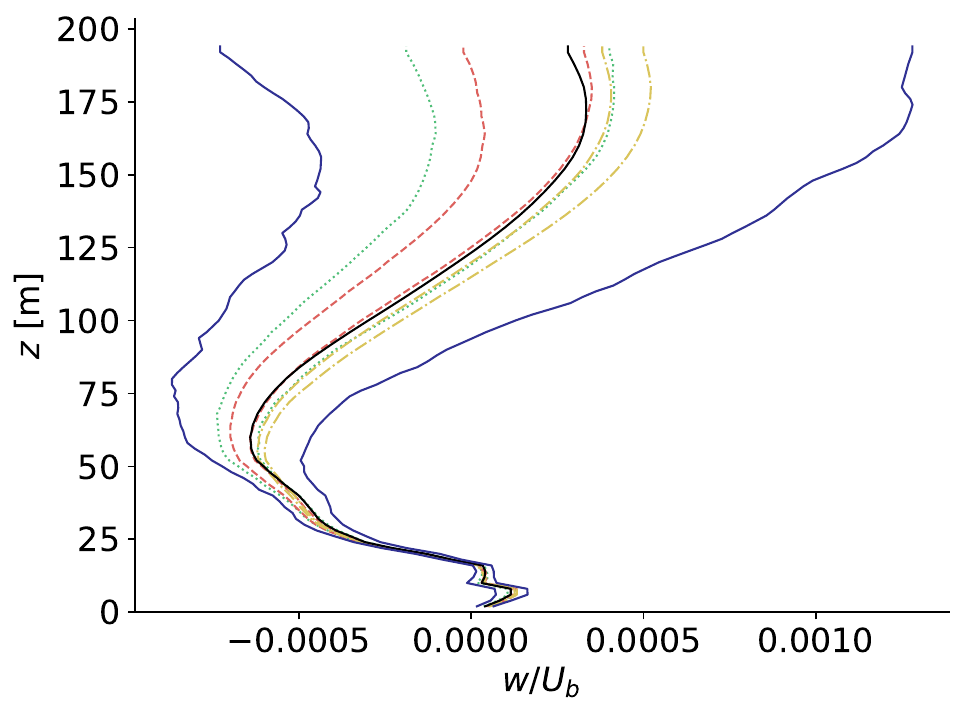}
  \includegraphics[width=0.37\textwidth]{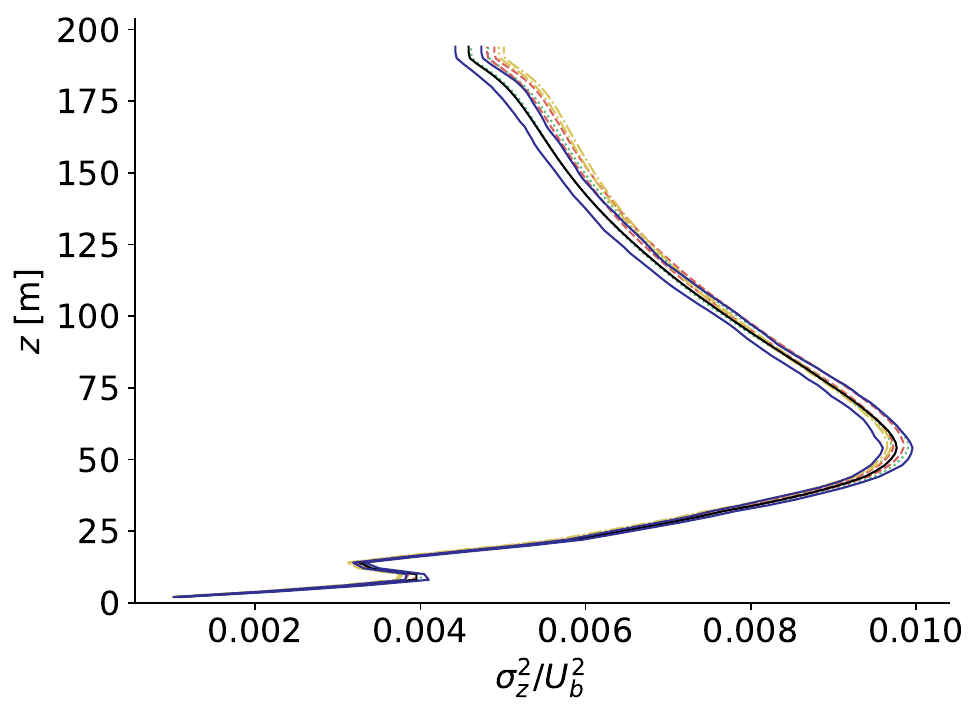} \\
  g) \hspace{0.4\textwidth} h) \\    
  \includegraphics[width=0.37\textwidth]{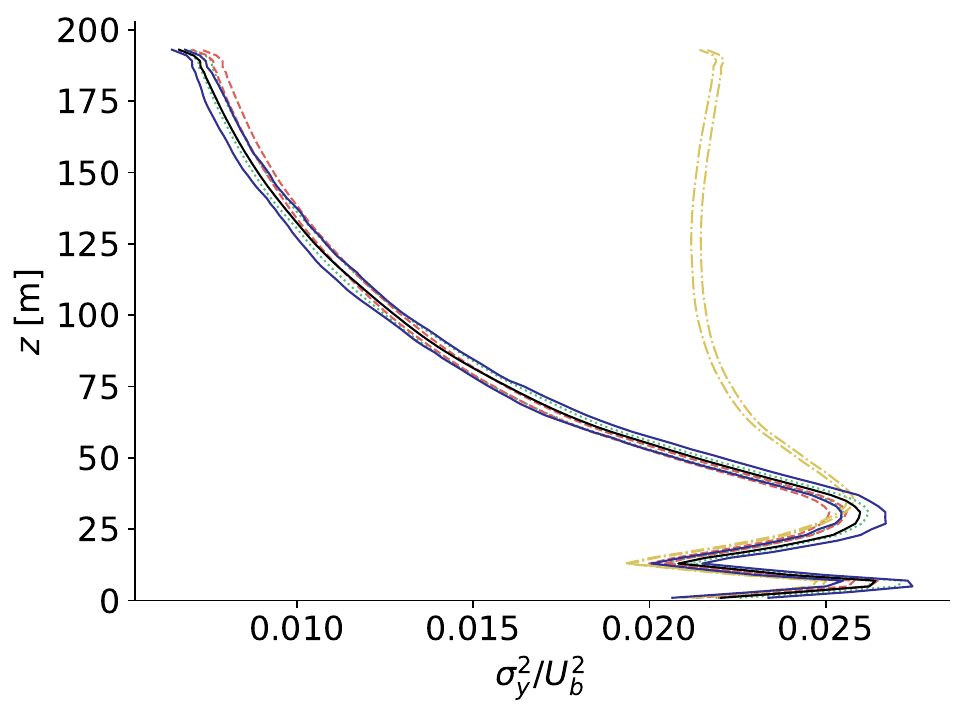}
  \includegraphics[width=0.37\textwidth]{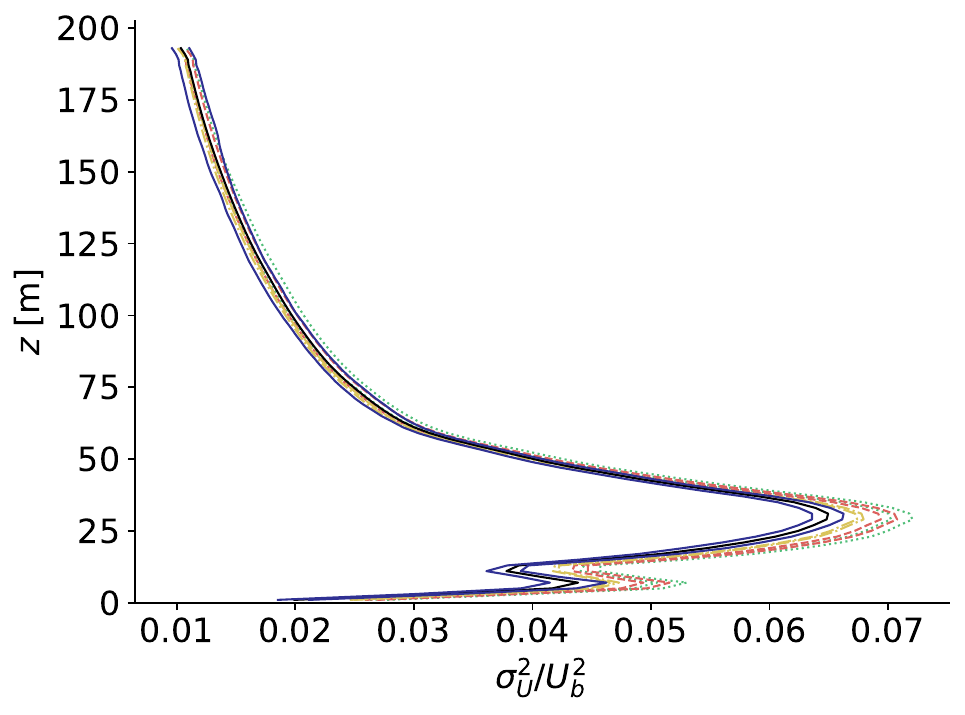}  
  \caption{\small Vertical wind profiles in the innermost domain in the case of the realistic urban flow.
    The mean and variances for all quantities have been accumulated over the horizontal directions and statistical accumulation over time is indicated with different lines.
    In time, all quantities are centred around the mid point of the turning pressure gradient simulation, at $ t/T_\Omega = 0.526 $.
    The black line indicates the ensemble mean.
    The coloured lines indicate the 25th and 75th percentiles of the ensemble where the specified time-averaging has been applied to each member.
    a) Velocity along $x$ axis ($u$), b) variance of $u$ ($\sigma_u^2$), c) velocity along $y$ axis ($v$), d) variance of $v$ ($\sigma_v^2$), e) velocity along $z$ axis ($w$), f) variance of $w$ ($\sigma_w^2$) g) horizontal velocity ($U$), h) variance of $U$ ($\sigma_U^2$)}
\label{Turkukaprofiilit}       % Give a unique label
\end{figure}

The vertical profiles of ensemble mean wind and variance for the innermost domain at $t/T_\omega =0.526$ since the start of pressure gradient turning are shown in Fig.\,\ref{Turkukaprofiilit} with black lines.
The variation within the ensemble is quantified using the 25th and 75th percentiles as well as their separation, the interquartile range.
As with the cube-array, additional averaging in the horizontal directions have been used in the calculation of all profiles.

A decrease in variation between the ensemble members, compared to the staggered cube array, can be seen also in the vertical profiles from the interquartile range.
This is clearest in the mean velocity along the $x$ direction where an almost full overlap of the ensemble mean (shown using a black line) and the limits of the interquartile range (shown using blue lines).
Other velocity components and the variances also display a narrower interquartile range than in the case of the cube array.
The spread between the ensemble members is largest at around the top of the roughness elements and to some extent also in upper end of the domain.
The latter can be explained by the large scale turbulence and other flow variation that is brought in from the larger domains.
The vertical extent of the roughness elements can be seen as a jump in the lower part of the mean velocity profiles and as well as in the variances profiles.

The effects of time averaging are similar to the cube array set-up for the vertical profiles.
In the velocity components, the ensemble average is mostly within the interquartile range of the $0.0438 T_\Omega$ time averaged simulations.
With $0.131 T_\Omega$ time averages, the ensemble mean is partially outside the interquartile range when the variances are considered.
The long, $0.920 T_\Omega$ time average fails to capture the ensemble average in almost all cases and especially at the upper parts of the domain.
The largest deviation from the ensemble mean is observed in the $v$ component and especially in the variance where catastrophic deterioration of the solution is seen for the long averaging time.

The profiles of the mean horizontal velocity $U$ are almost identical to those of $u$.
At the very upper end up the profile of the horizontal mean wind, the percentiles of from the long time average are slightly more off from the ensemble mean than in the case of the mean $u$.
However, the variances of the horizontal wind and $u$ display interesting differences. The time averaged horizontal variances are closer to the ensemble value than those of $\sigma_x^2$ at the upper parts of the profile.
In the roughness sublayer its the opposite: time averaged horizontal variances overestimate the ensemble profile while the same is not seen in the profiles of $\sigma_x^2$. 

\begin{figure}[tp]
  % Skripti R1_KT_NT.py
  a) \hspace{0.49\textwidth} b) \\
  \includegraphics[height=0.35\textwidth]{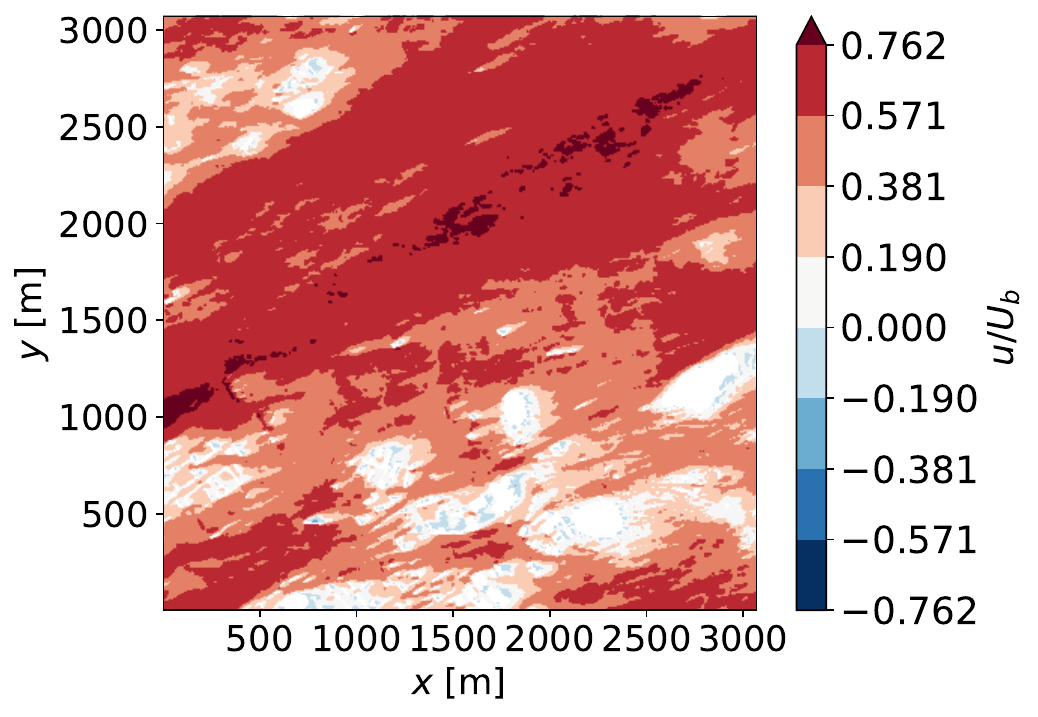}
  \includegraphics[height=0.35\textwidth]{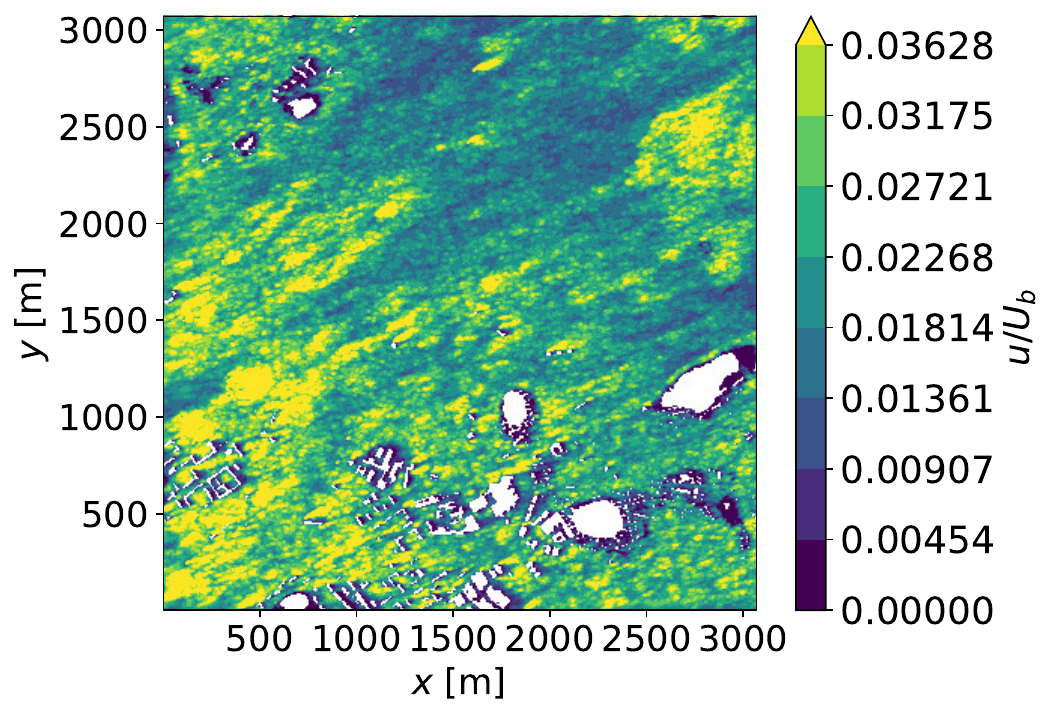} \\
  \caption{\small 
    Ensemble-averaged velocity component $u$ calculated using instantaneous values for a horizontal plane through the roughness elements at $z=\unit[40]{m}$ for the innermost domain in the case of the realistic urban flow.
    The coordinate axis indicate the relative distance in metres to the domain origin at 60°26'53.8"N    22°15'51.8"E.
    a) mean and b) variance}
  \label{TKUpoikki1}
\end{figure}

As expected, the ensemble of 50 members does not appear to be large enough for mean velocity and variance when instantaneous values are used.
Figure\,\ref{TKUpoikki1} displays the ensemble mean velocity and velocity variances for the $x$ component at $z=\unit[40]{m}$ in the smallest simulated domain.
The insufficient ensemble size is visible as the sharp variations in the mean velocity field and as static noise type patterns in the variance field.
Other velocity and variance components behave in a similar manner (not shown).

\begin{figure}[tp]
  % Skripti R1_KT_NT2.py
  a) \hspace{0.27\textwidth} b)  \hspace{0.27\textwidth} c) \\
  \includegraphics[height=0.23\textwidth]{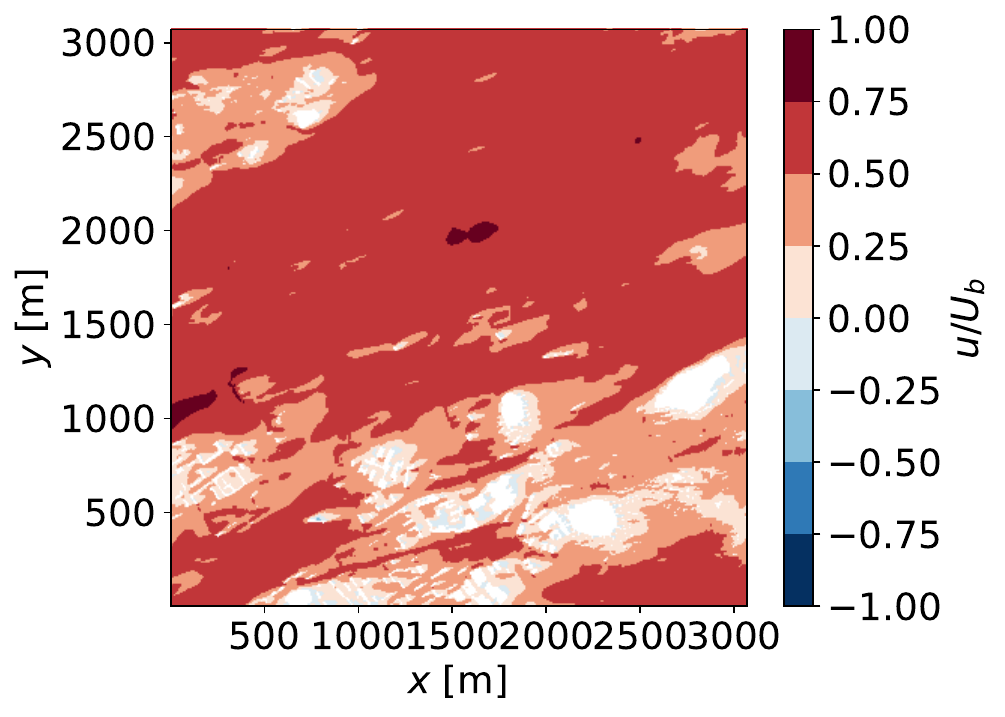}
  \includegraphics[height=0.23\textwidth]{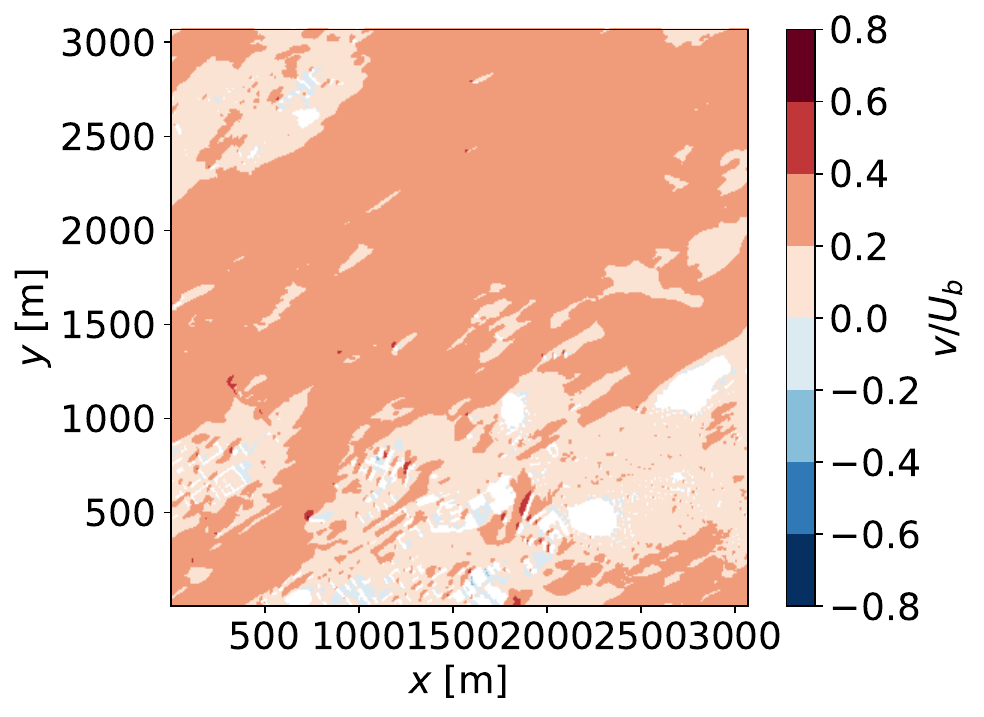} 
  \includegraphics[height=0.23\textwidth]{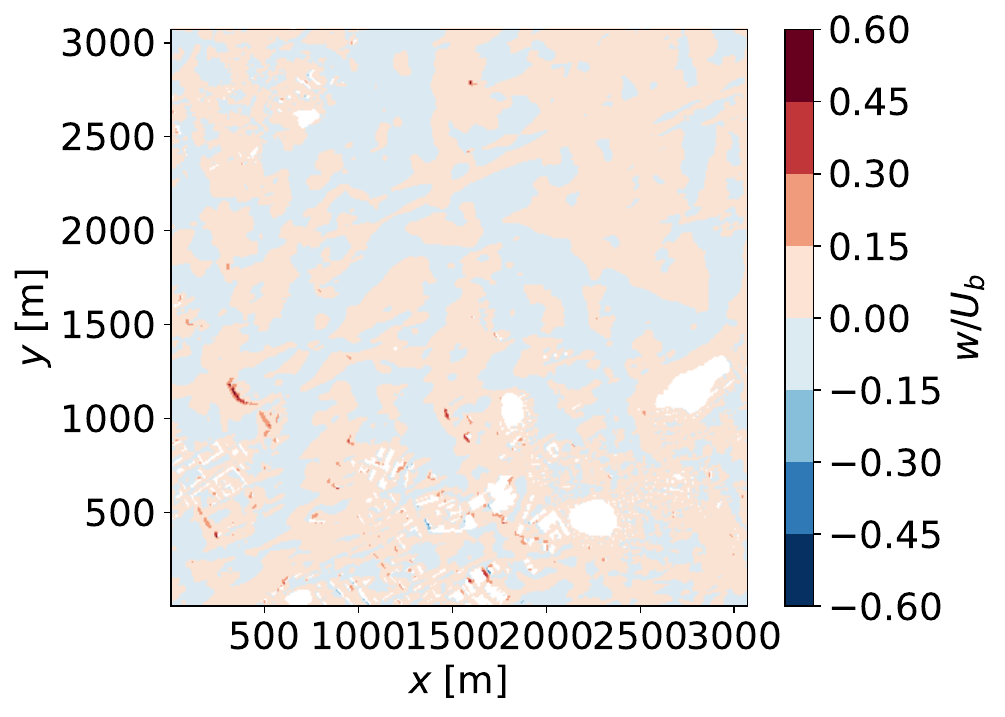} \\
  d) \hspace{0.27\textwidth} e)  \hspace{0.27\textwidth} f) \\
  \includegraphics[height=0.23\textwidth]{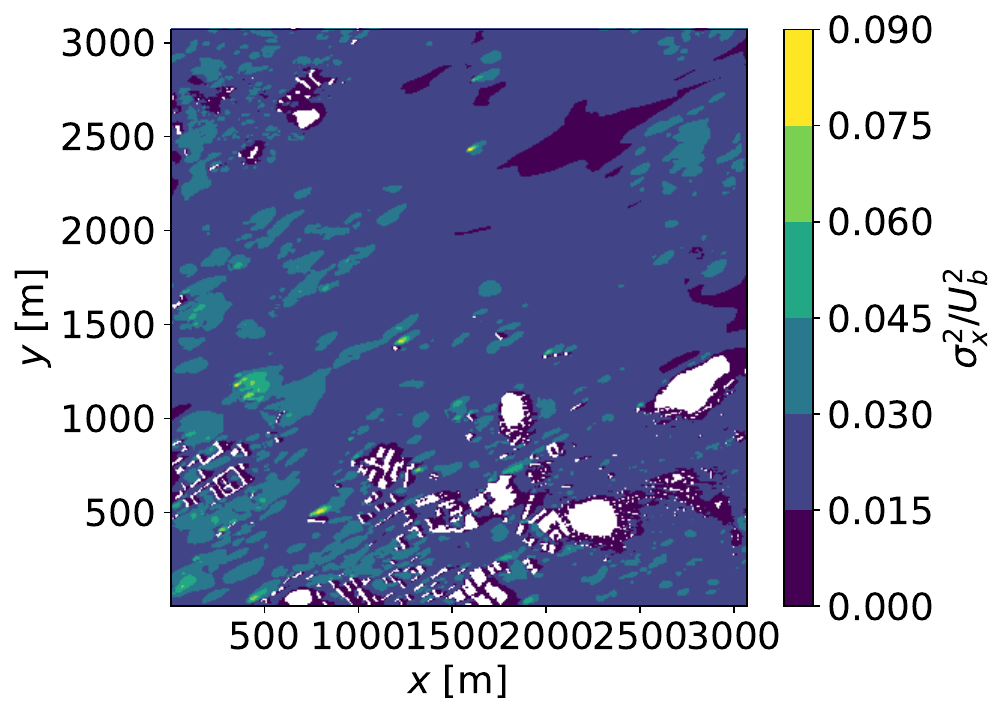} 
  \includegraphics[height=0.23\textwidth]{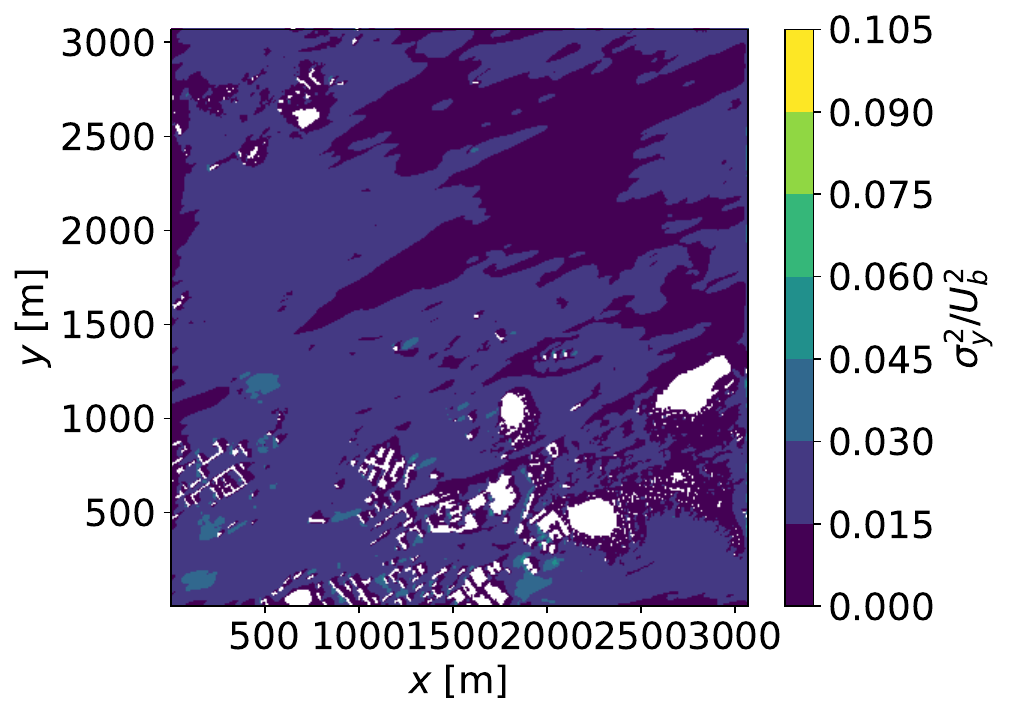} 
  \includegraphics[height=0.23\textwidth]{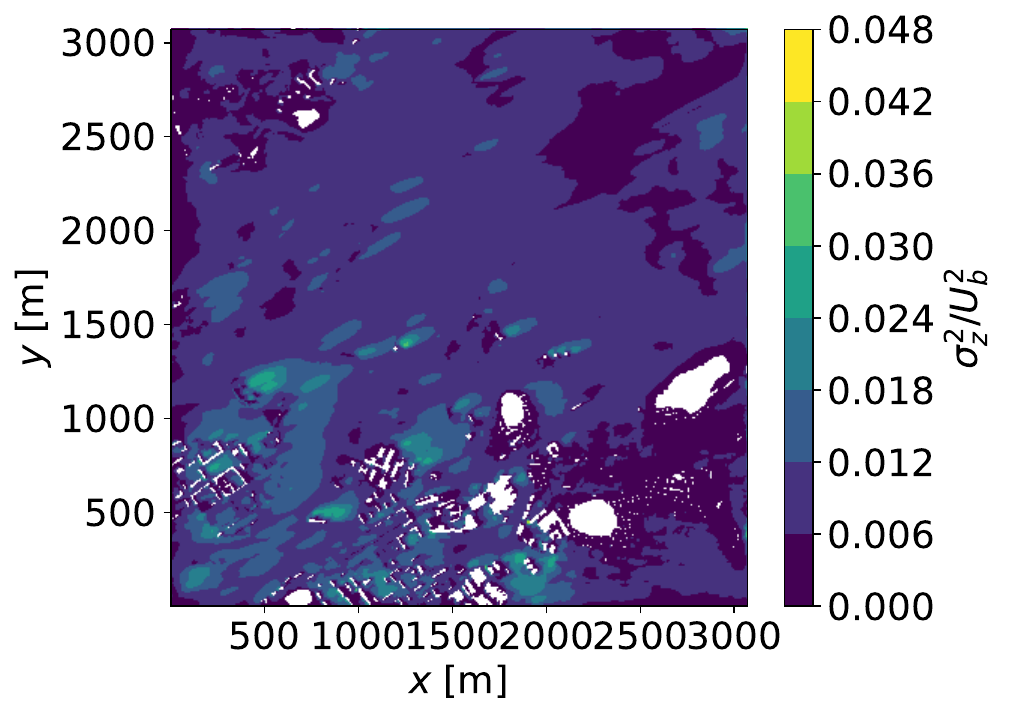} \\    
  g) \hspace{0.27\textwidth} h)  \hspace{0.27\textwidth}  \\
  \includegraphics[height=0.23\textwidth]{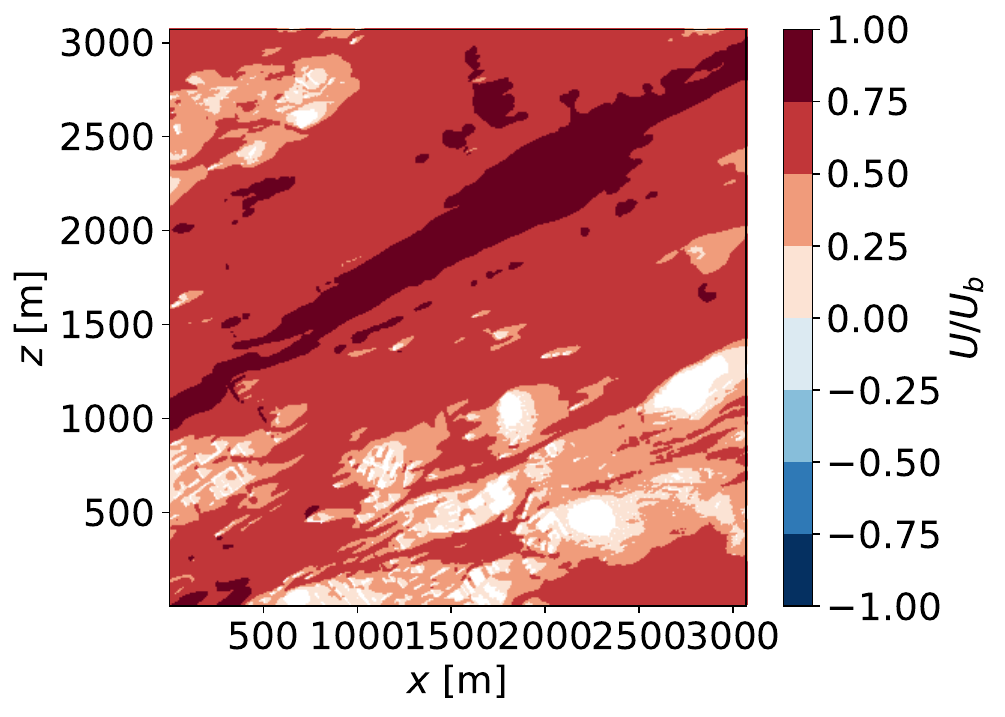} 
  \includegraphics[height=0.23\textwidth]{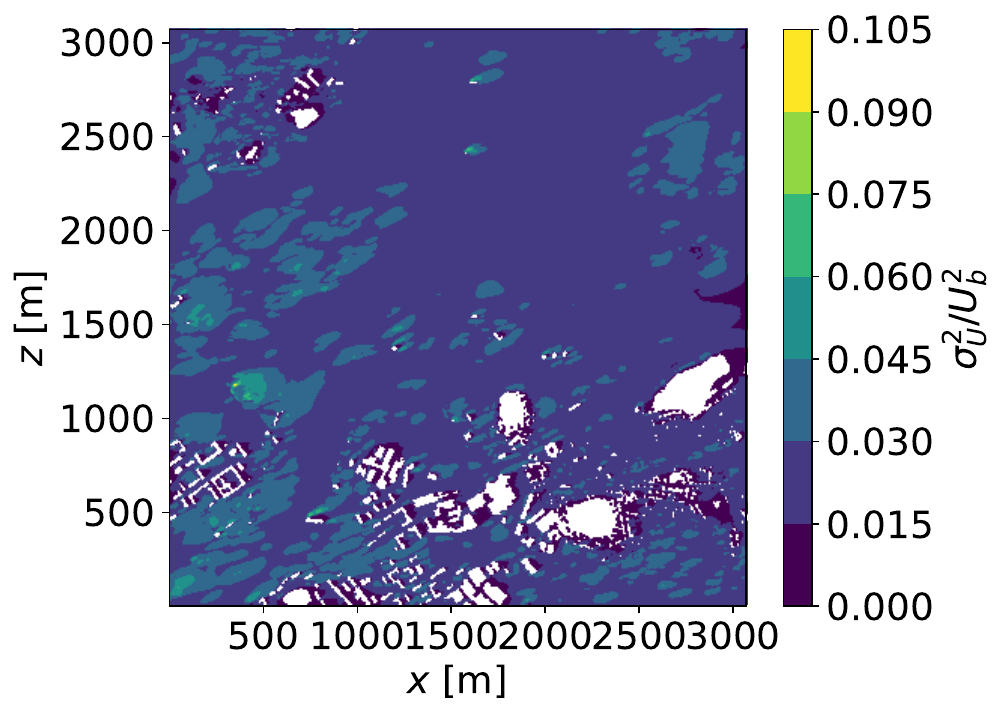} 
  \includegraphics[height=0.23\textwidth]{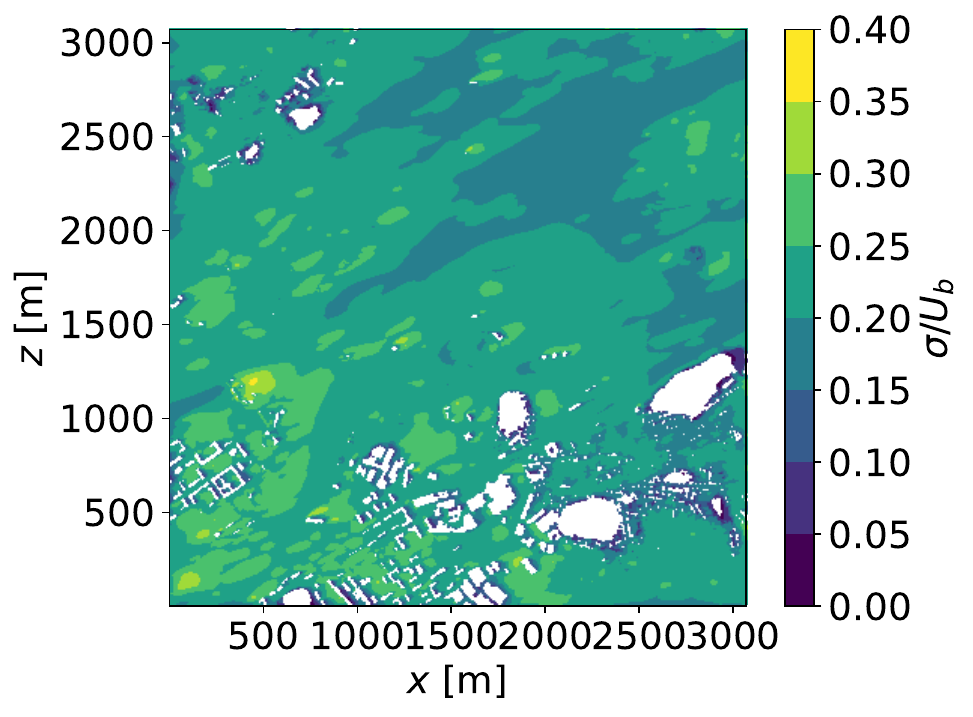} \\
  \caption{\small Ensemble-averaged velocity, calculated with $0.131 T_\Omega$ time averages) for a horizontal plane through the topography at $z=\unit[40]{m}$ for the innermost domain in the case of the realistic urban flow.
    The coordinate axis indicate the relative distance in metres to the domain origin at 60°26'53.8"N    22°15'51.8"E.
    a) mean of $u$, b) mean of $v$, c) mean of $w$, d) variance of $u$, e) variance of $v$, f) variance of $w$, mean of horizontal velocity $U$, g) variance of $U$, and h) turbulence intensity $\sigma/U_b$}
  \label{TKUpoikki2}
\end{figure}

As suggested by the results of the cube array, improved convergence of ensemble statistics can be achieved using statistics accumulated temporally over $0.131 T_\Omega$ instead of the instantaneous values.
This is confirmed from the ensemble mean velocity and velocity variances at $z=\unit[40]{m}$ in the smallest simulated domain, shown in Fig.\,\ref{TKUpoikki2}.
The velocity and the variance fields are not fully smooth but clearly smoother than when instantaneous values are used.
The static noise pattern has disappeared from the variance fields.
For the rest of the paper, we will utilize ensemble statistics calculated using statistics that have been temporally accumulated over the period of $0.131 T_\Omega$.

The overall flow field through the roughness elements is very typical for the considered urban flow.
Ensemble mean and variance for the realistic urban case are shown in Fig.\,\ref{TKUpoikki2}.
The plane cuts through some of the buildings and the orography, especially in the southeastern quadrant of the domain, and their effects on the flow are clearly visible.
There is strong, large-scale channelling along the less obstructed parts on the northwestern quarter of the domain, seen as high $u$, $v$, and $U$ velocities.
Channelling is visible also in the southeastern quadrant of the domain in street canyons and other minor openings that are aligned with the main flow direction.
The roughness elements affect the flow also by reducing the mean wind speed and increasing variances in their wakes through the turbulence they generate, especially in the southeastern quadrant of the domain where the terrain is more elevated and where there are more buildings.
The variances are strong in the wake especially when there are no further obstacles downstream.
The horizontal mean velocity and variance are similar to mean $u$ and $\sigma_x^2$ while the turbulence intensity nicely highlights the areas with high variance.

\begin{figure}[tp]
  % Skripti R1_KT_NT2.py
  a) \hspace{0.27\textwidth} b)  \hspace{0.27\textwidth} c) \\
  \includegraphics[height=0.23\textwidth]{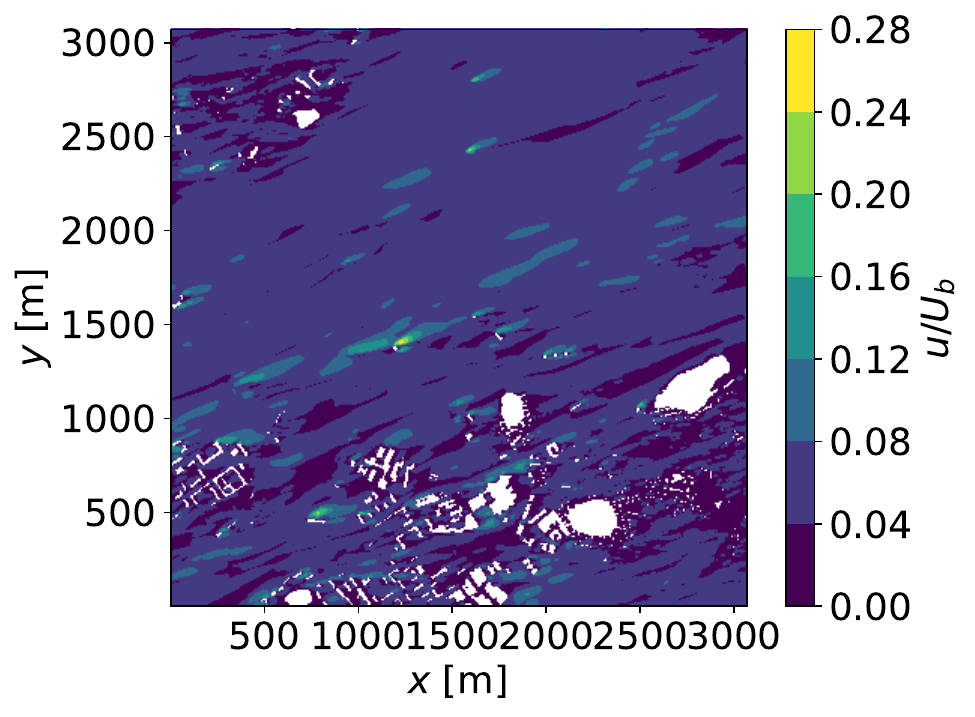}
  \includegraphics[height=0.23\textwidth]{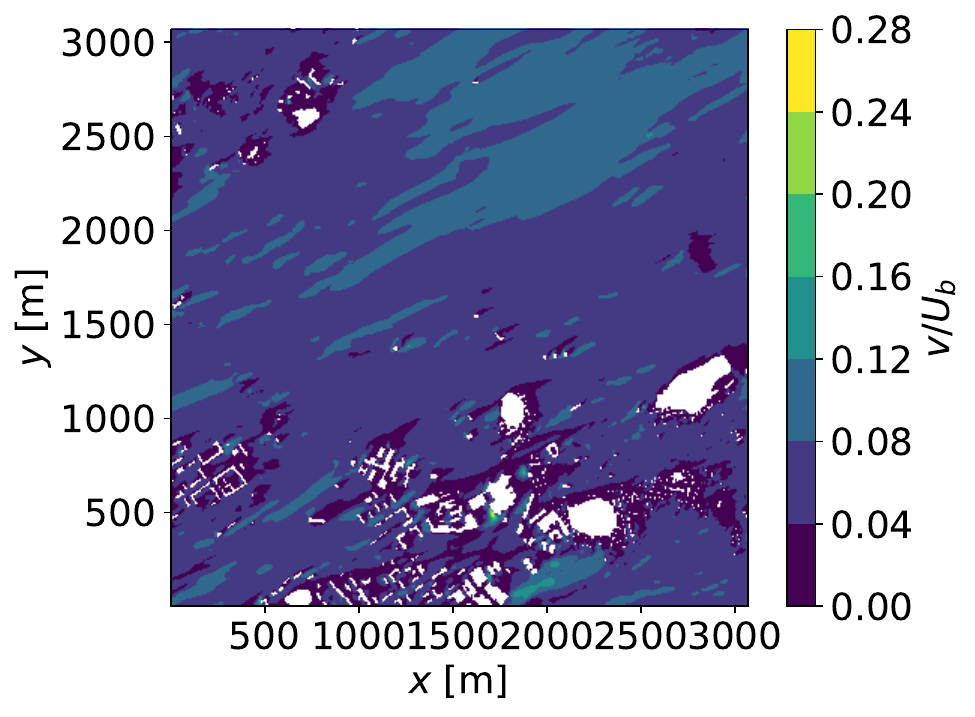} 
  \includegraphics[height=0.23\textwidth]{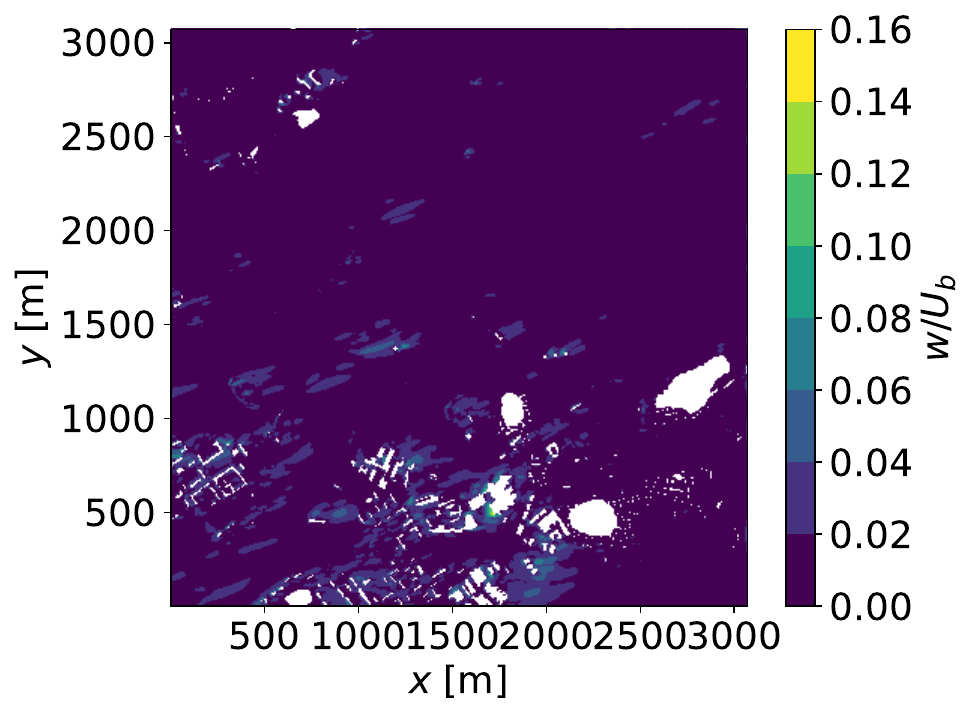} \\
  d) \hspace{0.27\textwidth} e)  \hspace{0.27\textwidth} f) \\
  \includegraphics[height=0.23\textwidth]{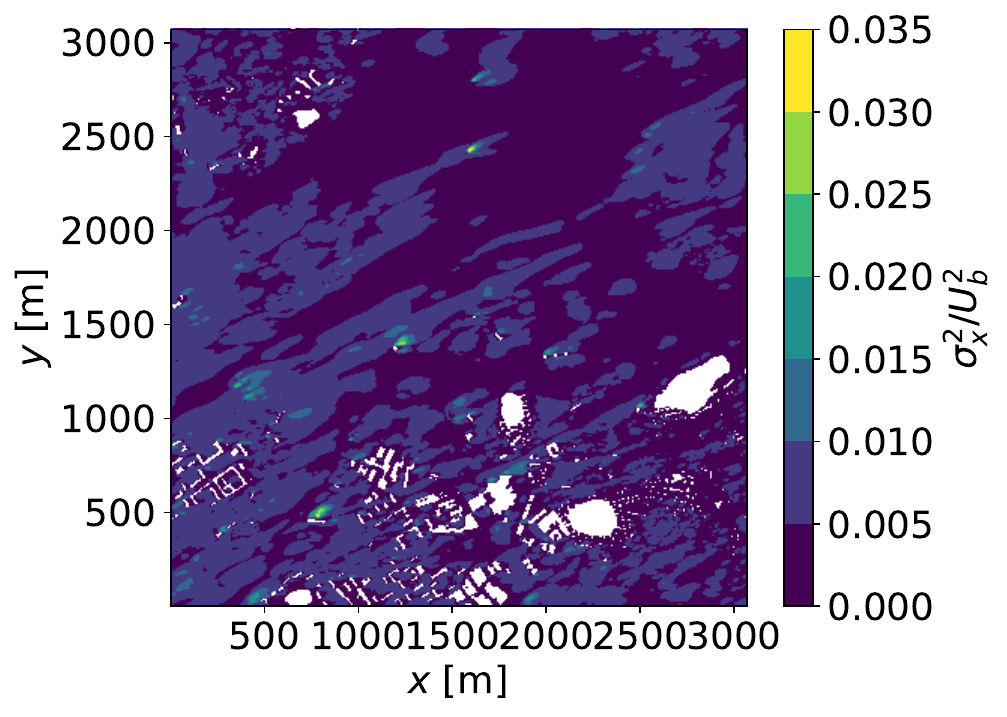} 
  \includegraphics[height=0.23\textwidth]{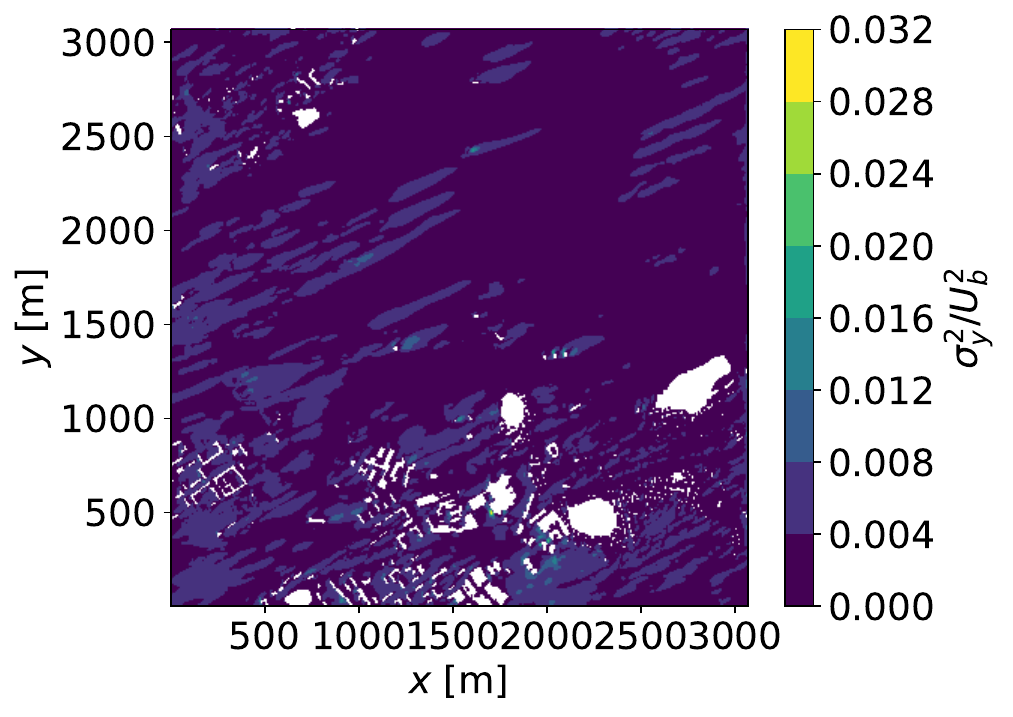} 
  \includegraphics[height=0.23\textwidth]{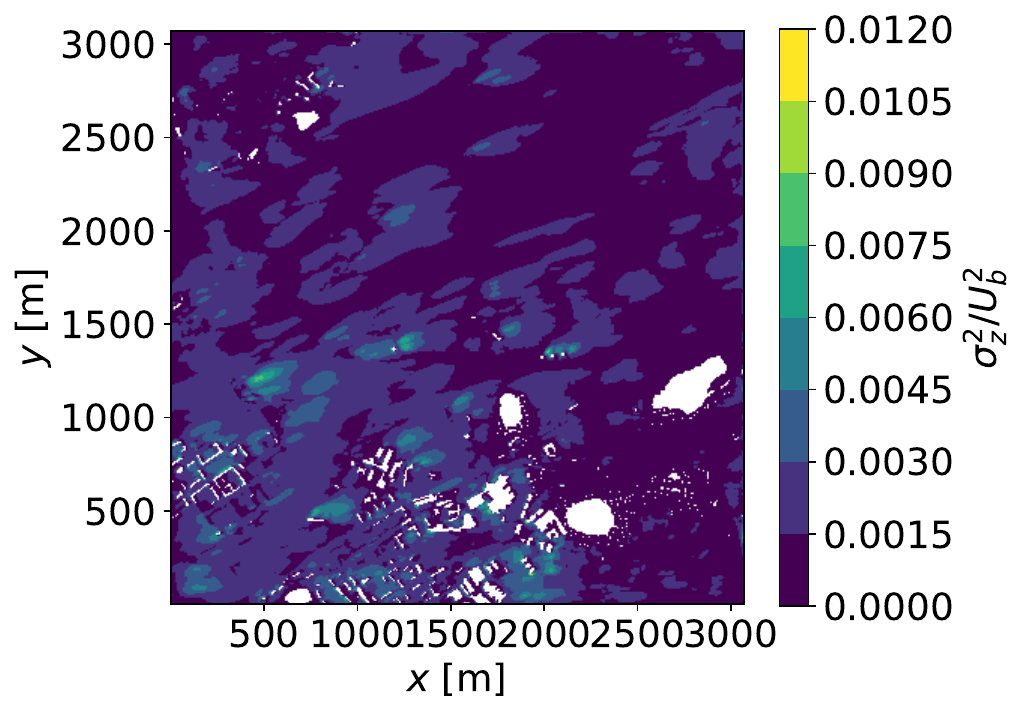} \\
  g) \hspace{0.27\textwidth} h)  \hspace{0.27\textwidth} i)  \\
  \includegraphics[height=0.23\textwidth]{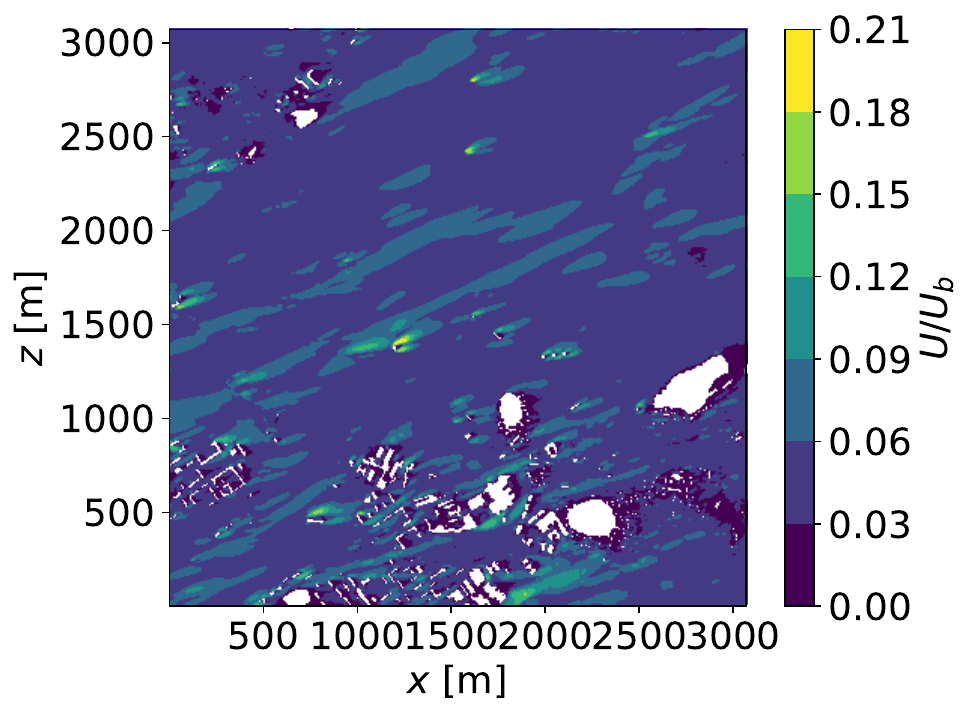} 
  \includegraphics[height=0.23\textwidth]{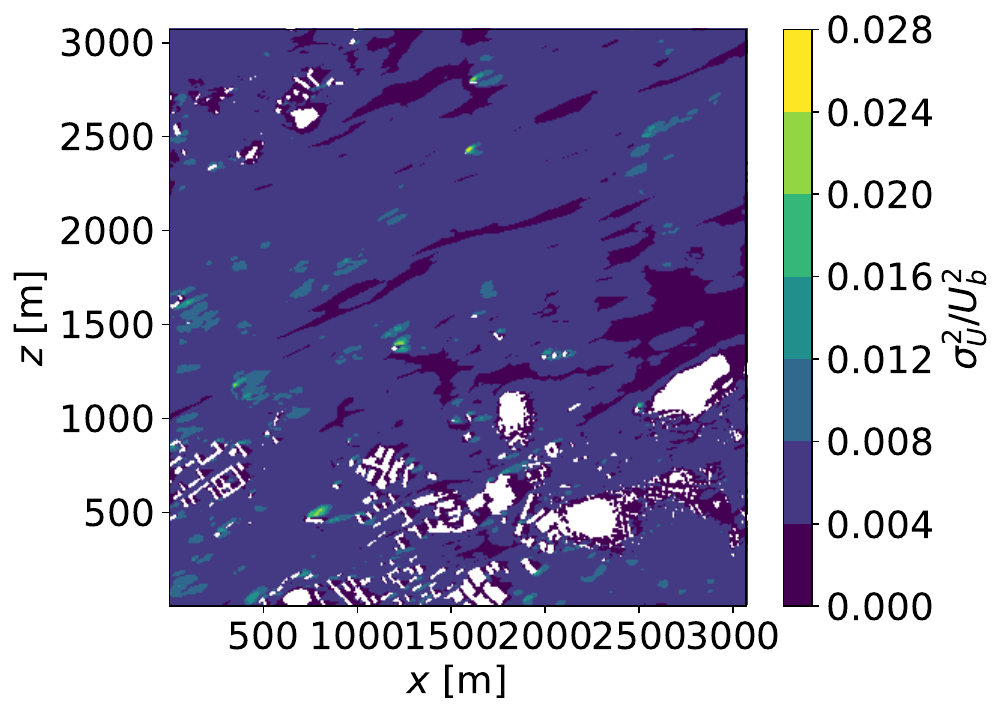} 
  \includegraphics[height=0.23\textwidth]{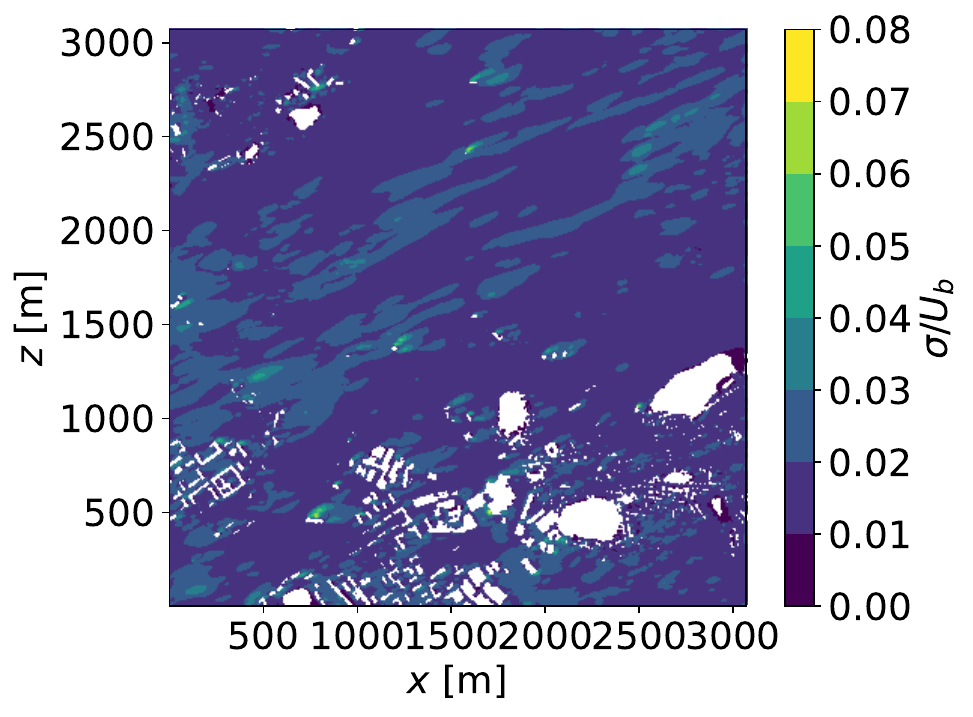} \\
  \caption{\small RMSE between the ensemble velocity calculated using $0.131 T_\Omega$ time average and the plain time average calculated over $0.920 T_\Omega$ for a horizontal plane through the roughness elements at $z=\unit[40]{m}$ for the innermost domain in the case of the realistic urban flow.
    The coordinate axis indicate the relative distance in metres to the domain origin at 60°26'53.8"N 22°15'51.8"E.
    a) mean of $u$, b) mean of $v$, c) mean of $w$, d) variance of $u$, e) variance of $v$, f) variance of $w$, g) mean of horizontal velocity $U$, h) variance of $U$, and  i) turbulence intensity $\sigma/U_b$}
  \label{TKUpoikki3}
\end{figure}

In order to study the difference between plain time average and ensemble-averages, we utilise the RMSE.
The RSME is calculated for each cell, then over all time averaged ($0.920 T_\Omega$) ensemble members, and plotted in Fig.\,\ref{TKUpoikki3}.
The direct utilisation of long time averaging can cause significant errors localized around the wakes of the roughness elements.
Depending on the quantity, different regions display the largest errors.
In the case of mean horizontal velocity components, the regions of the highest error form long structures in the wakes of the roughness elements.
We interpret this as the inability of the long averaging interval to accurately capture the changing direction of the wake.
The vertical mean velocity $w$, $\sigma_z^2$, and $\sigma_x^2$ have their largest errors in the immediate vicinity of the roughness elements.
This is probably due to the long avareging interval failing to capture the chaning direction of the waske as in the case of the horizontal mean velocities.
Another reason could be that the smoothing out of small time scale variation of the recirculation regions and other smaller flow features caused by the roughness elements.
Finally, the $v$ variance has its largest errors in the areas where the mean $v$ velocity is at its largest, at the open areas mostly located in the northern half of the domain.
The errors in the horizontal velocity $U$ are similar to but appear to be slightly stronger than those in $u$.

As in the case of the cube array, we utilize Taylor diagrams to evaluate the effect of averaging time on the simulation accuracy.
The error measures are calculated for each cell and these are then spatially averaged to produce a single value per ensemble member (simulation).
We consider again only the roughness sublayer, here defined using the average height of the roughness elements in the domain: $4h_\textrm{avg}=\unit[84]{m}$.
As a reference we value, we use the ensemble mean calculated with the $0.131 T_\Omega$ time average.

\begin{figure}[tp]
  % Skriptit R1_KT_TD1A.py ja R1_KT_TD1A_piirto.py
  a) \hspace{0.32\textwidth} b) \hspace{0.32\textwidth} c) \\
  \includegraphics[width=0.32\textwidth]{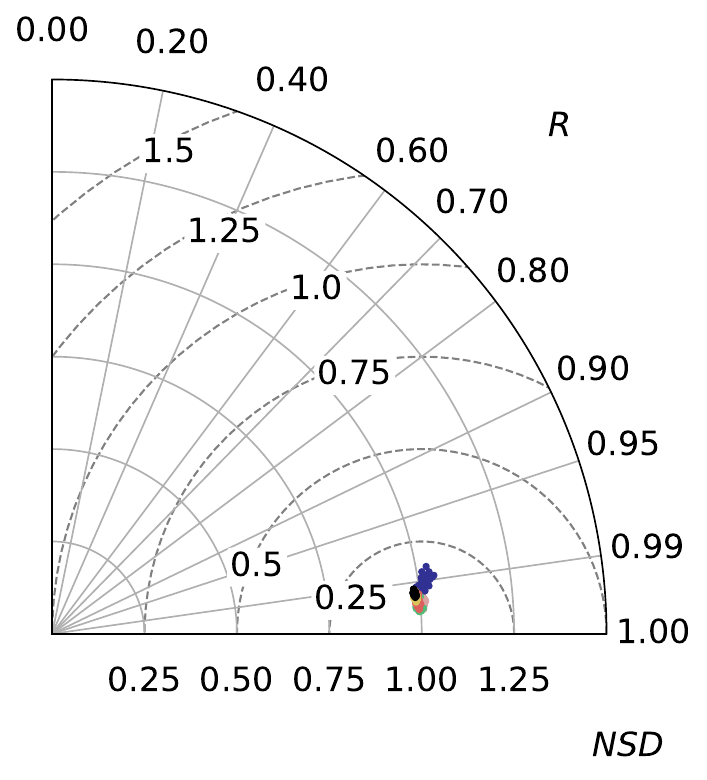}
  \includegraphics[width=0.32\textwidth]{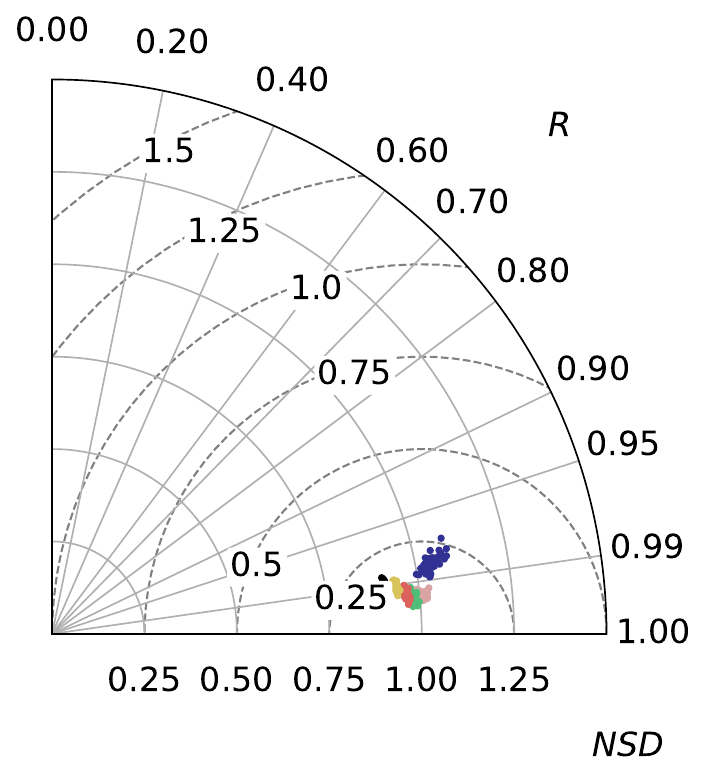}
  \includegraphics[width=0.32\textwidth]{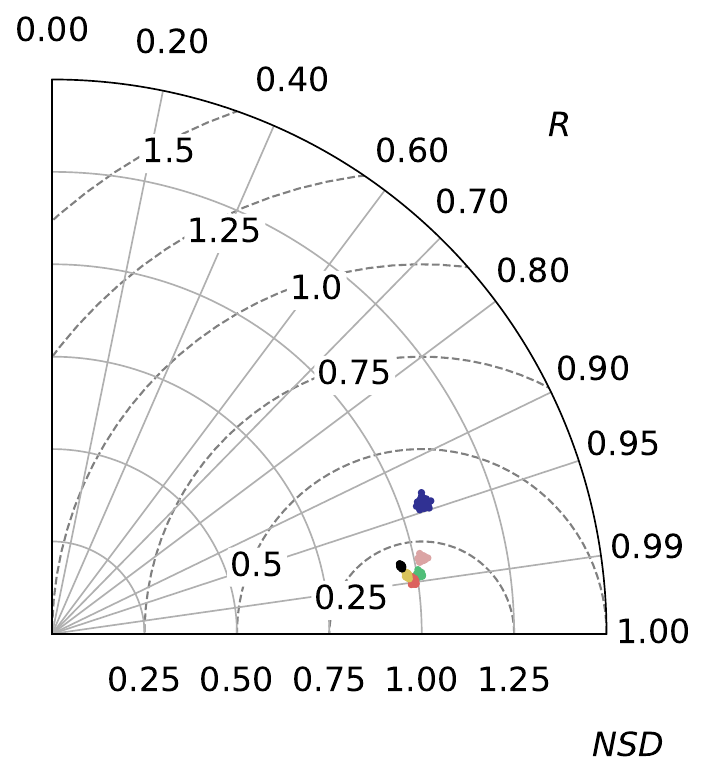} \\
  d) \hspace{0.32\textwidth} e) \hspace{0.32\textwidth} f) \\
  \includegraphics[width=0.32\textwidth]{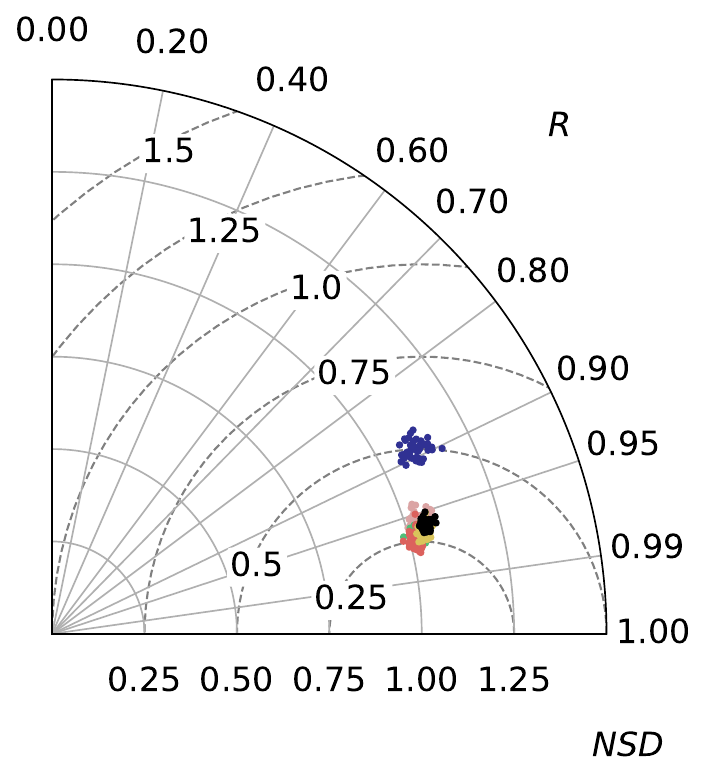}
  \includegraphics[width=0.32\textwidth,clip]{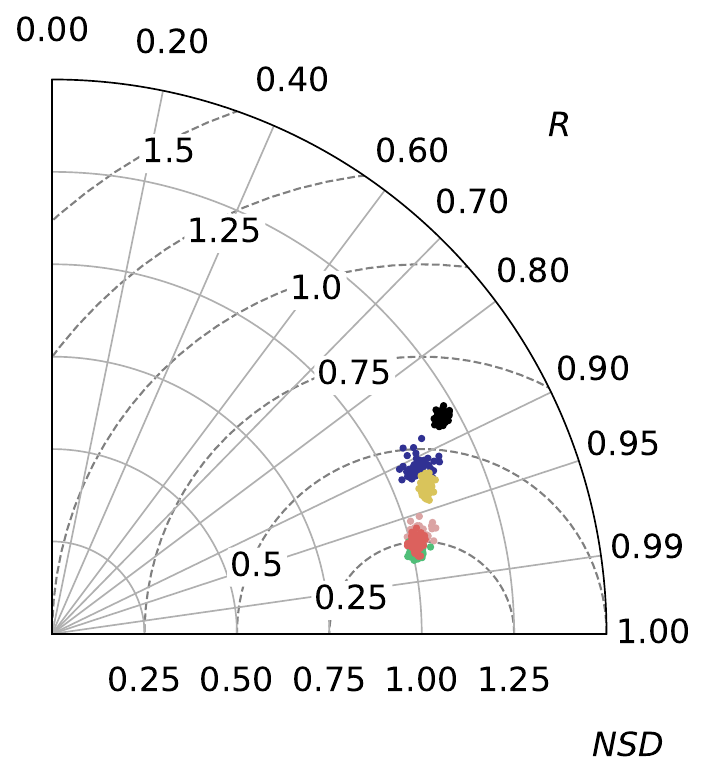} 
    \includegraphics[width=0.32\textwidth]{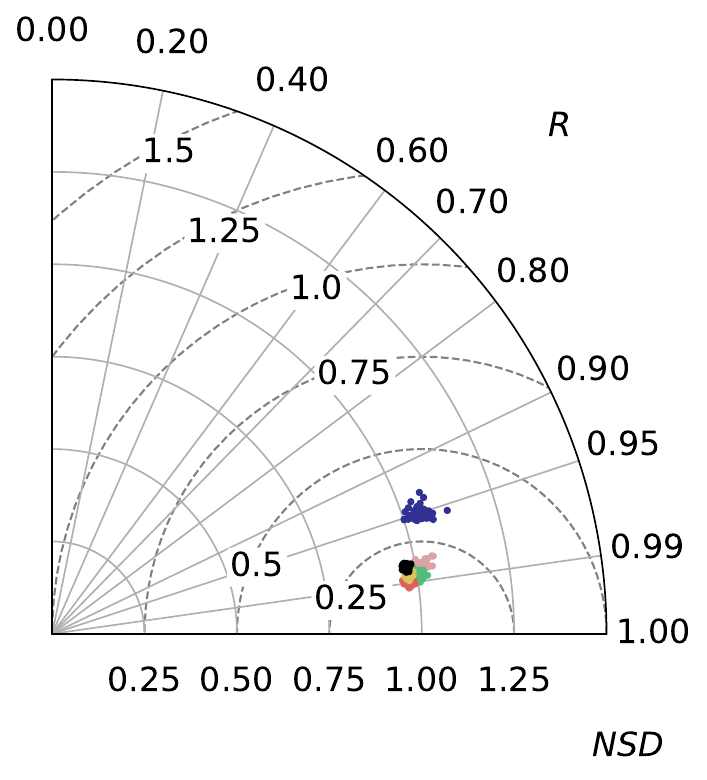}   \\
  g) \hspace{0.32\textwidth} h) \hspace{0.32\textwidth} \\
  \includegraphics[width=0.32\textwidth]{TD_Turku_ka_U.eps} 
  \includegraphics[width=0.32\textwidth]{TD_Turku_var_U.eps}   
  \caption{\small Taylor diagrams for the mean (a--c, g) and variance (d--f, h) of $u$, $v$, $w$, and $U$ components of velocity with different averaging intervals for $z < 4h_\textrm{avg}=\unit[84]{m}$ in the innermost domain in the case of the realistic urban flow.
    The ensemble mean calculated with $0.131 T_\Omega$ time averages is used as the reference.
    The error measures are calculated first cellwise and then averaged spatially to produce a single value per ensemble member.
    The polar plot has normalized standard deviation as the radial axis and correlation as the azimuthal coordinate.
    The circles drawn using a dashed lines indicate normalized root mean square error with a perfect model at point (1,1).
    Each dot indicates an ensemble member and the colours  indicate different averaging intervals for $0.0438 T_\Omega$ (blue), $0.131 T_\Omega$ (pink), $0.219 T_\Omega$ (green), $0.394 T_\Omega$ (red), $0.657 T_\Omega$ (yellow), and $0.920 T_\Omega$ (black)}
  \label{Turku_TD1}
\end{figure}

The low spread between the ensemble members that was observed earlier is clearly visible also in the Taylor diagrams in Fig.\,\ref{Turku_TD1} where each dot represents an ensemble member and the colours denote different averaging intervals.
Especially with mean velocity components, all ensemble members are located in tight bundles for all averaging intervals, indicating similar R and NSD values.
Similar error behaviour as with the cube array can be observed here with the realistic urban case with increasing averaging time: the error decreases with increasing averaging time until $0.131 T_\Omega$
or $0.219 T_\Omega$ depending on the quantity and increases for longer averaging times.
This effect is minor for all other quantities except for variance of $v$ where the $0.657 T_\Omega$ and $0.920 T_\Omega$ values displays a clearly poorer performance on the correlation.
However, one has to keep in mind that the Taylor diagram was created using a small ensemble that was calculated with $0.131 T_\Omega$ averages as the reference value.
This can be expected to result in improved performance for at least the $0.131 T_\Omega$ averaging time.
The horizontal velocity and variance behaves in a manner almost indistinguishable from mean $u$ and $\sigma_x^2$.

The interquartile ranges of fractioanl bias, shown in Table \ref{FB2}, mostly agree with the Taylor diagram in Fig.\,\ref{TD2}.
The best performance is mostly from the $0.131 T_\Omega$ time averages and the worst performance can be seen in the $0.657 T_\Omega$ and $0.920 T_\Omega$ variances of $v$.
However, the vertical variance $\sigma_z^2$ is an exception and improves in fractional bias until $0.219 T_\Omega$ with only minor changes after that.
Unlike in the staggered cube array, fractional bias does not fail in describing the systematic bias in mean $w$.
Nevertheless, fractional bias displays similar development on the reliability of mean $w$ as the measures in the Taylor diagram.
As in the Taylor diagram in Fig.\,\ref{TD2}, the horizontal mean velocity and variance have a very similar fractional bias as those of mean $u$ and $\sigma_x^2$.

\begin{table}
  % Skripti R1_KT_FBT.py
  \caption{\small 
    The interquartile ranges of the fractional bias for different averaging times for the case of the realistic urban flow.
    All values have been calculated within the roughness sublayer ($z < 4h_\textrm{avg}=\unit[84]{m}$) in the innermost domain.
    The fractional bias have been calculated in each cell for different averaging times using the $0.131 T_\Omega$ time-averaged ensemble average as the reference value.
    The cell values of fractional bias are then spatially averaged to produce a single value per ensemble member which are then used to calculate the shown interquartile range}
  \begin{center} \small
  \begin{tabular}{llcccc}
    \hline
    & $t/T_\Omega$ & $u$ & $v$ & $w$ & U  \\ \hline
 $\langle \cdot \rangle_t$ & 0.0438  & [-0.007,\,0.001] & [-0.018,\,0.000] & [-0.084,\,0.077] & [-0.008,\,0.000] \\
    &0.131  & [-0.003,\,0.003] & [-0.003,\,0.006] & [-0.052,\,0.044] & [-0.002,\,0.003]  \\
  &0.219   & [0.002,\,0.005] & [0.008,\,0.014] & [ -0.021,\,0.054] & [0.002,\,0.006]  \\
  &0.394   & [0.003,\,0.006] & [0.020,\,0.026] & [0.043,\,0.096] & [0.005,\,0.008]  \\
 &0.657   & [0.003,\,0.006] & [0.043,\,0.046] & [0.058,\,0.115] & [0.006,\,0.009]  \\
 &0.920  & [0.003,\,0.005] & [0.072,\,0.076] & [0.077,\,0.119] & [0.007,\,0.009] \\ \hline
 
$\sigma$ & 0.0438  & [0.046,\,0.073] & [0.027,\,0.053] & [-0.002,\,0-019] & [0.058,\,0.080] \\
    &0.131   & [0.009,\,0.023] & [-0.001,\,0.014] & [-0.052,\,0.044] & [0.011,\,0.023]  \\
 &0.219   & [0.001,\,0.013] & [-0.015,\,-0.001] & [0.003,\,0.013] & [0.001,\,0.017]  \\
 &0.394   & [-0.024,\,-0.013] & [-0.052,\,-0.042] & [0.007,\,0.012] & [-0.015,\,-0.004] \\
 & 0.657  & [-0.057,\,-0.046] & [-0.148,\,-0.140] & [0.007,\,0.013]& [-0.028,\,-0.018]  \\
 &0.920  & [-0.076,\,-0.067] & [-0.275,\,-0.266] & [0.007,\,0.012]& [-0.034,\,-0.025] \\ \hline
  \end{tabular}
  \end{center}
  \label{FB2}
\end{table}

\section{Conclusions} \label{johtop}

We have studied the differences between time-averaged and ensemble-averaged statistics on a flow with a temporally turning pressure gradient in the simplified case of a staggered cube array and in the case of a realistic urban flow past the city of Turku, Finland.
The staggered cube array consisted of a 648 repeating elements, which we used together with five individual simulations to create an ensemble of 3\,240 members.
In the case of the realistic urban flow, there were no repeating elements and hence the ensemble size was the same as the number of simulations: 50.
This ensemble size was not enough for proper analysis and hence the calculation of ensemble statistics was enhanced by the use of short time averaging. 

Our results indicate that, in the case of a turning pressure gradient, significant errors can contaminate the results if one disregards the statistically nonstationary nature of the flow and accumulates mean wind and variances throughout the simulation.
These errors become visible when the used averaging time is not small compared to the characteristic time scale of the turning of the pressure gradient $T_\Omega = 1/\Omega$.
The mean velocity of the main (strongest) wind components remains mostly unaffected but all other wind components and variances are affected.
The largest departures from the ensemble velocity were observed in the weaker wind component (here $v$) and especially in its variance.

Based on Taylor diagrams for the roughness sublayer, our results on the cube array indicate that the poor performance of plain time-averaged statistics can be mitigated by combining time averaging and ensemble averaging.
Time averaging up to approximately $0.219T_\Omega$ improved the results by bringing them to a better agreement with the ensemble statistics.
We then constructed new ensembles using simulations averaged over $0.131 T_\Omega$ and evaluated their performance against the instantaneous full ensemble using Taylor diagrams.
These indicated that an ensemble of 10--50 members appeared to provide a good correspondence against the full, instantaneous ensemble.
Above the roughness sublayer, spatial averaging in the horizontal direction can be utilized together with instantaneous fields to obtain sufficiently well-converged statistics, provided that the flow is otherwise homogeneous.

Using a 50 member ensemble together with $0.131 T_\Omega$ statistics, we analysed the spatial error distribution in the vicinity of the roughness elements in the case of the realistic urban flow in Turku, Finland.
When long time averaging procedures are used, the wake regions of the roughness elements show the biggest errors.
This is probably due to both the change of wind direction, and hence the wake direction, but also due to the small, unsteady features within the wakes.
Variance of the weaker wind component (here $v$) was predicted incorrectly by long time averaging in high-wind areas.

The results presented here are based on simulations carried out without the effects of the Coriolis force and the buoyancy.
These simplifications were made in order to concentrate on the effects due to the changing wind direction only.
The current results should be best applicable to cases with near-neutral stratification within short distances.
Such conditions are present in the urban roughness layer at the scale of individual buildings or blocks.
Further research with more realistic and complicated cases should be conducted in the future.
In order to gain a comprehensive view on the effects of changing conditions on the outcome of urban LES and for the requirements for urban LES ensembles, one should consider at least the Coriolis force, buoyancy, and the changing magnitude of the driving force in addition to the changing wind direction considered here.

\section{Data Availability}

The data produced and used in this publication are openly available through the research data repository of the Finnish Meteorological Institute \citep{Keskinen_Hellsten_2025}.

\section*{Acknowledgements}
  The research is a part of the RESPONSE project.
  This project has received funding from the European Union’s Horizon 2020 research and innovation programme under grant agreement No 957751.
  This research has also received funding from the European Union’s Horizon 2020 research and innovation programme under the Marie Skłodowska-Curie grant agreement No 872931 through the YADES project.
  The authors wish to acknowledge CSC – IT Center for Science, Finland, for generous computational resources.
  The authors also acknowledge the City of Turku and the National Land Survey of Finland for providing us with surface elevation data.

\bibliographystyle{spbasic_updated}     
\bibliography{bib} 

\begin{thebibliography}{46}
\providecommand{\natexlab}[1]{#1}
\providecommand{\url}[1]{{#1}}
\providecommand{\urlprefix}{URL }
\expandafter\ifx\csname urlstyle\endcsname\relax
  \providecommand{\doi}[1]{DOI~\discretionary{}{}{}#1}\else
  \providecommand{\doi}{DOI~\discretionary{}{}{}\begingroup
  \urlstyle{rm}\Url}\fi
\providecommand{\eprint}[2][]{\url{#2}}

\bibitem[{Ahn et~al.(2013)Ahn, Lee, and Sung}]{ahn_statistics_2013}
Ahn J, Lee JH, Sung HJ (2013) Statistics of the turbulent boundary layers over
  {3D} cube-roughened walls. International Journal of Heat and Fluid Flow
  44:394--402, \doi{10.1016/j.ijheatfluidflow.2013.07.010}

\bibitem[{Ahola et~al.(2022)Ahola, Raatikainen, Alper, Keskinen, Kokkola,
  Kukkurainen, Lipponen, Liu, Nordling, Partanen, Romakkaniemi, Räisänen,
  Tonttila, and Korhonen}]{ahola_technical_2022}
Ahola J, Raatikainen T, Alper ME, Keskinen JP, Kokkola H, Kukkurainen A,
  Lipponen A, Liu J, Nordling K, Partanen AI, Romakkaniemi S, Räisänen P,
  Tonttila J, Korhonen H (2022) Technical note: {Parameterising} cloud base
  updraft velocity of marine stratocumuli. Atmospheric Chemistry and Physics
  22(7):4523--4537, \doi{10.5194/acp-22-4523-2022}

\bibitem[{Blackman et~al.(2015)Blackman, Perret, and
  Savory}]{blackman_effect_2015}
Blackman K, Perret L, Savory E (2015) Effect of upstream flow regime on street
  canyon flow mean turbulence statistics. Environmental Fluid Mechanics
  15(4):823--849, \doi{10.1007/s10652-014-9386-8}

\bibitem[{Chang and Hanna(2004)}]{chang_air_2004}
Chang JC, Hanna SR (2004) Air quality model performance evaluation. Meteorology
  and Atmospheric Physics 87(1):167--196, \doi{10.1007/s00703-003-0070-7}

\bibitem[{Cheng and Porté-Agel(2015)}]{cheng_adjustment_2015}
Cheng WC, Porté-Agel F (2015) Adjustment of {Turbulent} {Boundary}-{Layer}
  {Flow} to {Idealized} {Urban} {Surfaces}: {A} {Large}-{Eddy} {Simulation}
  {Study}. Boundary-Layer Meteorology 155(2):249--270,
  \doi{10.1007/s10546-015-0004-1}

\bibitem[{Chow and Street(2009)}]{chow_evaluation_2009}
Chow FK, Street RL (2009) Evaluation of {Turbulence} {Closure} {Models} for
  {Large}-{Eddy} {Simulation} over {Complex} {Terrain}: {Flow} over {Askervein}
  {Hill}. Journal of Applied Meteorology and Climatology 48(5):1050--1065,
  \doi{10.1175/2008JAMC1862.1}

\bibitem[{Dauxois et~al.(2021)Dauxois, Peacock, Bauer, Caulfield, Cenedese,
  Gorlé, Haller, Ivey, Linden, Meiburg, Pinardi, Vriend, and
  Woods}]{dauxois_confronting_2021}
Dauxois T, Peacock T, Bauer P, Caulfield CP, Cenedese C, Gorlé C, Haller G,
  Ivey GN, Linden PF, Meiburg E, Pinardi N, Vriend NM, Woods AW (2021)
  Confronting {Grand} {Challenges} in environmental fluid mechanics. Physical
  Review Fluids 6(2):020,501, \doi{10.1103/PhysRevFluids.6.020501}

\bibitem[{Deardorff(1980)}]{deardorff_stratocumulus-capped_1980}
Deardorff JW (1980) Stratocumulus-capped mixed layers derived from a
  three-dimensional model. Boundary-Layer Meteorology 18(4):495--527,
  \doi{10.1007/BF00119502}

\bibitem[{Efron and Tibshirani(1993)}]{efron_introduction_1993}
Efron B, Tibshirani RJ (1993) An {Introduction} to the {Bootstrap}. No.~57 in
  Monographs on {Statistics} and {Applied} {Probability}, Chapman
  {\textbackslash}\& Hall, New York, U.S.A.

\bibitem[{Eyring et~al.(2016)Eyring, Bony, Meehl, Senior, Stevens, Stouffer,
  and Taylor}]{eyring_overview_2016}
Eyring V, Bony S, Meehl GA, Senior C, Stevens B, Stouffer RJ, Taylor KE (2016)
  Overview of the {Coupled} {Model} {Intercomparison} {Project} {Phase} 6
  ({CMIP6}) experimental design and organisation. Geoscientific Model
  Development 9:1937--1958, \doi{10.5194/gmdd-8-10539-2015}

\bibitem[{Giometto et~al.(2017)Giometto, Christen, Egli, Schmid, Tooke, Coops,
  and Parlange}]{giometto_effects_2017}
Giometto MG, Christen A, Egli PE, Schmid MF, Tooke RT, Coops NC, Parlange MB
  (2017) Effects of trees on mean wind, turbulence and momentum exchange within
  and above a real urban environment. Advances in Water Resources 106:154--168,
  \doi{10.1016/j.advwatres.2017.06.018}

\bibitem[{Gropp et~al.(1999)Gropp, Lusk, and Skjellum}]{gropp_MPI_1999}
Gropp W, Lusk E, Skjellum A (1999) Using {MPI}: Portable parallel programing
  with the message passing interface, 2nd edn. MIT Press

\bibitem[{Hackbusch(1985)}]{hackbusch_MG_1985}
Hackbusch W (1985) Multigrid methods and applications, Springer, p 378

\bibitem[{Hagishima et~al.(2009)Hagishima, Tanimoto, Nagayama, and
  Meno}]{hagishima_aerodynamic_2009}
Hagishima A, Tanimoto J, Nagayama K, Meno S (2009) Aerodynamic {Parameters} of
  {Regular} {Arrays} of {Rectangular} {Blocks} with {Various} {Geometries}.
  Boundary-Layer Meteorology 132(2):315--337, \doi{10.1007/s10546-009-9403-5}

\bibitem[{Hanna et~al.(2006)Hanna, Brown, Camelli, Chan, Coirier, Hansen,
  Huber, Kim, and Reynolds}]{hanna_detailed_2006}
Hanna SR, Brown MJ, Camelli FE, Chan ST, Coirier WJ, Hansen OR, Huber AH, Kim
  S, Reynolds RM (2006) Detailed {Simulations} of {Atmospheric} {Flow} and
  {Dispersion} in {Downtown} {Manhattan}: {An} {Application} of {Five}
  {Computational} {Fluid} {Dynamics} {Models}. Bulletin of the American
  Meteorological Society 87(12):1713--1726, \doi{10.1175/BAMS-87-12-1713}

\bibitem[{Harman et~al.(2016)Harman, Böhm, Finnigan, and
  Hughes}]{harman_spatial_2016}
Harman IN, Böhm M, Finnigan JJ, Hughes D (2016) Spatial {Variability} of the
  {Flow} and {Turbulence} {Within} a {Model} {Canopy}. Boundary-Layer
  Meteorology 160(3):375--396, \doi{10.1007/s10546-016-0150-0}

\bibitem[{Harms et~al.(2011)Harms, Leitl, Schatzmann, and
  Patnaik}]{harms_validating_2011}
Harms F, Leitl B, Schatzmann M, Patnaik G (2011) Validating {LES}-based flow
  and dispersion models. Journal of Wind Engineering and Industrial
  Aerodynamics 99(4):289--295, \doi{10.1016/j.jweia.2011.01.007}

\bibitem[{Hellsten et~al.(2021)Hellsten, Ketelsen, Sühring, Auvinen, Maronga,
  Knigge, Barmpas, Tsegas, Moussiopoulos, and Raasch}]{hellsten_nested_2021}
Hellsten A, Ketelsen K, Sühring M, Auvinen M, Maronga B, Knigge C, Barmpas F,
  Tsegas G, Moussiopoulos N, Raasch S (2021) A nested multi-scale system
  implemented in the large-eddy simulation model {PALM} model system 6.0.
  Geoscientific Model Development 14(6):3185--3214,
  \doi{10.5194/gmd-14-3185-2021}

\bibitem[{Kanani et~al.(2014)Kanani, Träumner, Ruck, and
  Raasch}]{kanani_what_2014}
Kanani F, Träumner K, Ruck B, Raasch S (2014) What determines the differences
  found in forest edge flow between physical models and atmospheric
  measurements? – {An} {LES} study. Meteorologische Zeitschrift pp 33--49,
  \doi{10.1127/0941-2948/2014/0542}

\bibitem[{Karttunen et~al.(2020)Karttunen, Kurppa, Auvinen, Hellsten, and
  Järvi}]{karttunen_large-eddy_2020}
Karttunen S, Kurppa M, Auvinen M, Hellsten A, Järvi L (2020) Large-eddy
  simulation of the optimal street-tree layout for pedestrian-level aerosol
  particle concentrations – {A} case study from a city-boulevard. Atmospheric
  Environment: X 6:100,073, \doi{10.1016/j.aeaoa.2020.100073}

\bibitem[{Keskinen and Hellsten(2025)}]{Keskinen_Hellsten_2025}
Keskinen JP, Hellsten A (2025) Simulated data for the manuscript “ensembles
  in urban large eddy simulations with changing wind direction” by {K}eskinen
  and {H}ellsten. \doi{10.57707/FMI-B2SHARE.4EA51ED8864D465FB7D4C43B8BCE8972}

\bibitem[{Keskinen et~al.(2011)Keskinen, Vuorinen, and
  Larmi}]{keskinen_large_2011}
Keskinen JP, Vuorinen V, Larmi M (2011) Large {Eddy} {Simulation} of {Flow}
  over a {Valve} in a {Simplified} {Cylinder} {Geometry}. SAE Technical Paper
  pp 2011--01--0843

\bibitem[{Keskinen et~al.(2025)Keskinen, Auvinen, and Hellsten}]{ITM39}
Keskinen JP, Auvinen M, Hellsten A (2025) High-resolution case study of
  pollutant dispersion in an urban environment using large-eddy simulation. In:
  Mensink C, Mathur R, Arunachalam S (eds) Air Pollution Modeling and its
  Application XXIX, Springer, Springer Proceedings in Complexity, pp 209--215,
  \doi{10.1007/978-3-031-70424-6\_26}

\bibitem[{Khan et~al.(2021)Khan, Banzhaf, Chan, Forkel, Kanani-Sühring,
  Ketelsen, Kurppa, Maronga, Mauder, Raasch, Russo, Schaap, and
  Sühring}]{khan_development_2021}
Khan B, Banzhaf S, Chan EC, Forkel R, Kanani-Sühring F, Ketelsen K, Kurppa M,
  Maronga B, Mauder M, Raasch S, Russo E, Schaap M, Sühring M (2021)
  Development of an atmospheric chemistry model coupled to the {PALM} model
  system 6.0: implementation and first applications. Geoscientific Model
  Development 14(2):1171--1193, \doi{https://doi.org/10.5194/gmd-14-1171-2021}

\bibitem[{Kurppa et~al.(2020)Kurppa, Roldin, Strömberg, Balling, Karttunen,
  Kuuluvainen, Niemi, Pirjola, Rönkkö, Timonen, Hellsten, and
  Järvi}]{kurppa_sensitivity_2020}
Kurppa M, Roldin P, Strömberg J, Balling A, Karttunen S, Kuuluvainen H, Niemi
  JV, Pirjola L, Rönkkö T, Timonen H, Hellsten A, Järvi L (2020) Sensitivity
  of spatial aerosol particle distributions to the boundary conditions in the
  {PALM} model system 6.0. Geoscientific Model Development 13(11):5663--5685,
  \doi{https://doi.org/10.5194/gmd-13-5663-2020}

\bibitem[{Leonardi and Castro(2010)}]{leonardi_channel_2010}
Leonardi S, Castro IP (2010) Channel flow over large cube roughness: a direct
  numerical simulation study. Journal of Fluid Mechanics 651:519--539,
  \doi{10.1017/S002211200999423X}

\bibitem[{Letzel et~al.(2008)Letzel, Krane, and Raasch}]{letzel_high_2008}
Letzel MO, Krane M, Raasch S (2008) High resolution urban large-eddy simulation
  studies from street canyon to neighbourhood scale. Atmospheric Environment
  42(38):8770--8784, \doi{10.1016/j.atmosenv.2008.08.001}

\bibitem[{Letzel et~al.(2012)Letzel, Helmke, Ng, An, Lai, and
  Raasch}]{letzel_les_2012}
Letzel MO, Helmke C, Ng E, An X, Lai A, Raasch S (2012) {LES} case study on
  pedestrian level ventilation in two neighbourhoods in {Hong} {Kong}.
  Meteorologische Zeitschrift pp 575--589, \doi{10.1127/0941-2948/2012/0356}

\bibitem[{Leutbecher and Palmer(2008)}]{leutbecher_ensemble_2008}
Leutbecher M, Palmer TN (2008) Ensemble forecasting. Journal of Computational
  Physics 227(7):3515--3539, \doi{10.1016/j.jcp.2007.02.014}

\bibitem[{Li and Giometto(2023)}]{li_mean_2023}
Li W, Giometto MG (2023) Mean flow and turbulence in unsteady canopy layers.
  Journal of Fluid Mechanics 974:A33, \doi{10.1017/jfm.2023.801}

\bibitem[{Lin et~al.(2021)Lin, Khan, Katurji, Bird, Faria, and
  Revell}]{lin_wrf4palm_2021}
Lin D, Khan B, Katurji M, Bird L, Faria R, Revell LE (2021) {WRF4PALM} v1.0: a
  mesoscale dynamical driver for the microscale {PALM} model system 6.0.
  Geoscientific Model Development 14(5):2503--2524,
  \doi{10.5194/gmd-14-2503-2021}

\bibitem[{Maronga and Raasch(2013)}]{maronga_large-eddy_2013}
Maronga B, Raasch S (2013) Large-{Eddy} {Simulations} of {Surface}
  {Heterogeneity} {Effects} on the {Convective} {Boundary} {Layer} {During} the
  {LITFASS}-2003 {Experiment}. Boundary-Layer Meteorology 146:17--44,
  \doi{10.1007/s10546-012-9748-z}

\bibitem[{Maronga et~al.(2015)Maronga, Gryschka, Heinze, Hoffmann,
  Kanani-Sühring, Keck, Ketelsen, Letzel, Sühring, and
  Raasch}]{maronga_parallelized_2015}
Maronga B, Gryschka M, Heinze R, Hoffmann F, Kanani-Sühring F, Keck M,
  Ketelsen K, Letzel MO, Sühring M, Raasch S (2015) The {Parallelized}
  {Large}-{Eddy} {Simulation} {Model} ({PALM}) version 4.0 for atmospheric and
  oceanic flows: model formulation, recent developments, and future
  perspectives. Geoscientific Model Development 8(8):2515--2551,
  \doi{10.5194/gmd-8-2515-2015}

\bibitem[{Maronga et~al.(2020)Maronga, Banzhaf, Burmeister, Esch, Forkel,
  Fröhlich, Fuka, Gehrke, Geletič, Giersch, Gronemeier, Groß, Heldens,
  Hellsten, Hoffmann, Inagaki, Kadasch, Kanani-Sühring, Ketelsen, Khan,
  Knigge, Knoop, Krč, Kurppa, Maamari, Matzarakis, Mauder, Pallasch, Pavlik,
  Pfafferott, Resler, Rissmann, Russo, Salim, Schrempf, Schwenkel, Seckmeyer,
  Schubert, Sühring, Tils, Vollmer, Ward, Witha, Wurps, Zeidler, and
  Raasch}]{maronga_overview_2020}
Maronga B, Banzhaf S, Burmeister C, Esch T, Forkel R, Fröhlich D, Fuka V,
  Gehrke KF, Geletič J, Giersch S, Gronemeier T, Groß G, Heldens W, Hellsten
  A, Hoffmann F, Inagaki A, Kadasch E, Kanani-Sühring F, Ketelsen K, Khan BA,
  Knigge C, Knoop H, Krč P, Kurppa M, Maamari H, Matzarakis A, Mauder M,
  Pallasch M, Pavlik D, Pfafferott J, Resler J, Rissmann S, Russo E, Salim M,
  Schrempf M, Schwenkel J, Seckmeyer G, Schubert S, Sühring M, Tils Rv,
  Vollmer L, Ward S, Witha B, Wurps H, Zeidler J, Raasch S (2020) Overview of
  the {PALM} model system 6.0. Geoscientific Model Development
  13(3):1335--1372, \doi{10.5194/gmd-13-1335-2020}

\bibitem[{Munters et~al.(2016)Munters, Meneveau, and
  Meyers}]{munters_shifted_2016}
Munters W, Meneveau C, Meyers J (2016) Shifted periodic boundary conditions for
  simulations of wall-bounded turbulent flows. Physics of Fluids 28(2):025,112,
  \doi{10.1063/1.4941912}

\bibitem[{Park et~al.(2015)Park, Baik, and Lee}]{park_impacts_2015}
Park SB, Baik JJ, Lee SH (2015) Impacts of {Mesoscale} {Wind} on {Turbulent}
  {Flow} and {Ventilation} in a {Densely} {Built}-up {Urban} {Area}. Journal of
  Applied Meteorology and Climatology 54(4):811--824,
  \doi{10.1175/JAMC-D-14-0044.1}

\bibitem[{Patnaik et~al.(2007)Patnaik, Boris, Young, and
  Grinstein}]{patnaik_large_2007}
Patnaik G, Boris JP, Young TR, Grinstein FF (2007) Large {Scale} {Urban}
  {Contaminant} {Transport} {Simulations} {With} {Miles}. Journal of Fluids
  Engineering 129:1524, \doi{10.1115/1.2801368}

\bibitem[{Patrinos and Kistler(1977)}]{patrinos_numerical_1977}
Patrinos AAN, Kistler AL (1977) A numerical study of the {Chicago} lake breeze.
  Boundary-Layer Meteorology 12:93--123, \doi{10.1007/BF00116400}

\bibitem[{Pope(2000)}]{pope_turbulent_2000}
Pope SB (2000) Turbulent {Flows}. Cambridge University Press

\bibitem[{Sagaut(2006)}]{sagaut_large_2006}
Sagaut P (2006) Large eddy simulation for incompressible flows: an
  introduction, 3rd edn. Scientific computation, Springer, Berlin ; New York

\bibitem[{Salizzoni et~al.(2011)Salizzoni, Marro, Soulhac, Grosjean, and
  Perkins}]{salizzoni_turbulent_2011}
Salizzoni P, Marro M, Soulhac L, Grosjean N, Perkins RJ (2011) Turbulent
  {Transfer} {Between} {Street} {Canyons} and the {Overlying} {Atmospheric}
  {Boundary} {Layer}. Boundary-Layer Meteorology 141:393--414,
  \doi{10.1007/s10546-011-9641-1}

\bibitem[{Taylor(2001)}]{taylor_summarizing_2001}
Taylor KE (2001) Summarizing multiple aspects of model performance in a single
  diagram. Journal of Geophysical Research: Atmospheres 106:7183--7192,
  \doi{10.1029/2000JD900719}

\bibitem[{Tseng et~al.(2006)Tseng, Meneveau, and
  Parlange}]{tseng_modeling_2006}
Tseng YH, Meneveau C, Parlange MB (2006) Modeling {Flow} around {Bluff}
  {Bodies} and {Predicting} {Urban} {Dispersion} {Using} {Large} {Eddy}
  {Simulation}. Environmental Science \& Technology 40(8):2653--2662,
  \doi{10.1021/es051708m}

\bibitem[{Wicker and Skamarock(2002)}]{wicker_time-splitting_2002}
Wicker LJ, Skamarock WC (2002) Time-{Splitting} {Methods} for {Elastic}
  {Models} {Using} {Forward} {Time} {Schemes}. Monthly Weather Review
  130(8):2088--2097,
  \doi{10.1175/1520-0493(2002)130\textless2088:TSMFEM\textgreater2.0.CO;2}

\bibitem[{Williamson(1980)}]{williamson_low-storage_1980}
Williamson JH (1980) Low-storage {Runge}-{Kutta} {Schemes}. Journal of
  Computational Physics 35:48--56

\bibitem[{Xie and Castro(2006)}]{xie_les_2006}
Xie Z, Castro IP (2006) {LES} and {RANS} for {Turbulent} {Flow} over {Arrays}
  of {Wall}-{Mounted} {Obstacles}. Flow, Turbulence and Combustion 76(3),
  \doi{10.1007/s10494-006-9018-6}

\end{thebibliography}

\end{document}